%% file: main.tex
\documentclass[journal]{IEEEtran}
\ifCLASSINFOpdf
\usepackage[pdftex]{graphicx}
\else
\usepackage[dvips]{graphicx}
\fi
\IEEEoverridecommandlockouts
\usepackage{cite}
\usepackage{amsmath,amssymb,amsfonts}
\usepackage{algorithmic}
\usepackage{graphicx}
\usepackage{textcomp}
\usepackage{xcolor}
\def\BibTeX{{\rm B\kern-.05em{\sc i\kern-.025em b}\kern-.08em
    T\kern-.1667em\lower.7ex\hbox{E}\kern-.125emX}}
\usepackage{placeins}
\usepackage{multirow}
\usepackage[ruled,vlined]{algorithm2e}

\usepackage{makecell}

\ifCLASSOPTIONcompsoc
\usepackage[caption=false,font=normalsize,labelfon
t=sf,textfont=sf]{subfig}
\else
\usepackage[caption=false,font=footnotesize]{subfig}
\fi

\IEEEaftertitletext{\vspace{-2\baselineskip}}

\begin{document}
\title{Hybrid ASIC-FPAA Fabric for Performance Security Trade-off}

\author{Ziyi Chen,
        Vaibhav Venugopal Rao,~\IEEEmembership{Student Member,~IEEE,}
        Kyle Juretus,~\IEEEmembership{Member,~IEEE,} \\
        and  Ioannis Savidis,~\IEEEmembership{Senior Member,~IEEE}
        }

\maketitle

\begin{abstract}
In this paper, a hybrid architecture that combines an application specific integrated circuit (ASIC) and a field-programmable analog array (ASIC-FPAA)  is proposed to address the performance-security trade-off in analog circuits. A programmable transistor pair (TP) is developed that obfuscates the topology of an analog circuit within an array of transistor pairs while minimizing the degradation in circuit performance due to the implemented security features. In addition, a technique that utilizes dummy transistor pairs is proposed that further masks the topology of a circuit by obfuscating the number of active transistors that comprise the target analog circuit.
A method to obfuscate the entire topology of an analog circuit is also developed. 
The FPAA provides the highest-level of security robustness as the entire circuit topology is programmed onto the transistor pair array, while trading-off performance.
 For large analog circuits requiring multiple configurable analog blocks (CABs), an algorithm is developed that identifies and maps the sub-blocks of the circuit best suited for obfuscation on the FPAA fabric, while accounting for the trade-off between performance and security. For the delta-sigma ($\Delta$$\Sigma$) modulator, the comparator module is selected for implementation on the FPAA fabric, which results in a reduction of less than 0.25\% in the effective number of bits (ENOB) and an increase of 28.1\% in the power consumption as compared to an unobfuscated $\Delta$$\Sigma$ modulator.
The security robustness of the proposed techniques are evaluated against the latest attack methodologies, which include the  satisfiability modulo theory (SMT) attack, the genetic algorithm attack, the monotonic response attack, and the direct current (DC) nodal analysis attack. 
A metric that evaluates the complexity of the topology after obfuscation is also proposed.
Results indicate that the proposed techniques provide a 5x improvement in security as measured by a developed metric without compromising performance.

\end{abstract}


% \fi
\IEEEpeerreviewmaketitle

\section{Introduction}
\IEEEPARstart{T}{he} increasing reliance on analog and mixed-signal integrated circuits (ICs) in emerging applications including next generation automotive systems (driverless cars), Internet of Things (IoT) devices, and communication networks, has heightened concerns regarding the security of such hardware systems\cite{Attack_Metric}. Logic locking techniques\cite{LogicLock, Logic_mixed, Logic_RF, Logic_SAT} effectively obfuscate the functionality of a digital circuit and the digital circuitry within an analog and mixed-signal (AMS) circuit or a radio-frequency (RF) circuit. However, securing analog circuits is a challenge due to the sensitivity of the performance parameters to parasitic impedances\cite{Secure_keybased}.
Current methodologies to obfuscate analog circuits secure either device parameters including the width\cite{Secure_keybased}, length\cite{Secure_length}, and threshold voltage\cite{Secure_vth} of a transistor or the biasing voltage and/or current\cite{Secure_body, Secure_biasANN} of an analog circuit.

The field-programmable analog array (FPAA) delivers a promising solution to secure analog circuit implementation by providing a reconfigurable fabric that allows for the redaction of circuit components\cite{Attack_FPAA}. The FPAA obfuscates device parameters, biasing conditions, and circuit topologies, which protects against intellectual property (IP) piracy.
In this paper, a hybrid ASIC-FPAA architecture is introduced that obfuscates the topology of an analog/mixed-signal circuit while also preserving circuit performance. The proposed obfuscation technique utilizes reconfigurable transistor pairs (TPs), where the width of the transistors and the topology of a grouped transistor pair are masked. 
The implementation of dummy transistor pairs further enhances the  security robustness of the circuit by obfuscating the true number of transistor pairs utilized within an analog circuit.
The TP-based topology obfuscation and the dummy TPs  are utilized to implement and secure a folded-cascode operational amplifier (op amp), a StrongARM comparator, and a successive-approximation (SAR)
analog-to-digital converter (ADC).  The results from the analysis of the TP-based security technique implemented on the folded-cascode op amp, StrongARM comparator, and SAR ADC provide characterization of the security-performance trade-off when a different number of transistor pairs are used to obfuscate the circuit.
In addition, a technique is proposed to fully obfuscate a circuit topology within the programmable FPAA fabric.
The FPAA-based obfuscation of the circuit topology  is implemented on a folded-cascode op amp and a delta-sigma modulator.
The security robustness of obfuscated analog circuits is evaluated on the latest analog attack algorithms, with three circuits proposed as benchmarks to evaluate the developed security techniques.
The primary contributions of the paper include
\begin{enumerate}
\item A transistor pair (TP)-based obfuscation technique that efficiently masks the device and performance parameters of an analog circuit.
\item The development of an algorithm to select the TP(s) to obfuscate and to determine the maximum number of obfuscated TPs based on target performance metrics.
\item The development of an algorithm to implement dummy transistor pairs that obfuscate the true number of transistors used within an analog circuit.
\item The use of the FPAA fabric to fully obfuscate device parameters, biasing conditions, and circuit topology. Utilizing the programmable TP array of the FPAA, the topology is obfuscated by providing flexible transistor-level connectivity and tunable transistor widths. 
\item An evaluation of the security provided by the developed obfuscation techniques to analog attack algorithms that include a topology attack\cite{Attack_FPAA}, an SMT attack\cite{Attack_SMT}, a genetic algorithm (GA) attack\cite{Attack_Gen}, a monotonic attack\cite{Attack_Mon}, and a DC nodal analysis (DNA) attack\cite{Attack_DNA}.  Developed performance metrics that account for the complexity of the obfuscated circuit topology are evaluated.
\end{enumerate}

Unlike prior analog-security techniques that mainly protect a single class of information, the proposed approach establishes a hybrid ASIC-FPAA framework that supports both partial and full topology obfuscation. The novelty of this work lies not in the design of a programmable analog fabric, but in the development of security methodologies and evaluation criteria that use the FPAA as one component of a hybrid obfuscation framework. The proposed framework jointly secures transistor-pair topology and sizing, uses dummy transistor pairs to mask the number of active devices, and enables performance-aware selection between low-overhead partial obfuscation and stronger full-circuit FPAA-based obfuscation.

The rest of the paper is organized as follows. Some prior work on analog obfuscation is presented in Section~\ref{AA}. The assumed threat model is described in Section~\ref{section_threat model}. An overview of the proposed topology-obfuscation techniques is provided in Section~\ref{BB}. An algorithm to select transistor pairs to obfuscate and an algorithm to insert dummy transistor pairs are presented in Section~\ref{CC}. The use of an FPAA to obfuscate the entire topology of an analog circuit is described in Section~\ref{DD}. The results from the characterization of the TP-based obfuscation technique that utilizes dummy TPs are presented in Section~\ref{EE}. A discussion of the results from the characterization of the technique that fully obfuscates the topology of an analog circuit on the FPAA fabric is provided in Section~\ref{FF}. The analysis of the non-ideality, robustness, and yield  of the FPAA fabric is presented in Section~\ref{section-non-ideal}. The security robustness provided by the proposed techniques is evaluated against the latest analog attack algorithms in Section~\ref{GG}. A comparison with prior analog security techniques is provided in Section~\ref{section-comparison-table}. Some concluding remarks are provided in Section~\ref{HH}.

\input{section2}
\input{section3}

\input{section4}
\input{section5}
\input{section6}

\input{section7}
\input{section8}

\input{section9}

\appendices

\ifCLASSOPTIONcaptionsoff
  \newpage
\fi



%\bibliographystyle{IEEEtran}


% \end{thebibliography}

\bibliographystyle{IEEEtran}

\bibliography{bibliography}

\begin{IEEEbiography}[{\includegraphics[width=.9in,height=1in,clip,keepaspectratio=true]{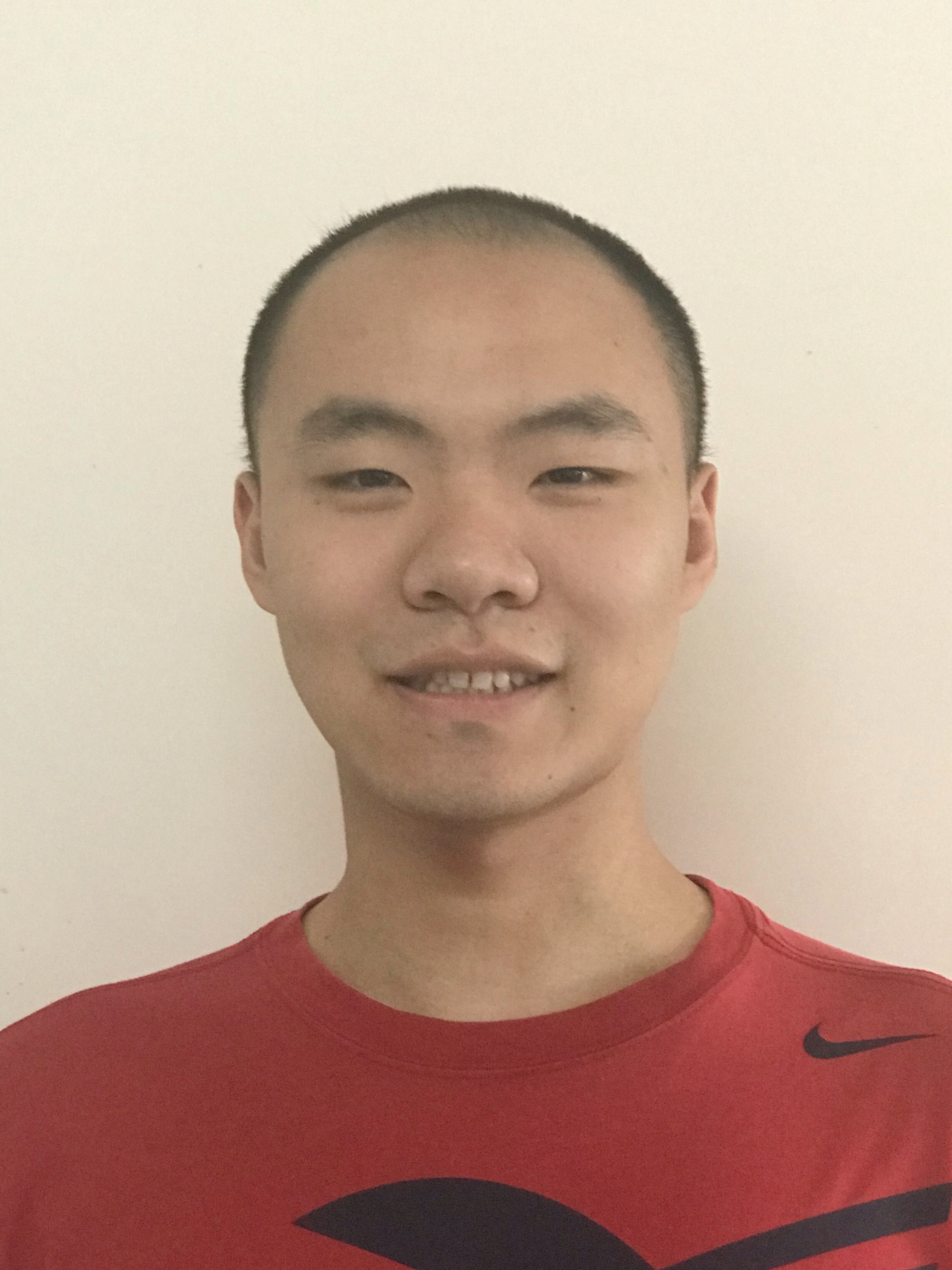}}]{Ziyi Chen}
received his Bachelor degree from the University of Electronic Science and Technology of China, Chengdu, in 2017 and his M.S. degree from Northeastern University, Boston, MA, in 2019. He  is  currently  pursuing  the  Ph.D.  degree from  the  Department  of  Electrical  and Computer Engineering, Drexel  University, Philadelphia, PA.  His research interests include field-programmable analog arrays (FPAAs), side-channel attacks on analog circuits, and analog obfuscation techniques using the FPAA.
\end{IEEEbiography}

\begin{IEEEbiography}[{\includegraphics[width=0.9in,height=1in,clip,keepaspectratio=true]{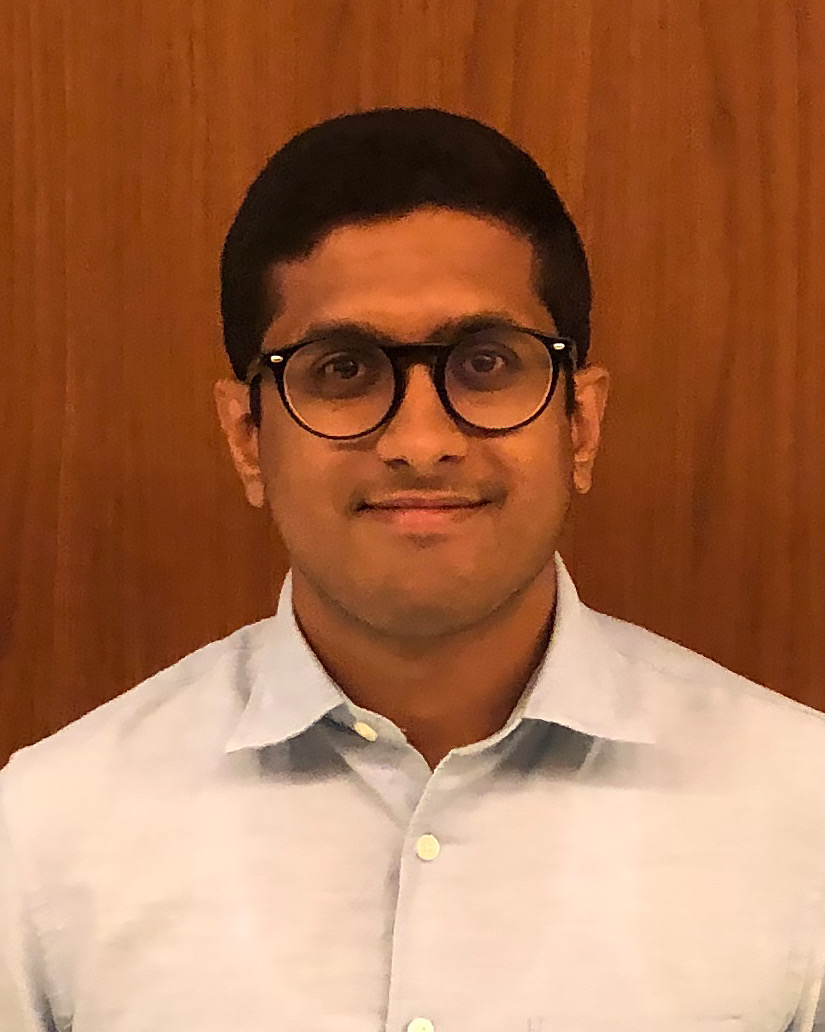}}]{Vaibhav Venugopal Rao}
received his B.E degree in electronics and communications from Visvesvaraya Technological University, India in 2015, and the M.Sc. and Ph.D. degrees from  the  Department  of  Electrical  and Computer Engineering, Drexel University, PA, USA, in 2017 and 2024, respectively. His current research interests include circuit level techniques to protect analog IP and variation aware automation of analog circuit design to reduce design time.
\end{IEEEbiography}

\begin{IEEEbiography}
[{\includegraphics[width=1in,height=1.1in,clip,keepaspectratio=true]{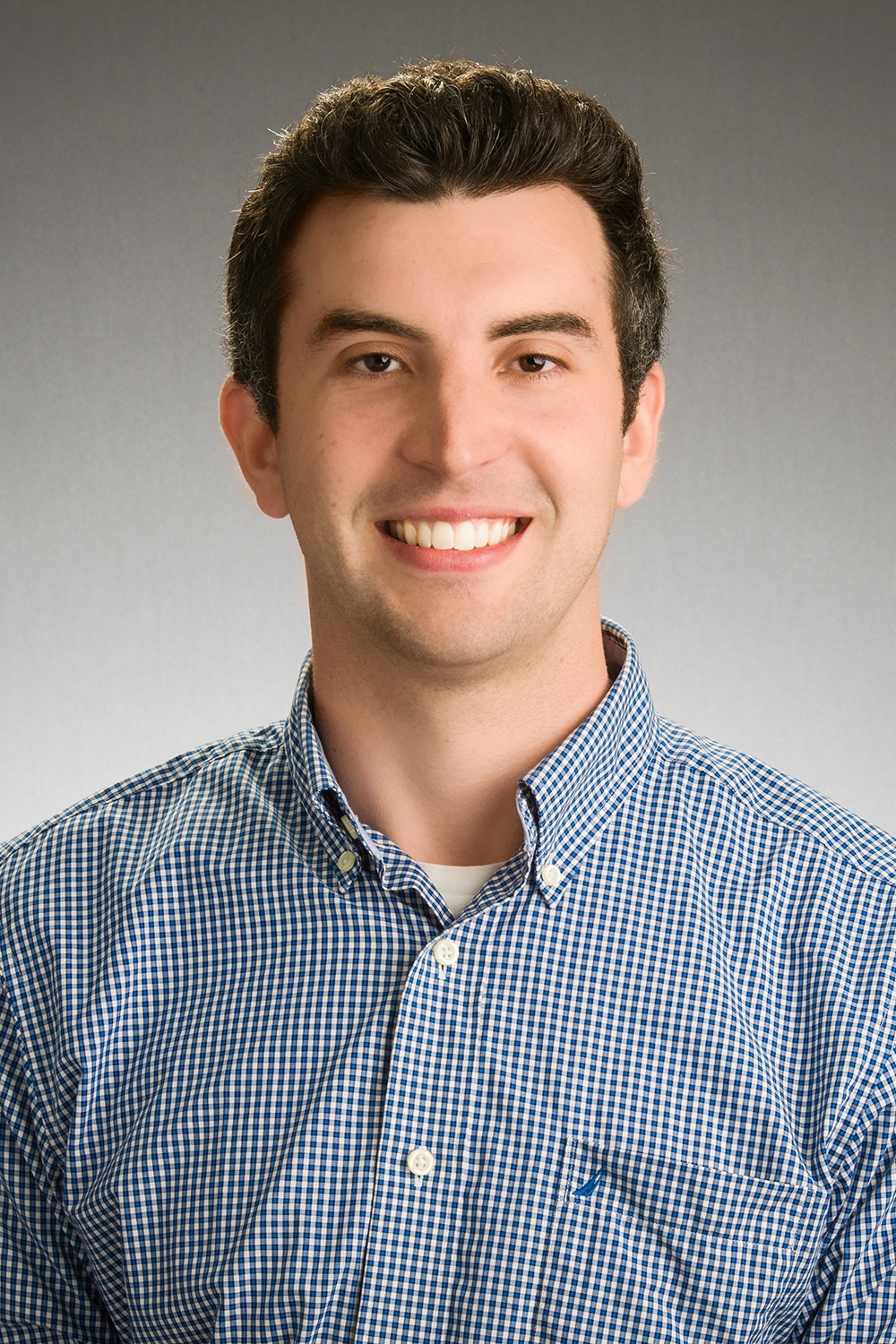}}]{Kyle Juretus} 
(Member, IEEE) Dr. Kyle Juretus is an assistant professor in the Electrical and Computer engineering department at Villanova University. He received his Ph.D. in Electrical Engineering from Drexel University in 2020. During his Ph.D. studies, Dr. Juretus was a National Defense Science and Engineering (NDSEG) fellow sponsored by Air Force Research Laboratory (AFRL) from 2016-2019. His research focuses on integrated circuit (IC) design with the objective to create faster, smaller, more efficient, and more secure ICs. To enable such changes with minimal design time overhead, Dr. Juretus focuses on the development of next generation electronic design automation (EDA) tools.  
\end{IEEEbiography}

\begin{IEEEbiography}[{\includegraphics[width=1.1in,height=1.1in,clip,keepaspectratio=true]{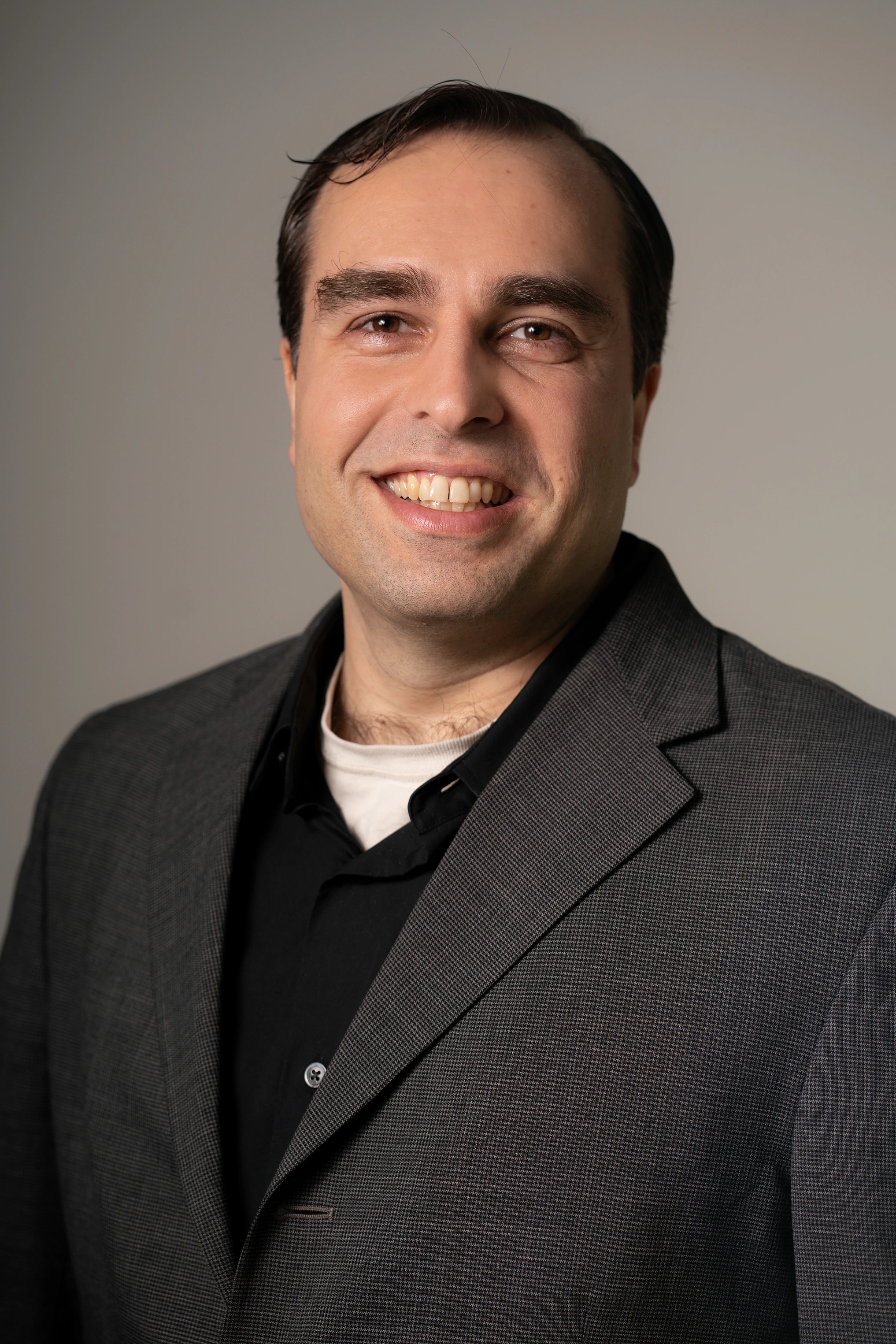}}]{Ioannis Savidis}
    (S'03-M'13-SM'18) received the B.S.E. degree in electrical and computer engineering and biomedical engineering from Duke University, Durham, NC, USA, in 2005, and the M.Sc. and Ph.D. degrees in electrical and computer engineering from the University of Rochester, Rochester, NY, USA, in 2007 and 2013, respectively.

He joined the Department of Electrical and Computer Engineering, Drexel University, Philadelphia, PA, USA, in 2013, where he is currently an Associate Professor and directs the Integrated Circuits and Electronics (ICE) Design and Analysis Laboratory. His current research interests include analysis, modeling, and design methodologies for high performance digital and mixed-signal integrated circuits, power management for SoC and microprocessor circuits, hardware security, including digital and analog obfuscation and side-channel analysis, and electrical and thermal modeling and characterization, signal and power integrity, and power and clock delivery for heterogeneous 2-D and 3-D circuits.

Dr. Savidis is a senior member of the Institute of Electrical and Electronic Engineers, and a member of the Association of Computing Machinery, the IEEE Circuits and Systems Society, the IEEE Communications Society, and the IEEE Electron Devices Society. 
He is a recipient of the 2018 National Science Foundation Early Faculty (CAREER) Award. 
He serves on the organizing committees of the IEEE International Symposium on Hardware Oriented Security and Trust~(HOST) and the IEEE International Symposium on Circuits and Systems~(ISCAS), and the steering committee of the ACM Great Lakes Symposium on VLSI~(GLSVLSI). 
He also serves on the editorial boards of the IEEE Transactions on Very Large Scale Integration Systems, the Microelectronics Journal, and the ACM Transactions on Design Automation of Electronic Systems. 
\end{IEEEbiography}

\end{document}

%% file: section2.tex
\section{Prior Work}\label{AA}
An overview of prior work in analog obfuscation techniques is presented in this section.
Logic locking techniques  are utilized in \cite{Logic_mixed,Logic_RF}, and \cite{Logic_mixed2} to obfuscate the digital blocks of an analog and mixed-signal (AMS) or RF circuit.
In \cite{Logic_mixed} and \cite{Logic_mixed2}, MixLock is proposed to lock the digital circuitry of a mixed-signal IC using stripped-functionality logic locking (SFLL). Although MixLock ensures seamless integration without affecting the performance of an AMS circuit, the analog portion of the circuit remains unsecured.
In \cite{Logic_RF}, SyncLock is proposed to protect an RF transceiver by locking the synchronization mechanism between the transmitter and receiver. A hard-coded error is embedded in the transmitted data frame, while a key-controlled correction mechanism is applied to restore accurate synchronization of the data.

Techniques that leverage algorithms for secure calibration\cite{Secure_calibration} and threshold voltage manipulation \cite{Secure_vth} have been proposed that require no additional key-based locking circuitry to secure an analog circuit.
In \cite{Secure_calibration}, the inherent programmability of an analog circuit is leveraged by treating the configuration of the biasing voltage and current as the secret key, which eliminates the need for additional locking circuitry. However, the technique is only applicable to AMS ICs with embedded programmability and calibration mechanisms, which limits the use of the locking technique to fixed-function analog circuits.
% In \cite{Secure_layout}, a key-less obfuscation methodology is introduced to mask the layout of an analog circuit by incorporating misleading (false) elements, which include inactive components, vias, and interconnects, during the last-level edit phase of fabrication. The technique prevents adversaries from extracting the correct circuit functionality while allowing the IP owner to restore the intended performance with focused ion beam (FIB) circuit editing, where false components are disconnected from the active circuits.
A technique  is proposed in \cite{Secure_vth} that utilizes multi-threshold voltage (VTH) to obfuscate the performance characteristics of a wide-swing cascode amplifier and a transconductance–capacitor (Gm-C) filter. By leveraging the sensitivity of analog circuit performance to variation in transistor threshold voltage (VTH), the transistors of the circuit with nominal threshold voltage (NVT) are replaced with low-VTH (LVT) and high-VTH (HVT) variants. The performance of the circuit meets specifications only when the correct VTH configuration is applied.

High-level synthesis (HLS) based methods have also been proposed for hardware security and for efficient circuit generation\cite{HLS1,HLS2,HLS3}. 
In \cite{HLS1}, functional locking through omission is introduced, where a portion of a behavioral design is extracted and programmed onto an embedded FPGA, with the eFPGA bitstream acting as the locking key. The use of HLS allows for the optimization of the secure circuit by splitting the circuit between the ASIC and eFPGA in order to reduce overhead while maintaining the original timing constraints.
In \cite{HLS2}, a graph analytics based high-level synthesis framework is proposed, where a compiler-assisted dependency graph is constructed using an intermediate representation from a low level virtual machine (LLVM) and optimized to generate efficient hardware accelerators from high-level programs.
In \cite{HLS3}, SHiELD is introduced, which is an obfuscation technique that utilizes HLS to secure digital signal processing (DSP) cores using a one-way random function, reconfigurable composite switching obfuscation cells, and security-aware design-space exploration. The method secures components of DSP cores generated by HLS against reverse engineering attacks.

Several key-based obfuscation techniques \cite{Secure_memristors,Attack_FPAA,Secure_body,Secure_keybased,Secure_Shared} have been proposed, where a key bitstream is required to enable the correct functionality of the locked circuit.
In \cite{Secure_memristors}, a memristive crossbar-based programming circuit is developed to obfuscate the operation of a sense amplifier. The method leverages memristors as security primitives by tuning the body bias of the transistors of the sense amplifier.
In \cite{Attack_FPAA}, an FPAA is used to mask the topology of analog circuits. The FPAA consists of a 6 x 6 matrix of configurable analog blocks (CABs), where each CAB is programmed into different circuit elements. The technique masks the topology of an analog circuit within the fixed FPAA fabric.
In \cite{Secure_body}, the body bias voltage of the transistors is obfuscated, with the correct key formed by concatenating the digital input codes of the DACs providing the body bias voltages.
Key-based obfuscation\cite{Secure_keybased}, including vector-based and mesh-based approaches, are proposed to mask the width and length of transistors. The correct performance and biasing voltage or current is only achieved when the correct key is applied.
In \cite{Secure_Shared}, a methodology is introduced that utilizes shared key dependencies, where the analog and digital blocks are jointly secured through interlinked key based circuit functionality. By establishing functional and behavioral dependencies between the analog and digital domains, the approach increases the key space an adversary must search and forces an attacker to consider both domains simultaneously.

% A DC nodal analysis (DNA) attack\cite{Attack_DNA} is proposed to determine the correct locked parameters using Kirchhoff’s Voltage Law (KVL) and Kirchhoff’s Current Law (KCL). DNA attack requires only the circuit netlist and the DC input-output response of an oracle IC.

\section{Threat Model}\label{section_threat model}
An untrusted foundry model is assumed, where an adversary has access to the fabricated integrated circuit and to the general architecture of the proposed TP-based and FPAA-based obfuscation techniques, including the set of possible transistor-pair topologies and the structure of the configurable analog blocks. The adversary is further assumed to possess reverse-engineering capabilities sufficient to identify the fixed fabric and device locations, while also having access to the I/O ports of the FPAA fabric. In addition, the adversary is capable of executing algorithmic attacks, which include the SMT-based, GA-based, monotonic, DNA, and topology attacks. However, the adversary does not have access to the correct programming bitstream, the keys that set the topology of the TP, or the keys that set the size of transistors. 
The location of active and dummy transistor pairs is also unknown to the adversary. For partial obfuscation, the attacker seeks to identify the obfuscated transistor pairs and recover a specification-compliant equivalent configuration. For full FPAA-based obfuscation, the attacker must jointly infer TP functionality, sizing, routing, and sub-block mapping within the programmable fabric. An attack is considered successful if the adversary recovers an equivalent circuit implementation that satisfies the target circuit specifications within the defined performance tolerance. The security objective of the proposed approach is, therefore, to maximize the effort required to recover a functionally equivalent and specification-compliant analog implementation.

%% file: section3.tex
\section{Techniques that Obfuscate Topology}\label{BB}
Two obfuscation techniques are developed, one that partially masks and another that fully masks the topology of an analog circuit. Considering the widespread use of differential-mode signaling in analog circuits, pairs of transistors are obfuscated using programmable transistor pairs. The programmable pairs allow for tunable transistor widths and topologies, which provides both improved security robustness and design flexibility when implementing partial topology obfuscation, as described in Section~\ref{CC}.
To account for the sensitivity of analog circuits to parasitic impedance, an algorithm is developed that optimally selects pairs of transistors to implement as programmable transistor pairs, which minimizes the degradation to the performance of the target circuit.
In addition, the programmable transistor pairs are utilized as a primary component of a field-programmable analog array, which provides full obfuscation of the topology of an analog circuit, as described in Section~\ref{DD}. The FPAA provides the highest level of security by obfuscating the entire topology of an analog circuit within the fixed structure of the programmable FPAA fabric. The remainder of this section describes the detailed structure of the programmable transistor pair utilized for partial topology obfuscation  in Section~\ref{BB_1}, and the architecture of the FPAA that fully masks the topology of an analog circuit in Section~\ref{BB_2}.

\subsection{Transistor Pair Topology Obfuscation}\label{BB_1}
\begin{figure}[htbp]
\centering
\subfloat[]{\includegraphics[width=.205\textwidth]{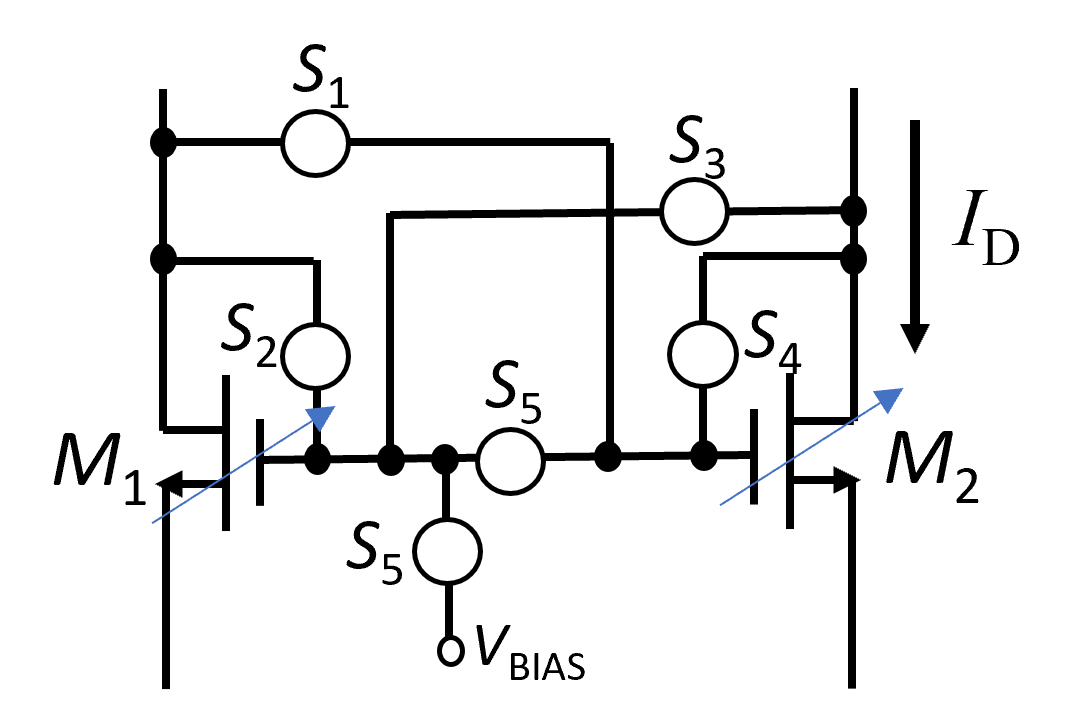}
\label{NP1_ARC}}
\hfil
\subfloat[]{\includegraphics[width=.202\textwidth]{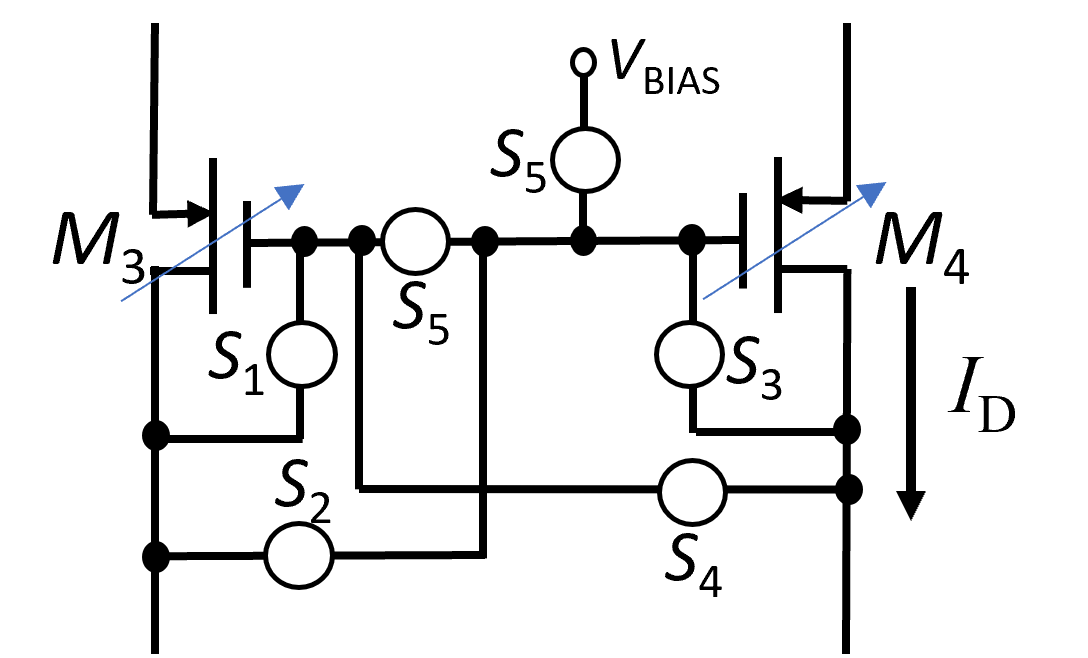}
\label{PP1_ARC}}
\hfil

\subfloat[]{\includegraphics[width=.45\textwidth]{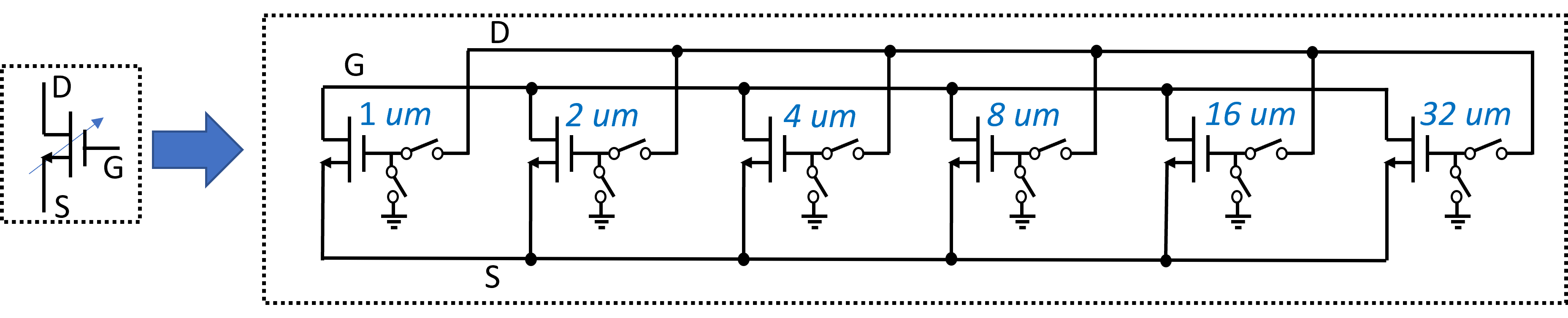}
\label{VT_ARC}}
\caption{Schematic representation of the (a) NMOS transistor pair (NP), (b) PMOS transistor pair (PP), and (c) reconfigurable transistor with a tunable width of 1 {\textmu}m to 63 {\textmu}m. The PMOS and NMOS transistor pair utilize switches \emph{S}\textsubscript{1} to \emph{S}\textsubscript{5} for topology configuration (current direction is also shown).}
\label{Program_TP1}
\vspace{-15pt}
\end{figure}

\begin{figure}[htbp]
\centering
\subfloat[]{\includegraphics[width=.11\textwidth]{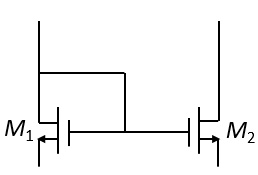}
\label{Graph_TP_first_case}}
\hfil
\subfloat[]{\includegraphics[width=.11\textwidth]{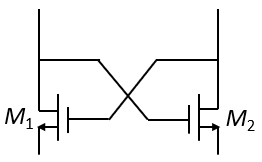}
\label{Graph_TP_second_case}}
\hfil
\subfloat[]{\includegraphics[width=.11\textwidth]{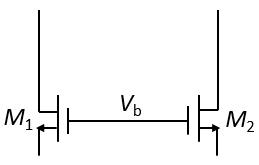}
\label{Graph_TP_third_case}}
\hfil
\subfloat[]{\includegraphics[width=.11\textwidth]{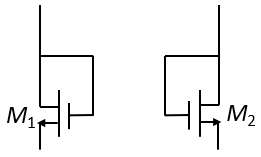}
\label{Graph_TP_fourth_case}}
\hfil
\caption{Possible topologies implemented with a single NMOS transistor pair include the (a) current mirror, (b) regeneration loop, (c) current source pair, and (d) diode-connected pair. The same topologies are implemented with PMOS transistor pairs.}
\label{Topology_TP}
\vspace{-15pt}
\end{figure}

The detailed structure of programmable transistor pairs is shown in Fig.~\ref{Program_TP1}. To secure the topology of a pair of NMOS transistors, a programmable NMOS transistor pair as shown in Fig.~\ref{NP1_ARC} is developed.
Switches \emph{S}\textsubscript{1} to \emph{S}\textsubscript{5} configure transistors \emph{M}\textsubscript{1} and \emph{M}\textsubscript{2} into four possible topologies, which include a current mirror, a regeneration loop, a current source, and a diode-connected pair, as illustrated in Fig.~\ref{Topology_TP}. 
% Switches \emph{S}\textsubscript{6} and \emph{S}\textsubscript{7} merge the current flow from both branches into the left branch, while switches \emph{S}\textsubscript{8} and \emph{S}\textsubscript{9} merge the current flow into the right branch. Additionally, switches \emph{S}\textsubscript{10} and \emph{S}\textsubscript{11} provide connections from the drain of transistor \emph{M}\textsubscript{1} to ground and the drain of transistor \emph{M}\textsubscript{2} to ground, respectively.
The switches are implemented as transmission gates, each controlled by a single-bit key. Consequently, each programmable transistor pair requires a 5-bit key that sets the topology to implement. In addition, transistors \emph{M}\textsubscript{1} and \emph{M}\textsubscript{2} provide tunable widths, which are implemented using six transistors with binary-weighted widths connected in parallel, as shown in Fig.~\ref{VT_ARC}. The width of a transistor is programmed between 1 {\textmu}m and 63 {\textmu}m using a six bit key. As a result, each programmable transistor pair requires a total of 17 key bits.
The function that sets the target topology of a programmable transistor pair is, therefore, given by
\begin{align}
f_p = \Phi(K_T,K_S),
\end{align}
where $\Phi$ defines the programmed analog function of a transistor pair, $K_T$ represents the 5-bit key that sets the topology, and $K_S$ represents the 12-bit key that sets the width of the programmable transistor pair.

\subsection{Topology Obfuscation Using an FPAA}\label{BB_2}
\begin{figure}[htbp]
\centering
\subfloat[]{\includegraphics[width=.35\textwidth]{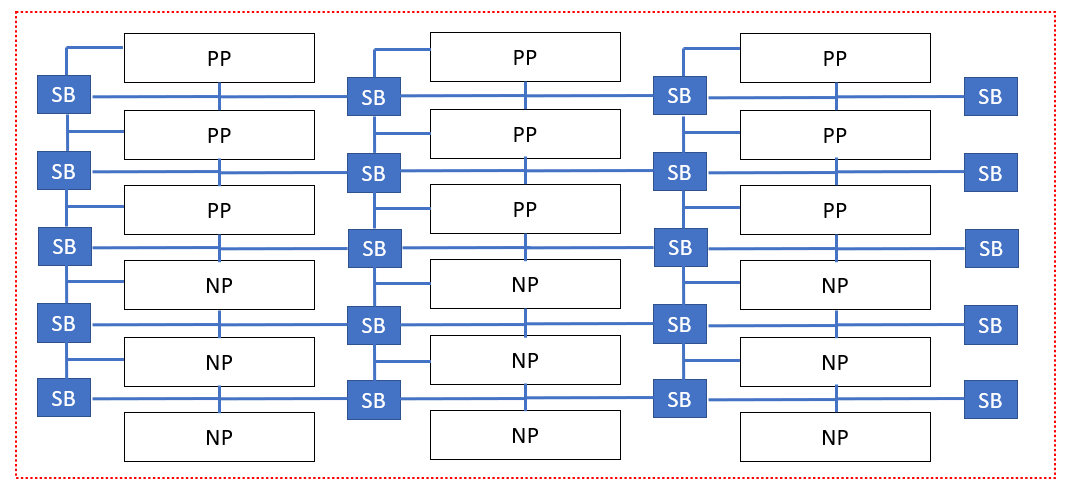}
\label{TP_ARC}}
\vspace{-10pt}
\hfil
\subfloat[]{\includegraphics[width=.23\textwidth]{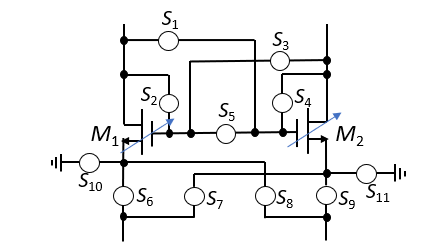}
\label{NP_ARC}}
\hfil
\subfloat[]{\includegraphics[width=.23\textwidth]{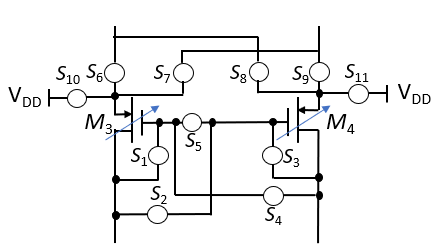}
\label{PP_ARC}}
\hfil
\caption{Structure of the (a) TP array of a single CAB that includes a 6 x 3 matrix of transistor pairs, (b) NMOS transistor pair NP that includes two NMOS transistors with tunable widths, eleven routing switches implemented as transmission gates, and routing paths to GND and (c) PMOS transistor pair PP that includes two PMOS transistors with tunable widths, eleven routing switches implemented as transmission gates, and routing paths to \emph{V}\textsubscript{DD}.}
\label{TP_ALL}
\end{figure}

To fully mask the topology of an analog circuit, an FPAA-based topology obfuscation technique is developed. The configurable analog block (CAB) of the FPAA\cite{FPAA_TP,FPAA_TP2} consists of a 6 x 3 transistor pair (TP) array, as shown in Fig.~\ref{TP_ARC}.
Each programmable transistor pair within the TP array includes six additional switches beyond the five used to program the transistor pairs, as shown in Fig.~\ref{NP_ARC}. Switches \emph{S}\textsubscript{6} and \emph{S}\textsubscript{7} merge the DC current into the left current branch, while switches \emph{S}\textsubscript{8} and \emph{S}\textsubscript{9} merge the current into the right current branch. In addition, switches \emph{S}\textsubscript{10} and \emph{S}\textsubscript{11} connect the source terminal of the NMOS transistor to ground and the source terminal of the PMOS transistor to \emph{V}\textsubscript{DD}.
The programmed functionality of the transistor pairs within the TP array is given by
\begin{align}
f_{tp} = \Theta(K_{TP},K_{SP}),
\end{align}
where $\Theta$ defines the analog functionality programmed on the TP array, $K_{TP}$ represents the 11-bit key that programs a topology on a transistor pair, and $K_{SP}$ represents the 12-bit key that sets the width of the two transistors (6-bit key for each).
In addition, each switch box (SB) includes 12 transmission-gate-based routing switches that enable connections between any two physical directions (xy-routing) in differential mode. The connections of the SB are defined by
\begin{align}
f_{SB} = \Psi(K_{SB}),
\end{align}
where $\Psi$ represents the function that programs the SB based on the 12-bit key \emph{K}\textsubscript{SB}.

Considering that analog circuits include a DC current path from \emph{V}\textsubscript{DD} to GND, the number of transistors that can be stacked vertically along the path is limited due to the voltage drop across the drain and source terminals of each stacked device (transistors and passives). Therefore, the proposed TP array allows a maximum of six transistor pairs to be stacked along the DC current path, which makes the 6 x 3 TP array sufficient to implement most small analog circuits.
A single CAB is, therefore, sufficient to implement  circuits that include operational amplifiers (op amps), comparators, and integrators. For larger analog circuits and complex analog systems, multiple TP arrays are required, which are pooled by utilizing multiple CABs.

The FPAA provides improved security robustness and allows for the obfuscation of the entire topology of an analog circuit or system. Since a TP array consists of 18 TPs and 20 SBs, the functionality set by the TP array is given by
\begin{align}
f_{Array} = (\prod_{i=1}^{18}\Theta_i(K_T,K_S))(\prod_{j=1}^{20}\Psi_j(K_{SB})),
\end{align}
where  \emph{i} and \emph{j} represent the index of the transistor pairs and switch boxes of the array, respectively.

The adversary must, therefore, determine the functionality of all the transistor pairs and switch boxes to decrypt the topology of the circuit implemented on the TP array. 
The search space when executing a brute force attack that iterates through all key combinations is 2\textsuperscript{18*23+12*20}, where 18 denotes the number of programmable transistor pairs in the TP array, 23 denotes the key length for programming each transistor pair (11-bit key that sets the topology and 12-bit key to set the width of the two transistors of each pair), 12 denotes the key length for programming each switch box, and 20 denotes the number of switch boxes in the TP array. Therefore, performing a brute-force attack is computationally infeasible.
In addition, for an analog system utilizing multiple CABs, the attacker must first determine the functionality of each individual CAB and then verify the performance of the overall circuit. Considering an analog circuit consisting of
\emph{N} sub-blocks, where each sub-block is mapped onto a TP array, the functionality of the system is given by
\begin{align}
f_{System} = \prod_{num=1}^{N}f_{Array_{num}},
\end{align}
where \emph{N} represents the number of CABs that are utilized.

%% file: section4.tex
\section{Partial Obfuscation of A Circuit Topology}\label{CC}
Obfuscating the topology of an analog circuit presents challenges due to the sensitivity of analog circuits to the parasitic impedance of internal nodes. 
An obfuscation technique is developed to effectively mask the topology of an analog circuit while meeting performance requirements.
An algorithm is developed that maximizes the number of obfuscated transistor pairs while limiting the degradation in the performance of the circuit to a set tolerable level.
The initial transistor sizes are set based on the unobfuscated version of the target circuit. When a transistor pair is replaced by a programmable transistor pair, the pair is first configured to match the original topology, and the transistor widths are set to the closest realizable width supported by the 6-bit binary-weighted programmability of the FPAA fabric.
In addition, a technique that utilizes dummy transistor pairs is proposed that masks the true number of stacked transistors between \emph{V}\textsubscript{DD} and GND.

\subsection{Selection of Transistor Pairs for Obfuscation}\label{section-dummy-algo}
Although analog circuits are sensitive to parasitic impedance, the effect of parasitics on different transistors and transistor pairs varies. The offset in circuit performance is most significant due to the effect of parasitic impedances on the input and output transistors of a circuit, while parasitic effects on transistors that provide DC voltage or current biasing typically result in less degradation in circuit performance.
Therefore, a selection algorithm is proposed to identify the transistors and transistor pairs best suited for obfuscation, which maximizes the security robustness of the circuit while minimizing the degradation in performance. The pseudocode for selecting transistors for partial obfuscation is provided as Algorithm~\ref{Partial}.

The algorithm first parses the netlist of the circuit, identifies transistor pairs, and extracts a list of transistor pairs, \emph{L}\textsubscript{pair}. Each transistor pair, \emph{T}\textsubscript{pair}, is then replaced with a PMOS or NMOS programmable transistor pair, which are shown in Fig.~\ref{NP1_ARC} and Fig.~\ref{PP1_ARC}, respectively. The programmable transistor pair is configured to match the topology of the replaced pair of transistors.
A SPICE simulation is then performed to calculate the degradation in performance, \emph{L}\textsubscript{Pm}. By sorting \emph{L}\textsubscript{Pm} in ascending order, the transistor pairs that result in the least degradation in circuit performance are identified, and an ordered list of transistor pairs to obfuscate, \emph{O}\textsubscript{pair}, is determined.
The algorithm continues to add transistor pairs for obfuscation to \emph{O}\textsubscript{pair} until the  degradation in the performance of the circuit, as determined by SPICE simulation, reaches the set tolerance threshold, \emph{D}\textsubscript{T}. Since for each executing iteration of Algorithm~\ref{Partial} only one transistor pair is replaced by an obfuscated transistor pair, while the simulation load for a given iteration remains approximately equal, the overall runtime of the SPICE simulations scales linearly with the number of candidate transistor pairs found in a circuit.

\begin{algorithm}
\footnotesize
\setlength{\algomargin}{0.6em}
\SetAlCapSkip{0.2em}
\DontPrintSemicolon
\caption{Pseudocode describing the selection and order of transistor pairs to obfuscate}
\label{Partial}
\textbf{Input}: circuit netlist $\phi$, performance metric $P_m$, degradation tolerance $D_{T}$ \;
\textbf{Output}: list of selected transistor pairs $S_{pair}$ \;
$L_{pair}$= NetlistParse($\phi$)\;
$L_{Pm}$ = [] \;
\For{$T_{pair}$ in $L_{pair}$}{
$\phi_N$ = ReplaceTranPair($T_{pair}$) \;
$P_I$ = SpiceSim($\phi_N$) \;
$L_{Pm}$.append($\lvert P_I - P_m \rvert$) \;
}
$O_{Pm}, O_{pair}$ = SortAscending($L_{Pm}$, $L_{pair}$) \;
$O_{Pm}, O_{pair}$ = RemoveOverlapping($O_{Pm}, O_{pair}$) \;
$S_{pair}$ = [] \;
\For{i in range(len($O_{Pm}$))}{
$S_{pair}$.append(SelectTranPairs(i)) \;
$\phi_{obfus}$ = ReplaceTranPairs($S_{pair}$[i]) \;
$P_{obfus}$ = SpiceSim($\phi_{obfus}$) \;
\If{$(P_{obfus} - P_m)/P_m > D_{T}$}{
\textbf{return} $S_{pair}$[i-1] \;
}

}

\textbf{return} $S_{pair}$[len($S_{pair}$)]\;
\end{algorithm}

\subsection{Dummy Transistor Pairs}\label{Dummy_theory}
An obfuscation technique that utilizes dummy transistor pairs is proposed to mask the topology of an analog circuit by adding redundant programmable transistor pairs.
Dummy transistor pairs are introduced to create redundant off-state current paths with the same vertical stack depth as the primary current path, which masks both the true circuit topology and the number of active stages while increasing the effective key space.
The architecture of the dummy transistor pair is identical to the programmable transistor pair, which is shown in Fig.~\ref{Program_TP1}. The dummy transistor pairs remain off when the obfuscated analog circuit operates with the correct key.
The pseudocode for the insertion of dummy transistor pairs into an analog circuit is provided as Algorithm~\ref{Algo_Dummy}. 

Considering the DC current path from \emph{V}\textsubscript{DD} to GND, transistors directly connected to \emph{V}\textsubscript{DD} are labeled as part of the lowest vertical level, while transistors directly connected to GND are labeled as part of the highest vertical level. Consequently, the maximum vertical configuration of an analog circuit depends on the number of transistors along the DC current path from \emph{V}\textsubscript{DD} to GND. After identifying the transistors at each vertical level, a primary current path from \emph{V}\textsubscript{DD} to GND containing the largest number of transistors is determined.
The algorithm then iterates through the transistor pairs at each vertical level. If a transistor pair is not part of the primary current path, a dummy current path that includes the target transistor pairs is formed with the same number of stacked transistors as the primary current path. The target transistor pair is placed in the originally determined vertical level within the dummy current path. The algorithm continues through all vertical levels to complete the implementation of the dummy current path(s). Dummy transistor pairs are then added to the circuit to ensure that the dummy DC current path includes the same number of vertical levels as the primary current path.

To stack dummy NMOS transistor pairs to a PMOS transistor pair, a programmable NMOS transistor pair as shown in Fig.~\ref{NP1_ARC} is used, with the drain of the NMOS transistor connected to the drain of the PMOS transistor. One or more additional NMOS transistor pairs are then stacked beneath the programmable NMOS transistor pair. The GND node is reached by connecting the source of the NMOS transistor at the highest vertical level to GND.
Similarly, to stack PMOS dummy transistor pairs to an NMOS transistor pair, a programmable PMOS transistor pair as shown in Fig.~\ref{PP1_ARC} is utilized, with the drain of the PMOS transistor connected to the drain of the NMOS transistor. One or more additional PMOS transistor pairs are then stacked above the programmable PMOS transistor pair. The \emph{V}\textsubscript{DD} node is reached by connecting the source of the PMOS transistor at the lowest vertical level to \emph{V}\textsubscript{DD}.
Note that the dummy transistor pairs remain in the off state during the operation of the analog circuit. Therefore, the additional transistors stacked below the programmable NMOS transistor pair or above the programmable PMOS transistor pair do not affect the functionality of the circuit. The insertion of dummy transistor pairs obfuscates both the topology of the circuit and the number of stages in an analog circuit by introducing a dummy current path from \emph{V}\textsubscript{DD} to GND, while also increasing the key space.

\begin{algorithm}
\footnotesize
\setlength{\algomargin}{0.6em}
\SetAlCapSkip{0.2em}
\DontPrintSemicolon
\caption{Pseudocode to implement dummy transistor pairs}
\label{Algo_Dummy}
\textbf{Input}: circuit netlist $\phi_{Raw}$ \;
\textbf{Output}: circuit netlist inserted with dummy transistor pairs  $\phi_{Dummy}$ \;
$\phi_{Dummy}$ =  copy($\phi_{Raw}$)\;
maxVertiLevel = getVertiLevelInfo($\phi_{Raw}$)\;
PrimCurrentPath = []\;
\For{i from 0 to maxVertiLevel}{
mosVertiLevel[i] = findMosVertiLevel(i)\;
\For{mos in mosVertiLevel[i]}{
\If{find(mos.config == regenLoop)}{PrimCurrentPath.append(mos)\;
mosVertiLevel[i].pop(mos)}
\ElseIf{find(mos.config == mirror)}{PrimCurrentPath.append(mos)\;
mosVertiLevel[i].pop(mos)}
\ElseIf{find(mos.config == source)}{PrimCurrentPath.append(mos)\;
mosVertiLevel[i].pop(mos)}
\ElseIf{find(mos.config == diode)}{PrimCurrentPath.append(mos)\;
mosVertiLevel[i].pop(mos)}
}
}
\While{mosVertiLevel  is not empty}{
DummyPath = List[len(PrimCurrentPath)]\;
\For{i from 0 to maxVertiLevel}{
\If{mosVertiLevel[i] is not empty}{
\For{mos in mosVertiLevel[i]}{
DummyPath = replace(i, mos)}
}
}
FillWithDummy(DummyPath)\;
UpdateNetlist(DummyPath,$\phi_{Dummy}$)
}
\textbf{return} $\phi_{Dummy}$\;

\end{algorithm}

%% file: section5.tex
\section{Full Obfuscation of Circuit Topology}\label{DD}
For applications with a high tolerance to parasitic impedance, typically for analog ICs operating at lower frequencies, an  obfuscation technique is developed to mask the entire topology of an analog circuit. An FPAA that includes a programmable transistor pair (TP) array in each CAB is utilized to conceal the architecture of an analog circuit within the fixed fabric of the FPAA. The FPAA remains non-functional until a programming bitstream is applied that properly configures the transistor pairs and routing switches of the FPAA fabric.

\subsection{Utilizing the FPAA Fabric}
The FPAA fabric is utilized to implement full topological obfuscation. Since routing switches are required to configure the FPAA fabric, and the FPAA fabric remains non-functional without a programming bitstream, the security robustness provided by the technique depends on the complexity of the FPAA fabric.
The proposed FPAA allows for transistor-level reconfiguration, which obfuscates both the width of the implemented transistors and the routing path between the transistors. Unlike the partial obfuscation of the circuit topology described in Section~\ref{CC}, which obfuscates only a subset of transistors, full obfuscation maps the entire analog circuit onto the FPAA fabric. As a result, critical information regarding the implemented analog circuit, such as the number of transistors, the size of the utilized transistors, and the internal wire connections between transistors remain unknown to the attacker.

\subsection{Implementing an Analog Circuit on a Single CAB}\label{Obfus_FPAA_Implementation}
For small analog circuits, a single TP array is sufficient to fully obfuscate the topology of the circuit. In the proposed structure, each TP array includes a 6 x 3 matrix of programmable transistor pairs as shown in Fig.~\ref{TP_ARC}, where the first three rows are implemented as PMOS transistor pairs, and the last three rows are implemented as NMOS transistor pairs. Considering the DC voltage drop across the drain and source terminals of the transistors, adding more transistor pairs along the DC current path reduces the voltage headroom available to implement analog circuits.
Therefore, each CAB enables full topological obfuscation of most small analog circuits, supporting a maximum of three differential stages  and six stacked transistor pairs between \emph{V}\textsubscript{DD} and GND. Transistor sizing is determined by a dedicated module that algorithmically sizes devices\cite{FPAA_TP2}. The sizing module performs DC nodal analysis of the target circuit, includes the on-resistance of the routing switches on the mapped current paths, and adjusts the widths of the programmable transistor pairs to match the DC operating point and current of the original unobfuscated netlist.

In addition to the transistor pair array, which obfuscates the topology of active devices, programmable resistors and capacitors are utilized to mask passive components. To allow for differential mode operation of the circuit, each TP array includes two sets of programmable resistors and capacitors available for the differential inputs and two sets of programmable resistors and capacitors available for the feedback paths. The proposed CAB architecture, therefore, efficiently implements circuits for analog signal processing including integration and switched-capacitor sampling.

\subsection{Implementing Higher Complexity Analog Circuits on Multiple CABs}
To obfuscate a large analog circuit, multiple CABs of the FPAA are utilized. Each sub-block of an analog system is mapped onto a single CAB within the FPAA, with connections between sub-blocks established through global routing switches. The minimum number of CABs required to obfuscate an analog circuit depends on the number of sub-modules that the circuit includes. For a complex analog circuit consisting of \emph{N} sub-blocks, at least \emph{N} CABs are required to implement the circuit.

Therefore, the proposed technique obfuscates the entire topological architecture of the analog system. The attacker only has access to the structure of the fixed FPAA fabric, which provides no usable function prior to programming. Although the total number of CABs included in the FPAA fabric is exposed through reverse engineering and netlist extraction, the programmable functionality of each CAB remains completely unknown to the attacker.
In addition, similar to adding dummy transistor pairs, the unused CABs of an FPAA act as dummy CABs that obfuscate the number of correctly set sub-blocks mapped from the original analog circuit.

\subsection{Securing the Bias Voltage of an Analog Circuit}
The proposed FPAA architecture not only masks the topology of an implemented analog circuit but also secures the DC biasing voltages and currents. A 5-bit resistor-ladder digital-to-analog converter (DAC) is included in each TP array to provide DC bias voltages to the gate terminal of the transistors. The gate terminal of each transistor pair within the TP array is connected to the DAC through a switch implemented as a transmission-gate. As a result, the attacker has no prior knowledge of the biasing voltage and the location of the biasing transistors before the programming of the FPAA is completed.
To successfully configure the FPAA into a functional circuit, the attacker must activate and select the correct transistor pairs within the array, program the transistor pairs into the correct topology, set the transistor widths of the pair appropriately, and program the DAC to provide the required DC biasing voltage(s).
Therefore, attacking a circuit implemented on the FPAA is equivalent to designing the circuit from scratch.

%% file: section6.tex
\section{Characterization of Partial Obfuscation of Circuit Topology}\label{EE}
In this section, the results from the characterization  of circuits secured utilizing the partial topology obfuscation technique are described. The obfuscated analog circuits include a folded-cascode op amp, a StrongARM comparator, and a successive-approximation analog-to-digital converter (SAR ADC).

% \begin{figure}[htbp]
% \centerline{\includegraphics[width=.49\textwidth]{./Fold_Opamp.png}}
% \caption{ The schematic of a folded-cascode op amp.} 
% \label{Folded_ARC}
% \end{figure}

% \begin{figure}[htbp]
% \centerline{\includegraphics[width=.49\textwidth]{./Fold_Opamp_Dummy.png}}
% \caption{ The schematic of folded-cascode op amp with dummy transistor pair implemented.} 
% \label{Folded_ARC_Dummy}
% \end{figure}

\vspace{-5pt}
\begin{figure}[htbp]
\centering
\subfloat[]{\includegraphics[width=.23\textwidth]{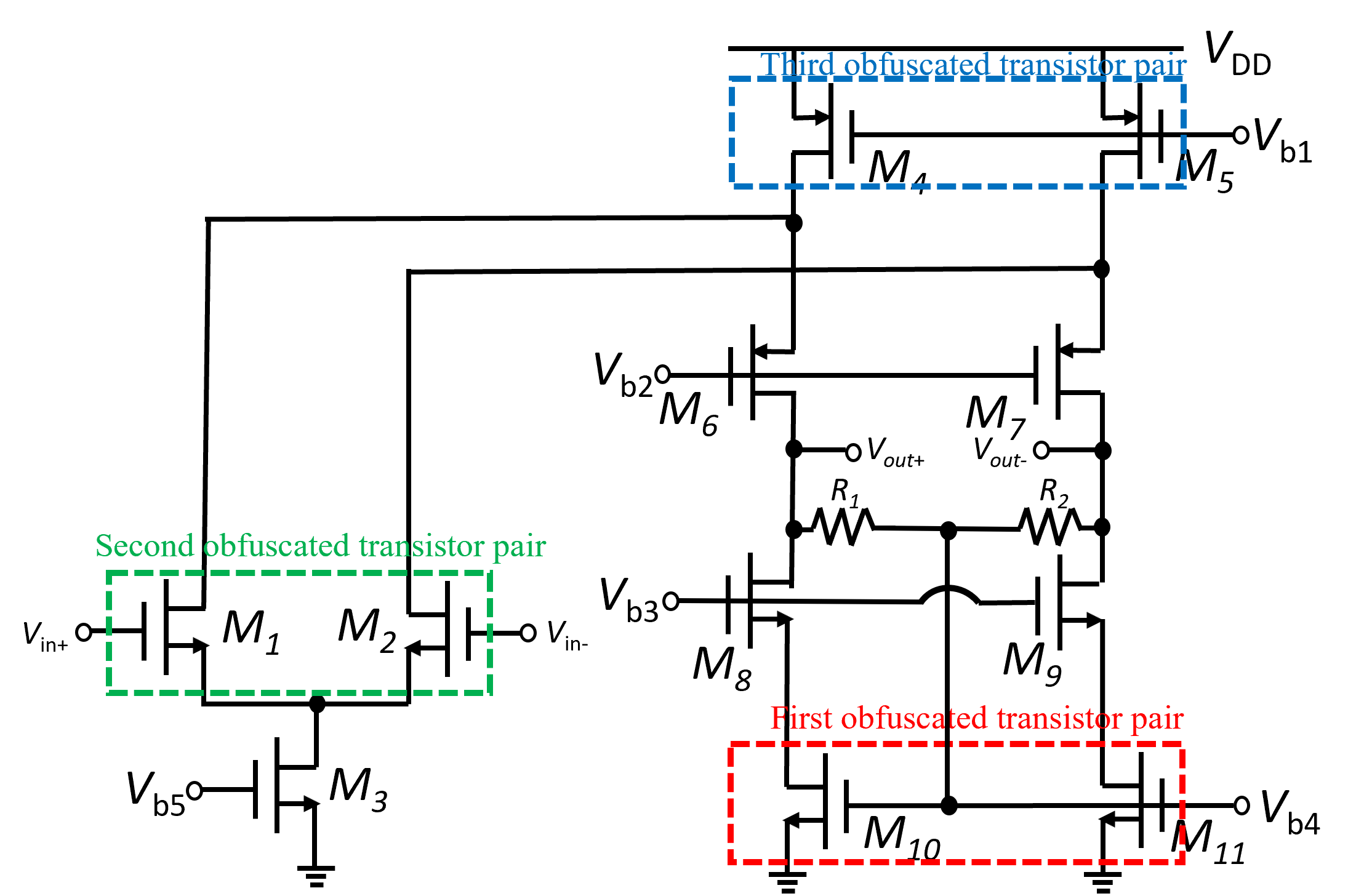}
\label{Folded_ARC}}
\hfil
\subfloat[]{\includegraphics[width=.23\textwidth]{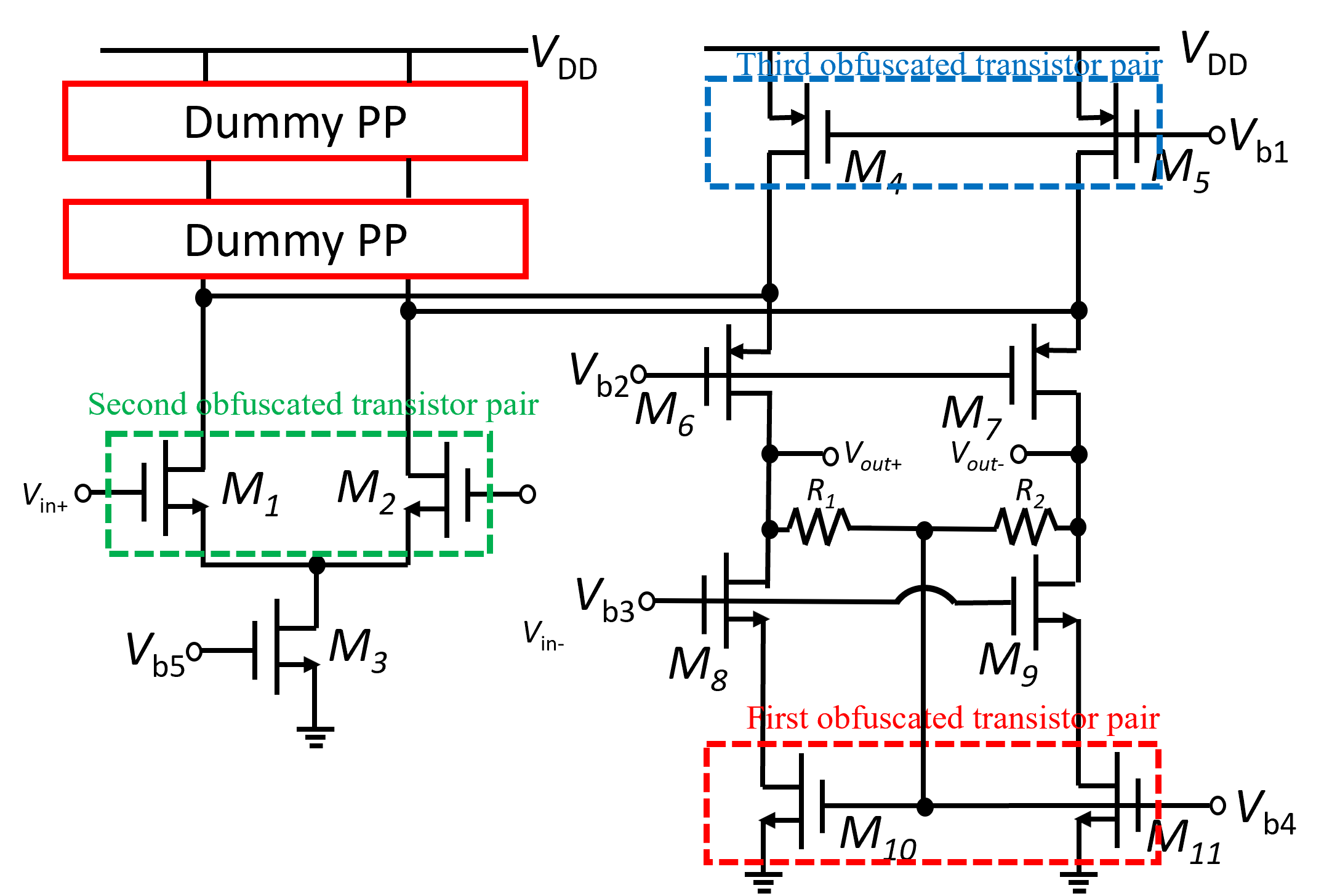}
\label{Folded_ARC_Dummy}}
\hfil
\caption{Schematic representation of a (a) folded-cascode op amp and (b) folded-cascode op amp with dummy PMOS transistor pairs (Dummy PPs).  Unobfuscated transistors are connected with fixed interconnects.}
\label{Folded_security}
\end{figure}

\begin{figure}[htbp]
\centerline{\includegraphics[width=.35\textwidth]{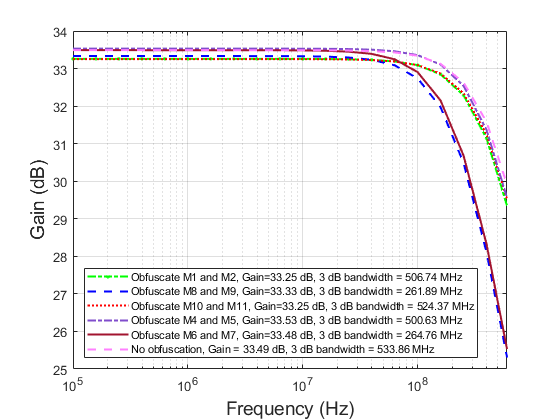}}
\caption{AC response of a folded-cascode op amp with transistor pairs \emph{M}\textsubscript{1}-\emph{M}\textsubscript{2}, \emph{M}\textsubscript{8}-\emph{M}\textsubscript{9}, \emph{M}\textsubscript{10}-\emph{M}\textsubscript{11}, \emph{M}\textsubscript{4}-\emph{M}\textsubscript{5}, and \emph{M}\textsubscript{6}-\emph{M}\textsubscript{7} obfuscated.} 
\label{folded_AC}
\vspace{-10pt}
\end{figure}

% \usepackage{multirow}
\begin{table*}[]
\caption{Characterization of the Gain, 3-dB Bandwidth, GBW, Power, and Area of an Extracted View (includes parasitics from layout) of a Folded-cascode Op Amp After Implementing Partial Topology Obfuscation. An Unobfuscated Op Amp is also characterized for Comparison.} 
\label{table_opamp}
\resizebox{\textwidth}{!}{%
\begin{tabular}{|c|c|cc|cc|cc|cc|}
\hline
\multirow{2}{*}{Parameter} &
  \multirow{2}{*}{Unobfuscated} &
  \multicolumn{2}{c|}{\begin{tabular}[c]{@{}c@{}}One Transistor \\ Pair Obfuscated\\ (M1, M2)\end{tabular}} &
  \multicolumn{2}{c|}{\begin{tabular}[c]{@{}c@{}}One Transistor \\ Pair Obfuscated\\ (M10, M11)\end{tabular}} &
  \multicolumn{2}{c|}{\begin{tabular}[c]{@{}c@{}}Two Transistor \\ Pairs Obfuscated\\ (M1, M2; M10, M11)\end{tabular}} &
  \multicolumn{2}{c|}{\begin{tabular}[c]{@{}c@{}}Three Transistor \\ Pairs Obfuscated\\ (M1, M2; M4, M5; M10, M11)\end{tabular}} \\ \cline{3-10} 
 &
   &
  \multicolumn{1}{c|}{\begin{tabular}[c]{@{}c@{}}No \\ Dummy\end{tabular}} &
  \begin{tabular}[c]{@{}c@{}}With\\ Dummy\end{tabular} &
  \multicolumn{1}{c|}{\begin{tabular}[c]{@{}c@{}}No \\ Dummy\end{tabular}} &
  \begin{tabular}[c]{@{}c@{}}With\\ Dummy\end{tabular} &
  \multicolumn{1}{c|}{\begin{tabular}[c]{@{}c@{}}No \\ Dummy\end{tabular}} &
  \begin{tabular}[c]{@{}c@{}}With\\ Dummy\end{tabular} &
  \multicolumn{1}{c|}{\begin{tabular}[c]{@{}c@{}}No \\ Dummy\end{tabular}} &
  \begin{tabular}[c]{@{}c@{}}With\\ Dummy\end{tabular} \\ \hline
Gain &
  30.21 dB &
  \multicolumn{1}{c|}{30.27 dB} &
  30.27 dB &
  \multicolumn{1}{c|}{30.38 dB} &
  30.39 dB &
  \multicolumn{1}{c|}{30.49 dB} &
  30.49 dB &
  \multicolumn{1}{c|}{30.84 dB} &
  30.83 dB \\ \hline
\begin{tabular}[c]{@{}c@{}}3-dB \\ Bandwidth\end{tabular} &
  247.06 MHz &
  \multicolumn{1}{c|}{240.2 MHz} &
  236.8 MHz &
  \multicolumn{1}{c|}{236.9 MHz} &
  232.7 MHz &
  \multicolumn{1}{c|}{233.6 MHz} &
  230.89 MHz &
  \multicolumn{1}{c|}{205.7 MHz} &
  204.06 MHz \\ \hline
GBW &
  7.46 GHz &
  \multicolumn{1}{c|}{7.27 GHz} &
  7.16 GHz &
  \multicolumn{1}{c|}{7.16 GHz} &
  7.07 GHz &
  \multicolumn{1}{c|}{7.12 GHz} &
  7.04 GHz &
  \multicolumn{1}{c|}{6.34 GHz} &
  6.3 GHz \\ \hline
Power &
  1.63 mW &
  \multicolumn{1}{c|}{1.65 mW} &
  1.65 mW &
  \multicolumn{1}{c|}{1.66 mW} &
  1.66 mW &
  \multicolumn{1}{c|}{1.72 mW} &
  1.72 mW &
  \multicolumn{1}{c|}{1.78 mW} &
  1.78 mW \\ \hline
Area &
  1925.93 {\textmu}m\textsuperscript{2} &
  \multicolumn{1}{c|}{2300.86 {\textmu}m\textsuperscript{2}} &
  3126.19 {\textmu}m\textsuperscript{2} &
  \multicolumn{1}{c|}{2337.66 {\textmu}m\textsuperscript{2}} &
  3162.99 {\textmu}m\textsuperscript{2} &
  \multicolumn{1}{c|}{2597.33 {\textmu}m\textsuperscript{2}} &
  3419.05 {\textmu}m\textsuperscript{2} &
  \multicolumn{1}{c|}{2932.37 {\textmu}m\textsuperscript{2}} &
  3754.09 {\textmu}m\textsuperscript{2} \\ \hline
\end{tabular}
}
\label{table_opamp}
\end{table*}

\subsection{Folded-cascode Op Amp}\label{section_opamp}

The operational amplifier (op amp) is widely used across different circuits including active filters, integrators, and analog-to-digital converters (ADC).
The proposed technique that partially obfuscates a circuit is implemented on a folded-cascode op amp. The schematic of the folded-cascode amplifier is shown in Fig.~\ref{Folded_ARC}. 
Algorithm~\ref{Partial} is executed to obfuscate the folded-cascode op amp without impacting circuit performance. The algorithm first iterates through each transistor pair. In each executed iteration of the algorithm, the target transistor pair is replaced with a programmable transistor pair, which is shown in Fig.~\ref{Program_TP1}. The parameters of the target transistor pair, including the width and the configured topology, are extracted. The extracted parameters are then used to configure the programmable transistor pair that corresponds to the replaced target transistor pair.
A SPICE simulation is performed to characterize the performance parameters of the circuit, including the gain and 3-dB bandwidth. Each transistor pair is then labeled with the corresponding degradation in the gain-bandwidth (GBW) product, which are then sorted in ascending order of degradation.
The AC response of the folded-cascode amplifier after executing Algorithm~\ref{Partial} is shown in Fig.~\ref{folded_AC}. Obfuscating the transistor pair \emph{M}\textsubscript{1} and \emph{M}\textsubscript{2} shown in Fig.~\ref{Folded_ARC} provides the highest GBW product, followed by obfuscating transistor pair \emph{M}\textsubscript{10} and \emph{M}\textsubscript{11}, and then obfuscating transistor pair \emph{M}\textsubscript{4} and \emph{M}\textsubscript{5}.
\emph{M}\textsubscript{6} and \emph{M}\textsubscript{7} and \emph{M}\textsubscript{8} and \emph{M}\textsubscript{9} are identified as parasitic-sensitive transistor pairs due to the significant reduction in the 3-dB bandwidth of the circuit when obfuscated.

\begin{figure}[htbp]
\centering
\subfloat[]{\includegraphics[width=.45\textwidth]{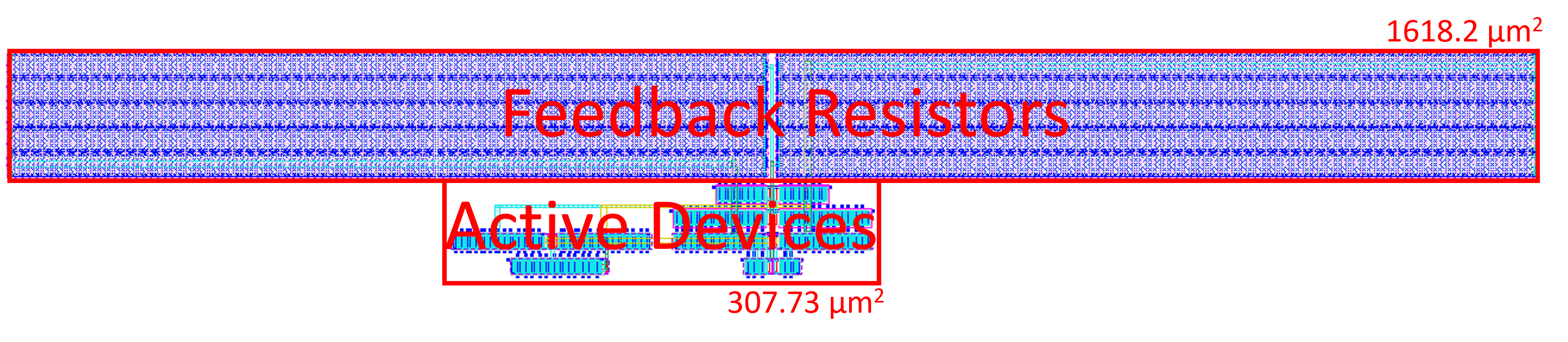}\label{opamp_layout_baseline}}
\vspace{-5pt}
\subfloat[]{\includegraphics[width=.45\textwidth]{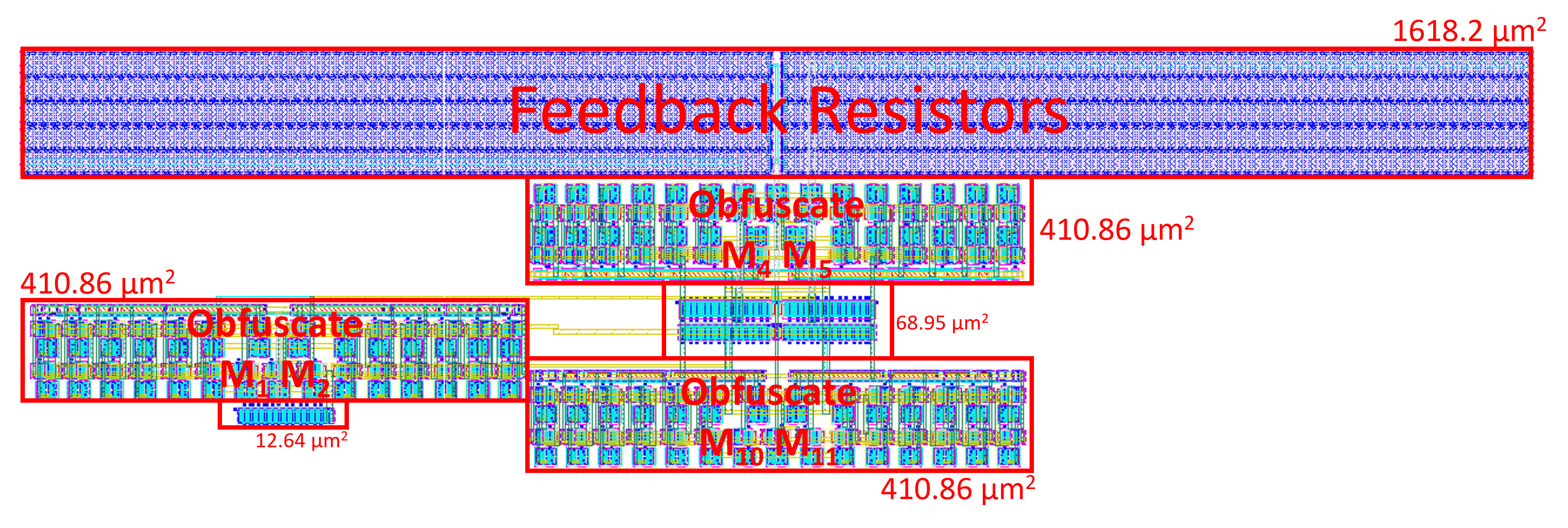}\label{opamp_layout_obfus}}
\hfil
\caption{Layout view of the (a) unobfuscated folded-cascode op amp and (b) folded-cascode op amp with three transistor pairs obfuscated.}
\label{opamp_layout}
\end{figure}

The results from the characterization of a folded-cascode op amp with  one, two, and three transistor pairs obfuscated are listed in Table~\ref{table_opamp}. The unobfuscated op amp provides a DC gain of 30.21 dB, a 3-dB bandwidth of 247.06 MHz, and a GBW of 7.46 GHz.
The maximum GBW product is achieved when one transistor pair, \emph{M}\textsubscript{1} and \emph{M}\textsubscript{2}, is obfuscated. 
The layout view of the op amp
when three transistor pairs, \emph{M}\textsubscript{1} and \emph{M}\textsubscript{2}, \emph{M}\textsubscript{10} and \emph{M}\textsubscript{11}, and \emph{M}\textsubscript{4} and \emph{M}\textsubscript{5}, are obfuscated, is shown in Fig.~\ref{opamp_layout_obfus}, where the GBW is reduced by 15\%, while the area increases by 52.3\%. The variation in power consumption remains within 3\% across the four obfuscation scenarios as listed in Table~\ref{table_opamp}.
The results from the characterization of amplifiers that include dummy transistors are also listed in Table~\ref{table_opamp}. Including dummy transistor pairs introduces an additional parasitic impedance to the internal nodes of the op amp. As a result, the 3-dB bandwidth is slightly reduced, while the DC gain remains the same as that of the op amp without dummy transistors. When one, two, and three transistor pairs are obfuscated, the GBW of the folded-cascode op amp with dummy transistor pairs is reduced by 1.5\%, 1.11\%, and 0.63\%, respectively, of that of an obfuscated folded-cascode op amp without dummy transistor pairs.
\vspace{-5pt}
\vspace{-5pt}

\begin{figure}[htbp]
\centering
\subfloat[]{\includegraphics[width=.23\textwidth]{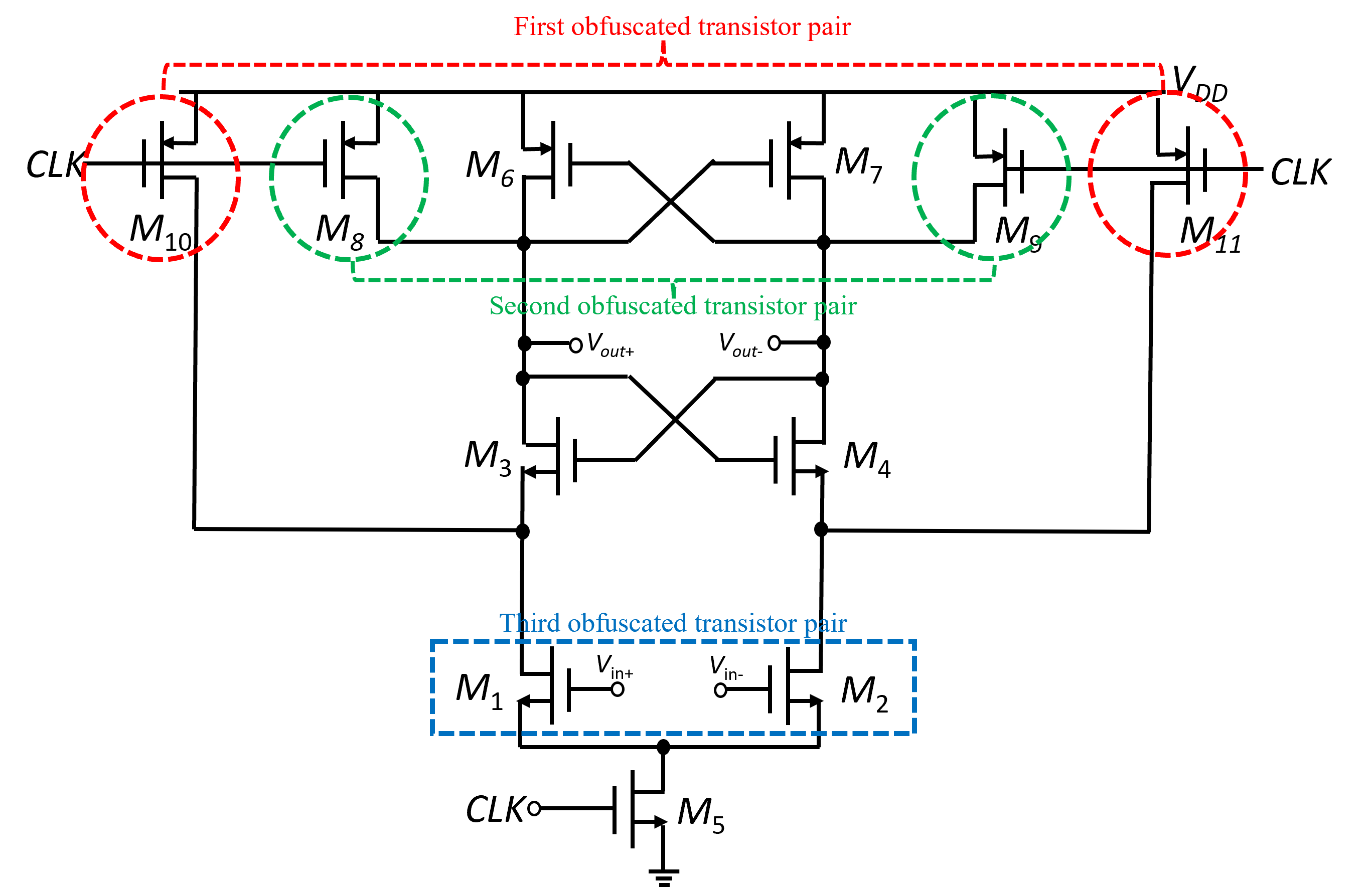}
\label{Comp_ARC}}
\hfil
\subfloat[]{\includegraphics[width=.23\textwidth]{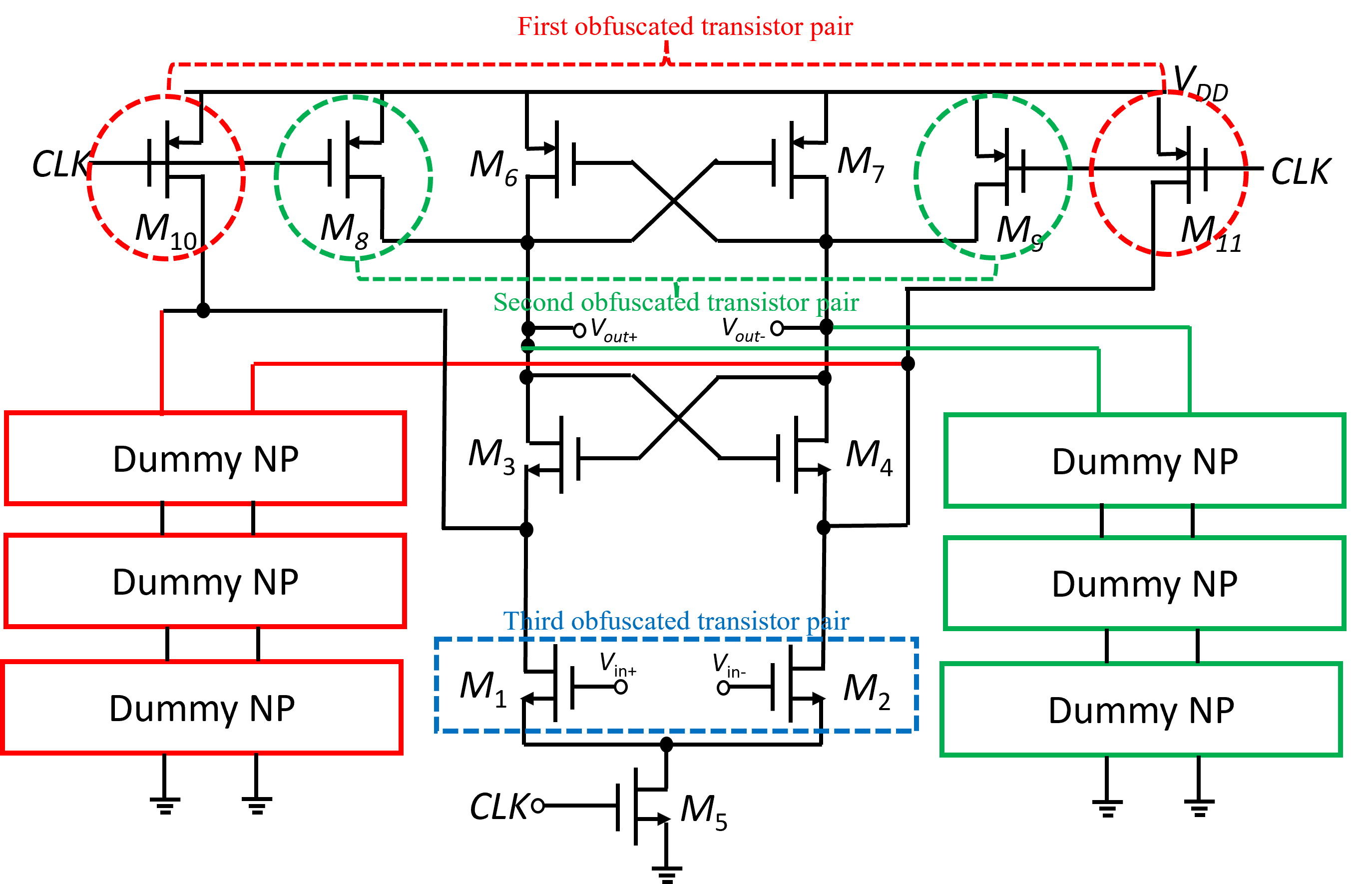}
\label{Comp_ARC_Dummy}}
\hfil
\caption{Schematic representation of a (a) StrongARM comparator and (b) StrongARM comparator with dummy NMOS transistor pairs (Dummy NP).}
\label{Comp_secure}
\end{figure}
\vspace{-5pt}
% \begin{figure}[htbp]
% \centerline{\includegraphics[width=.6\textwidth]{./foldedOpamp_AC1.png}}
% \caption{AC response of obfuscated op amp} 
% \label{folded_AC}
% \end{figure}
\begin{table*}[]
\caption{Characterization of the rise time, fall time, input-referred offset, regeneration time, power, and area of a StrongARM comparator after implementing partial topology obfuscation. An unobfuscated StrongARM comparator is characterized for comparison.} 
\label{table_comparator}
\resizebox{\textwidth}{!}{%
\begin{tabular}{|c|c|cc|cc|cc|cc|}
\hline
\multirow{2}{*}{Parameter} &
  \multirow{2}{*}{Unobfuscated} &
  \multicolumn{2}{c|}{\begin{tabular}[c]{@{}c@{}}One Transistor \\ Pair Obfuscated\\ (M10, M11)\end{tabular}} &
  \multicolumn{2}{c|}{\begin{tabular}[c]{@{}c@{}}One Transistor \\ Pair Obfuscated\\ (M1, M2)\end{tabular}} &
  \multicolumn{2}{c|}{\begin{tabular}[c]{@{}c@{}}Two Transistor \\ Pairs Obfuscated\\ (M8, M9; M10, M11)\end{tabular}} &
  \multicolumn{2}{c|}{\begin{tabular}[c]{@{}c@{}}Three Transistor \\ Pairs Obfuscated\\ (M1, M2; M8, M9; M10, M11)\end{tabular}} \\ \cline{3-10} 
 &
   &
  \multicolumn{1}{c|}{\begin{tabular}[c]{@{}c@{}}No \\ Dummy\end{tabular}} &
  \begin{tabular}[c]{@{}c@{}}With\\ Dummy\end{tabular} &
  \multicolumn{1}{c|}{\begin{tabular}[c]{@{}c@{}}No \\ Dummy\end{tabular}} &
  \begin{tabular}[c]{@{}c@{}}With\\ Dummy\end{tabular} &
  \multicolumn{1}{c|}{\begin{tabular}[c]{@{}c@{}}No \\ Dummy\end{tabular}} &
  \begin{tabular}[c]{@{}c@{}}With\\ Dummy\end{tabular} &
  \multicolumn{1}{c|}{\begin{tabular}[c]{@{}c@{}}No \\ Dummy\end{tabular}} &
  \begin{tabular}[c]{@{}c@{}}With\\ Dummy\end{tabular} \\ \hline
Rise Time &
  0.029 ns &
  \multicolumn{1}{c|}{0.0294 ns} &
  0.0307 ns &
  \multicolumn{1}{c|}{0.031 ns} &
  0.034 ns &
  \multicolumn{1}{c|}{0.0299 ns} &
  0.034 ns &
  \multicolumn{1}{c|}{0.033 ns} &
  0.035 ns \\ \hline
Fall Time &
  0.167 ns &
  \multicolumn{1}{c|}{0.175 ns} &
  0.213 ns &
  \multicolumn{1}{c|}{0.21 ns} &
  0.26 ns &
  \multicolumn{1}{c|}{0.2 ns} &
  0.242 ns &
  \multicolumn{1}{c|}{0.233 ns} &
  0.3 ns \\ \hline
\begin{tabular}[c]{@{}c@{}}Input-referred\\ Offset\end{tabular} &
  0.0952 mV &
  \multicolumn{1}{c|}{0.1386 mV} &
  0.1143 mV &
  \multicolumn{1}{c|}{0.7752 mV} &
  0.6352 mV &
  \multicolumn{1}{c|}{0.1442 mV} &
  0.1277 mV &
  \multicolumn{1}{c|}{0.1552 mV} &
  0.1352 mV \\ \hline
\begin{tabular}[c]{@{}c@{}}Regeneration\\ Time\end{tabular} &
  0.1469 ns &
  \multicolumn{1}{c|}{0.1563 ns} &
  0.1955 ns &
  \multicolumn{1}{c|}{0.1791 ns} &
  0.233 ns &
  \multicolumn{1}{c|}{0.1675 ns} &
  0.2262 ns &
  \multicolumn{1}{c|}{0.217 ns} &
  0.2799 ns \\ \hline
Power &
  68.72 {\textmu}W &
  \multicolumn{1}{c|}{78.47 {\textmu}W} &
  102 {\textmu}W &
  \multicolumn{1}{c|}{81.11 {\textmu}W} &
  104.5 {\textmu}W &
  \multicolumn{1}{c|}{90.44 {\textmu}W} &
  106.2 {\textmu}W &
  \multicolumn{1}{c|}{99.13 {\textmu}W} &
  123.6 {\textmu}W \\ \hline
Area &
  197.75 {\textmu}m\textsuperscript{2} &
  \multicolumn{1}{c|}{588.18 {\textmu}m\textsuperscript{2}} &
  3061.46 {\textmu}m\textsuperscript{2} &
  \multicolumn{1}{c|}{566.12 {\textmu}m\textsuperscript{2}} &
  3039.4 {\textmu}m\textsuperscript{2} &
  \multicolumn{1}{c|}{994.07 {\textmu}m\textsuperscript{2}} &
  3467.35 {\textmu}m\textsuperscript{2} &
  \multicolumn{1}{c|}{1299.37 {\textmu}m\textsuperscript{2}} &
  3772.65 {\textmu}m\textsuperscript{2} \\ \hline
\end{tabular}
}
\label{table_comparator}
\end{table*}

\subsection{Comparator}\label{section_comparator}
\begin{figure}[htbp]
\centering
\subfloat[]{\includegraphics[width=.23\textwidth]{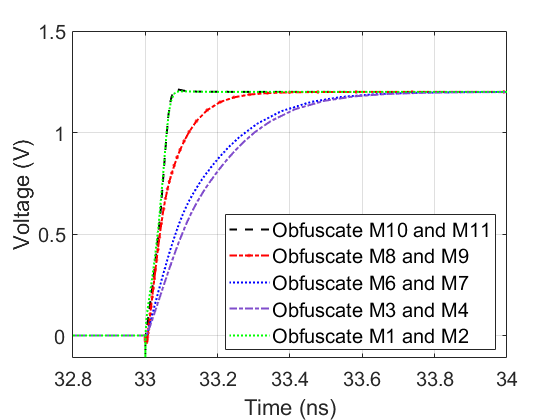}}
\label{riseTime}
\hfil
\subfloat[]{\includegraphics[width=.23\textwidth]{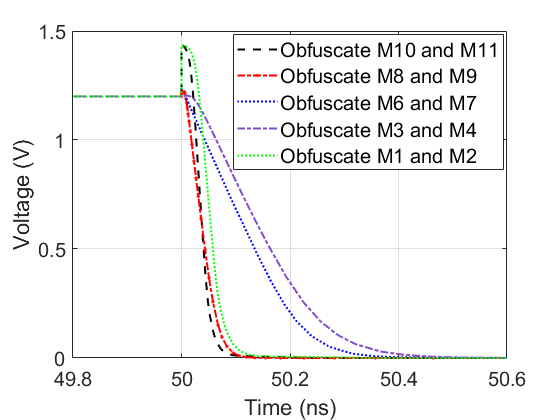}}
\label{fallTime}
\hfil

\caption{Analysis of the transient response to characterize the (a) rise time and (b) fall time of a comparator as a function of the obfuscated transistor pairs.}
\label{comparator_tran}
\end{figure}

The comparator is widely used in AMS applications including for SAR ADCs and delta-sigma ADCs. The schematic of a StrongARM comparator is shown in Fig.~\ref{Comp_ARC}.
% The comparator consists of an input stage with transistors \emph{M}\textsubscript{1} and \emph{M}\textsubscript{2}, while two cross-coupled pairs, \emph{M}\textsubscript{3} and \emph{M}\textsubscript{4}, and \emph{M}\textsubscript{6} and \emph{M}\textsubscript{7}, amplify the voltage difference at the internal nodes. Transistors \emph{M}\textsubscript{8}, \emph{M}\textsubscript{9}, \emph{M}\textsubscript{10}, and \emph{M}\textsubscript{11} reset the internal node voltages to \emph{V}\textsubscript{DD}.
Algorithm~\ref{Partial} is executed on each transistor pair of the StrongARM comparator, replacing a selected pair with a programmable transistor pair. A SPICE simulation is performed on the extracted view of the comparator to characterize the rise time and fall time of the comparator response. The results characterizing the StrongARM comparator after executing Algorithm~\ref{Partial} are shown in Fig.~\ref{comparator_tran}.
The summation of the rise time and fall time is used as the metric to determine the prioritized order of transistors selected for obfuscation. The preferred order of transistor pairs selected for obfuscation is \emph{M}\textsubscript{10} and \emph{M}\textsubscript{11}, followed by \emph{M}\textsubscript{8} and \emph{M}\textsubscript{9}, and finally \emph{M}\textsubscript{1} and \emph{M}\textsubscript{2}. Transistor pairs \emph{M}\textsubscript{3} and \emph{M}\textsubscript{4} and \emph{M}\textsubscript{6} and \emph{M}\textsubscript{7} are identified as parasitically-sensitive due to the significant increase in the rise time and fall time when obfuscated.

\begin{figure}[htbp]
\vspace{-5pt}
\centering
\subfloat[]{\includegraphics[width=.24\textwidth]{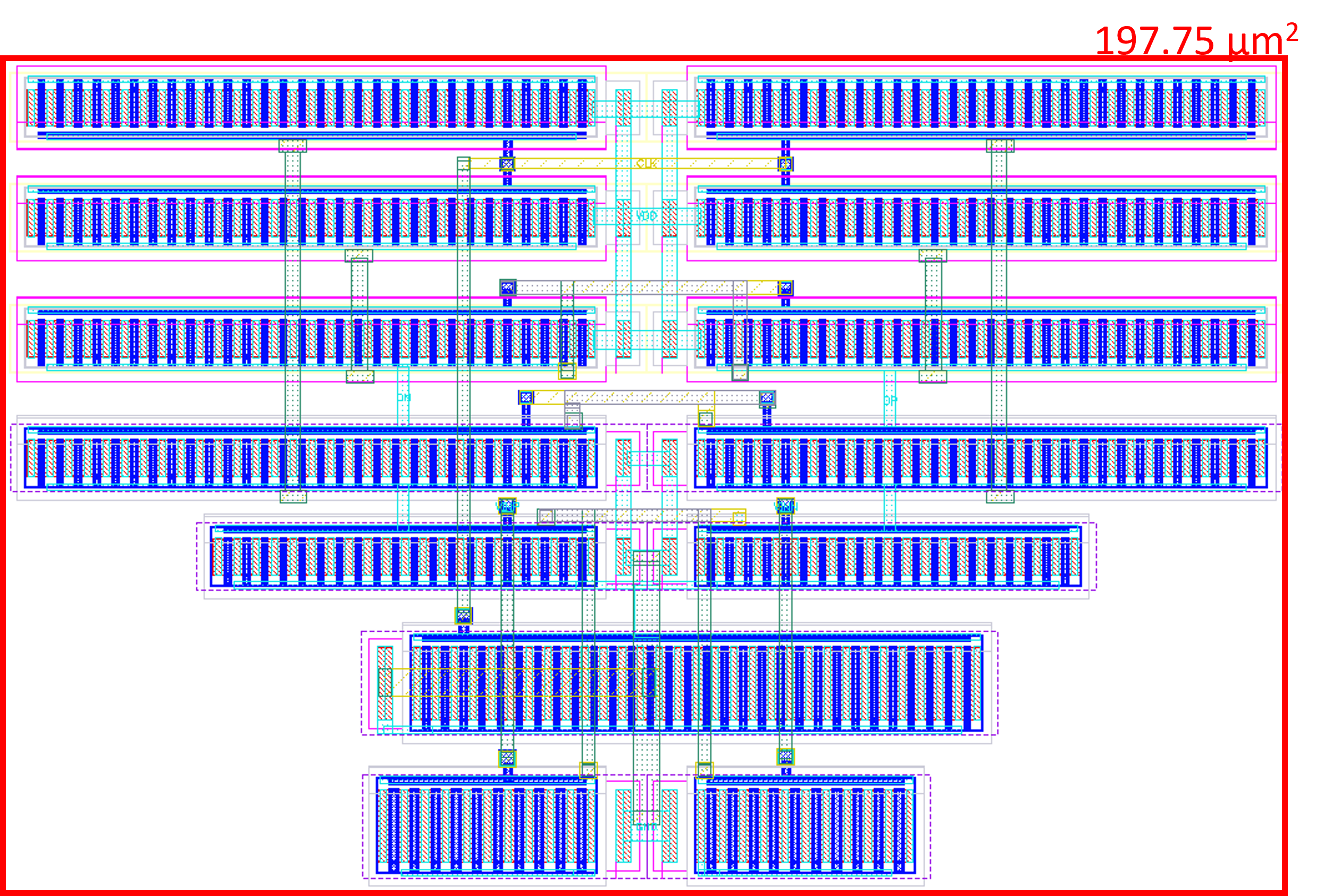}
\label{comparator_layout_baseline}}
\hfil
\subfloat[]{\includegraphics[width=.23\textwidth]{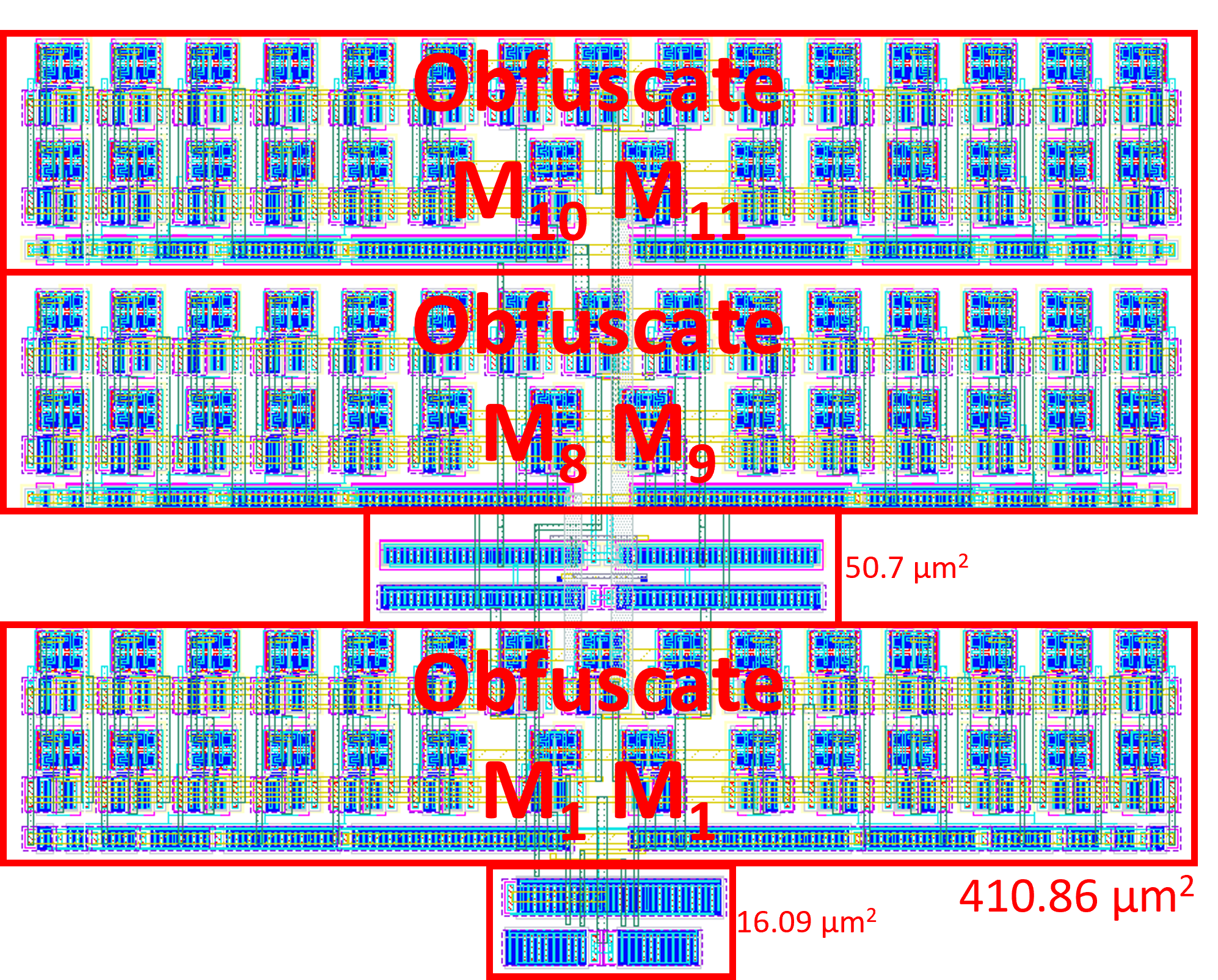}
\label{comparator_layout_obfus}}
\hfil
\caption{Layout representation of the (a) unobfuscated comparator and (b) comparator with three transistor pairs obfuscated.}
\label{comparator_layout}
\vspace{-5pt}
\end{figure}

The results from the characterization of a StrongARM comparator with the obfuscation of one, two, and three transistor pairs are listed in Table~\ref{table_comparator}. When one, two, and three transistor pairs are obfuscated, the fall time increases by 4.8\%, 20\%, and 39.5\%, respectively.
The rise time increases by 13.8\% when three transistor pairs are obfuscated.
The rising transition occurs during the precharge phase of the StrongARM comparator, where the outputs are charged to \emph{V}\textsubscript{DD} through switches M8 and M9. In contrast, the falling transition occurs during regeneration, where the outputs discharge with the application of a small differential input voltage. The time to complete the regeneration phase depends strongly on the time constant at the output node. The added parasitic impedance, therefore, has a larger impact on the fall time, which results in a larger increase in the fall time as compared to the rise time.
The power consumption  increases by 44.2\% when three transistor pairs are obfuscated. 
The layout view of the comparator when three transistor pairs are obfuscated is shown in Fig.~\ref{comparator_layout_obfus}. Due to the additional circuitry needed to obfuscate the StrongARM comparator, the area increases by 6.57x as compared to an unobfuscated version of the same comparator topology.
The results from the characterization of a StrongARM comparator with and without  
dummy transistor pairs are also listed in Table~\ref{table_comparator}. 
Up to six dummy NPs are added to the comparator as shown in Fig.~\ref{Comp_ARC_Dummy}. When three transistor pairs are obfuscated, the rise time, fall time, input-referred offset, regeneration time, and power increase by 20.7\%, 80\%, 42\%, 90.5\%, and 79.8\%, respectively.

% \usepackage{multirow}
\begin{table*}[]
\caption{Characterization of the ENOB, Power, and Area of a SAR ADC After Implementing Partial Topology Obfuscation on Different Sets of Transistor Pairs with and without Dummy Transistors. An Unobfuscated SAR ADC is Characterized for Comparison.} 
\label{table_sar}
\resizebox{\textwidth}{!}{%
\begin{tabular}{|c|c|cc|cc|cc|cc|}
\hline
\multirow{2}{*}{Parameter} &
  \multirow{2}{*}{Unobfuscated} &
  \multicolumn{2}{c|}{\begin{tabular}[c]{@{}c@{}}One Transistor \\ Pair Obfuscated\\ (M10, M11)\end{tabular}} &
  \multicolumn{2}{c|}{\begin{tabular}[c]{@{}c@{}}Two Transistor \\ Pairs Obfuscated\\ (M8, M9;\\ M10, M11)\end{tabular}} &
  \multicolumn{2}{c|}{\begin{tabular}[c]{@{}c@{}}Three Transistor \\ Pairs Obfuscated\\ (M6, M7; M8, M9;\\  M10, M11)\end{tabular}} &
  \multicolumn{2}{c|}{\begin{tabular}[c]{@{}c@{}}Four Transistor \\ Pairs Obfuscated\\ (M3, M4; M6, M7; \\ M8, M9; M10, M11)\end{tabular}} \\ \cline{3-10} 
 &
   &
  \multicolumn{1}{c|}{\begin{tabular}[c]{@{}c@{}}No \\ Dummy\end{tabular}} &
  \begin{tabular}[c]{@{}c@{}}With\\ Dummy\end{tabular} &
  \multicolumn{1}{c|}{\begin{tabular}[c]{@{}c@{}}No \\ Dummy\end{tabular}} &
  \begin{tabular}[c]{@{}c@{}}With\\ Dummy\end{tabular} &
  \multicolumn{1}{c|}{\begin{tabular}[c]{@{}c@{}}No \\ Dummy\end{tabular}} &
  \begin{tabular}[c]{@{}c@{}}With\\ Dummy\end{tabular} &
  \multicolumn{1}{c|}{\begin{tabular}[c]{@{}c@{}}No \\ Dummy\end{tabular}} &
  \begin{tabular}[c]{@{}c@{}}With\\ Dummy\end{tabular} \\ \hline
ENOB &
  8.62 bits &
  \multicolumn{1}{c|}{8.59 bits} &
  8.6 bits &
  \multicolumn{1}{c|}{8.58 bits} &
  8.69 bits &
  \multicolumn{1}{c|}{8.57 bits} &
  8.59 bits &
  \multicolumn{1}{c|}{8.52 bits} &
  8.76 bits \\ \hline
Power &
  1.41 mW &
  \multicolumn{1}{c|}{1.52 mW} &
  2.06 mW &
  \multicolumn{1}{c|}{1.71 mW} &
  2.7 mW &
  \multicolumn{1}{c|}{2.45 mW} &
  3.62 mW &
  \multicolumn{1}{c|}{4.55 mW} &
  5.23 mW \\ \hline
Area &
  0.1302 mm\textsuperscript{2} &
  \multicolumn{1}{c|}{0.1306 mm\textsuperscript{2}} &
  0.1327 mm\textsuperscript{2} &
  \multicolumn{1}{c|}{0.1311 mm\textsuperscript{2}} &
  0.1332 mm\textsuperscript{2} &
  \multicolumn{1}{c|}{0.1315 mm\textsuperscript{2}} &
  0.1336 mm\textsuperscript{2} &
  \multicolumn{1}{c|}{0.1319 mm\textsuperscript{2}} &
  0.134 mm\textsuperscript{2} \\ \hline
\end{tabular}
}
\end{table*}

\subsection{SAR ADC}\label{section_sar}

% \begin{figure}[htbp]
% \centerline{\includegraphics[width=.49\textwidth]{./FFT_SAR_obfus5tran.png}}
% \caption{Characterization of the FFT spectrum of the obfuscated SAR ADC. The frequency of the input is 93.75 kHz and the sampling frequency is 3 MHz. The ENOB is 7.19 bits.} 
% \label{SAR_FFT}
% \end{figure}

\begin{figure}[htbp]
\vspace{-5pt}
\centering
\subfloat[]{\includegraphics[width=.28\textwidth]{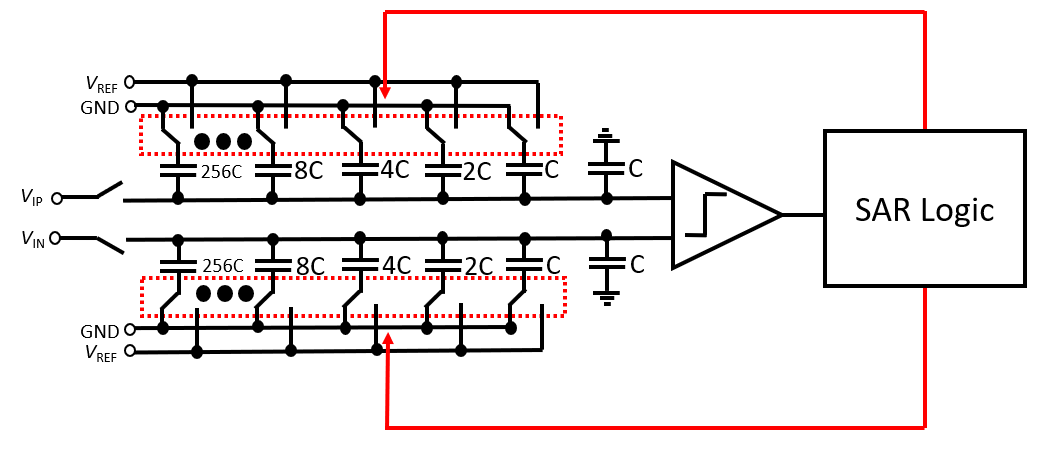}
\label{SAR_ARC}}
\hfil
\subfloat[]{\includegraphics[width=.19\textwidth]{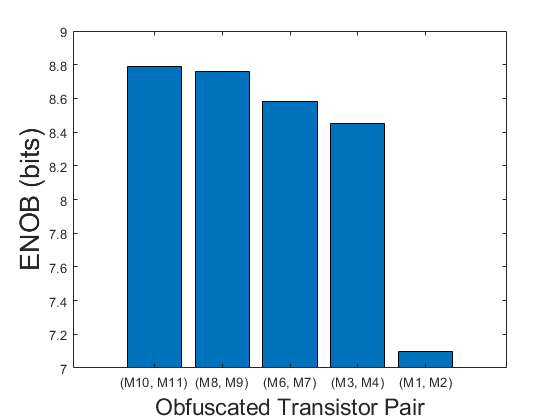}
\label{SAR_BAR}}
\caption{9-bit SAR ADC (a) circuit representation and (b) characterization of the ENOB for different pairs of obfuscated transistors.}
\label{SAR_combined}
\vspace{-5pt}
\end{figure}

The SAR ADC is widely used to quantize analog signals with mid-to-low bandwidth while producing high-resolution digital signals as output. The schematic of a SAR ADC is shown in Fig.~\ref{SAR_ARC}, which consists of a StrongARM comparator, a SAR logic module, and a switched-capacitor bank. The architecture of the implemented comparator is shown in Fig.~\ref{Comp_ARC}.
While the sequential logic locking technique described in \cite{LogicLock} is well-suited to obfuscate the digital logic circuit of the SAR ADC, this work targets obfuscation of the analog components of the ADC, specifically the StrongARM comparator, utilizing partial topology obfuscation.

The implemented 9-bit SAR ADC consists of a capacitor bank where the unit capacitance \emph{C} is 30 fF. 
The SAR ADC samples an input signal of 93.75 kHz frequency at a rate of 3 MHz.
Algorithm~\ref{Partial} is executed on the SAR ADC. The algorithm iterates through each transistor pair of the comparator, replacing each selected transistor pairs with a corresponding programmable transistor pair. A SPICE simulation is then performed, and the calculated effective number of bits (ENOB) of the ADC is determined, which is utilized as the criteria that determines the priority of transistor pairs to obfuscate.
Results from SPICE simulation characterizing the ENOB of the obfuscated ADC are shown in Fig.~\ref{SAR_BAR}.
The transistor pair \emph{M}\textsubscript{1} and \emph{M}\textsubscript{2} is the most sensitive to parasitic impedance, resulting in an ENOB of 7.1 if obfuscated. Conversely, the transistor pair \emph{M}\textsubscript{10} and \emph{M}\textsubscript{11} is the least sensitive to parasitics, resulting in an ENOB of 8.79 if obfuscated. The slight increase in the ENOB for the obfuscated SAR ADC, as compared to the 8.62 bits provided by an unobfuscated SAR ADC, is attributed to the sensitivity of the calculation of the FFT-based ENOB, in particular when the output waveforms are nearly identical. The computed ENOB varies with the record length, sampling coherence, FFT segment selection, and  perturbations in spur magnitude due to the additional parasitic impedances resulting from implementation of the obfuscation technique.

% \begin{figure}[htbp]
% \centerline{\includegraphics[width=.4\textwidth]{./SAR_ARC.PNG}}
% \caption{Circuit representation of a 9-bit SAR ADC.} 
% \label{SAR_ARC}
% \end{figure}

% \begin{figure}[htbp]
% \centerline{\includegraphics[width=.3\textwidth]{./ENOB.png}}
% \caption{Characterization of the ENOB of a SAR ADC for different pairs of obfuscated transistors.} 
% \label{SAR_BAR}
% \end{figure}

Therefore, execution of Algorithm~\ref{Partial} results in transistor pair \emph{M}\textsubscript{10} and \emph{M}\textsubscript{11} being obfuscated first, followed by transistor pairs \emph{M}\textsubscript{8} and \emph{M}\textsubscript{9}, \emph{M}\textsubscript{6} and \emph{M}\textsubscript{7}, and finally \emph{M}\textsubscript{3} and \emph{M}\textsubscript{4}.
The results from characterization of the ENOB, power, and area of the obfuscated SAR ADC are listed in Table~\ref{table_sar}. The variation in the ENOB remains within 1.16\% of an unobfuscated SAR ADC when up to four transistor pairs are obfuscated.
When one, two, three, and four transistor pairs are obfuscated, the power consumption increases by 7.8\%, 21.3\%, 73.8\%, and 222.6\%, respectively, as compared to an unobfuscated SAR ADC. The increase in power consumption is primarily due to the additional parasitic impedances introduced by the programmable transistor pairs, particularly when replacing a regeneration loop. However, due to the large area occupied by the capacitor bank, the area increases by only 1.3\% when obfuscating four transistor pairs and not including any dummy transistor pairs.
The inclusion of dummy transistor pairs in the SAR ADC further increases the power consumed as compared to an obfuscated SAR ADC that includes no dummy transistor pairs. When one, two, three, and four transistor pairs are obfuscated, including six dummy transistor pairs increases the power consumption by 35.6\%, 57.9\%, 47.8\%, and 14.9\%, respectively, as compared to a SAR ADC with the same corresponding transistor pairs obfuscated but with no dummy transistor pairs included. 
However, the increase in the overall area  remains small due to the large capacitor bank. A maximum increase of 3\% in the area occupied by the circuit is observed when four transistor pairs are obfuscated and dummy transistors are utilized. The case study of the SAR ADC allows for the analysis of the system-level impact of obfuscating an analog block within a larger mixed-signal circuit, indicating that the performance and area overhead observed at the block level for the comparator results in a substantially smaller overhead when considering the full ADC circuit with only the analog portion obfuscated.

%% file: section7.tex
\section{Characterization of Full Obfuscation of Circuit Topology}\label{FF}
In this section, characterization is performed on a folded-cascode op amp and a first-order delta-sigma modulator secured with full topology obfuscation. The folded-cascode op amp is masked within a single CAB of the FPAA, and the delta-sigma modulator is obfuscated utilizing two CABs.
\subsection{Fully Obfuscated Folded-cascode Op Amp}\label{section_fpaa_opamp}
Full topology obfuscation is applied to a folded-cascode op amp, the schematic of which is shown in Fig.~\ref{Folded_ARC}. Full obfuscation differs from the partial topology obfuscation technique described in Section~\ref{CC}, where only a sub-set of transistors are replaced with reconfigurable transistor pairs. 
For full topology obfuscation, all transistors of a circuit are replaced and masked by the programmable transistor pairs of a CAB, as shown in Fig.~\ref{TP_ARC}. The total area of a single configurable CAB is 0.278 mm\textsuperscript{2}.
The placement engine\cite{FPAA_TP} parses the netlist of the folded-cascode op amp and assigns the vertical and horizontal location of each transistor. Considering the DC current path from \emph{V}\textsubscript{DD} to GND, the transistor directly connected to \emph{V}\textsubscript{DD} is assigned to the smallest vertical level. The transistors assigned to the first column of the TP array are labeled as part of horizontal level 1, the middle column as part of horizontal level 2, and the third column as part of horizontal level 3. The routing engine, which is based on  Dijkstra's algorithm, automatically routes through the switch boxes according to the connections defined in the netlist.

\begin{figure}[htbp]
\centerline{\includegraphics[width=.3\textwidth]{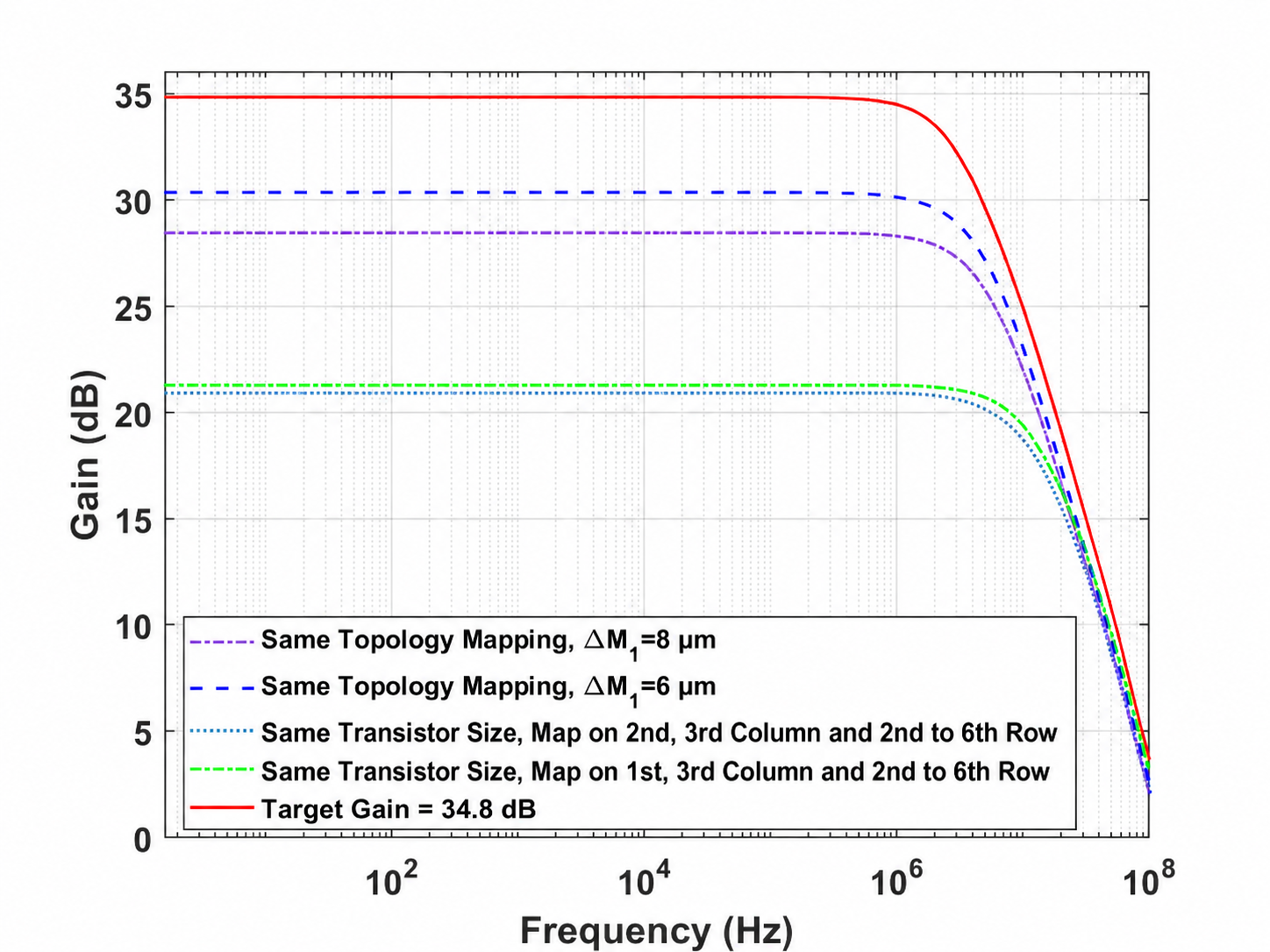}}
\caption{AC response of a folded-cascode op amp implemented on the FPAA fabric with transistor sizing and topology mapping.} 
\label{FPAAOpampAC}
\vspace{-10pt}
\end{figure}

The results from SPICE characterization of the gain of a folded-cascode op amp implemented on the FPAA fabric using the parasitically-extracted netlist are shown in Fig.~\ref{FPAAOpampAC}.
The input and output parasitic capacitances of the op amp implemented on the FPAA fabric with full topology obfuscation are 0.568 pF and 0.862 pF, respectively, whereas the baseline op amp implemented as an ASIC provides 8.26 fF and 22.6 fF of input and output node capacitance. The substantially larger parasitic impedance introduced by the FPAA fabric degrades the high-frequency response and, consequently, limits the bandwidth of the configured op amp. As a result, the configured implementation of the op amp exhibits a DC gain of 34.8 dB and a 3-dB bandwidth of 3.3 MHz.

The reduction in the 3-dB bandwidth of the folded-cascode op amp implemented on the FPAA fabric is primarily due to the routing switches and programmable interconnects at the CAB, which introduce additional parasitic impedance at internal nodes of the circuit. The penalty is inherent to full-topology obfuscation. Although the configured op amp exhibits lower bandwidth than the ASIC baseline, the achieved performance remains suitable for moderate-bandwidth analog and mixed-signal applications. As reported in \cite{FPAA_TP2}, the FPAA fabric supports the implementation of a 15~MHz pipeline ADC and a first-order delta-sigma modulator, which are applicable to audio-oriented signal-processing systems, where analog signals are converted into digital form for subsequent processing. In addition, the reconfigurability of the FPAA enables adaptive filtering and noise shaping after conversion at audio frequencies. Therefore, the fully obfuscated FPAA implementation is intended for secure and reconfigurable moderate-bandwidth analog and mixed-signal systems rather than replacing high bandwidth ASIC circuits. Note that the CAB provides no functionality prior to programming. An adversary is only able to determine the structure of the CAB after executing a reverse engineering attack. The transistor pairs are placed such that a 6 x 3 matrix is formed within the CAB. Therefore, the attacker has no information on the number of transistors used to implement an analog circuit. In addition, the programmable transistor pairs allow for the  configuration of topologies and the setting of transistor widths. Both the configuration of the transistor pairs and the size of the transistors is, therefore, unknown to the attacker. 
Reverse engineering a circuit implemented on the FPAA  requires the activation of the same TPs and the setting of identical transistor sizes to match the parameters of the original analog circuit. A slight deviation in the sizes of the transistors or the mapping of the circuit results in significant degradation in the performance of the op amp, as shown in Fig.~\ref{FPAAOpampAC}.

\subsection{Fully Obfuscated Delta-Sigma Modulator}\label{section_delta}
\begin{figure}[htbp]
\centering{\includegraphics[width=.3\textwidth]{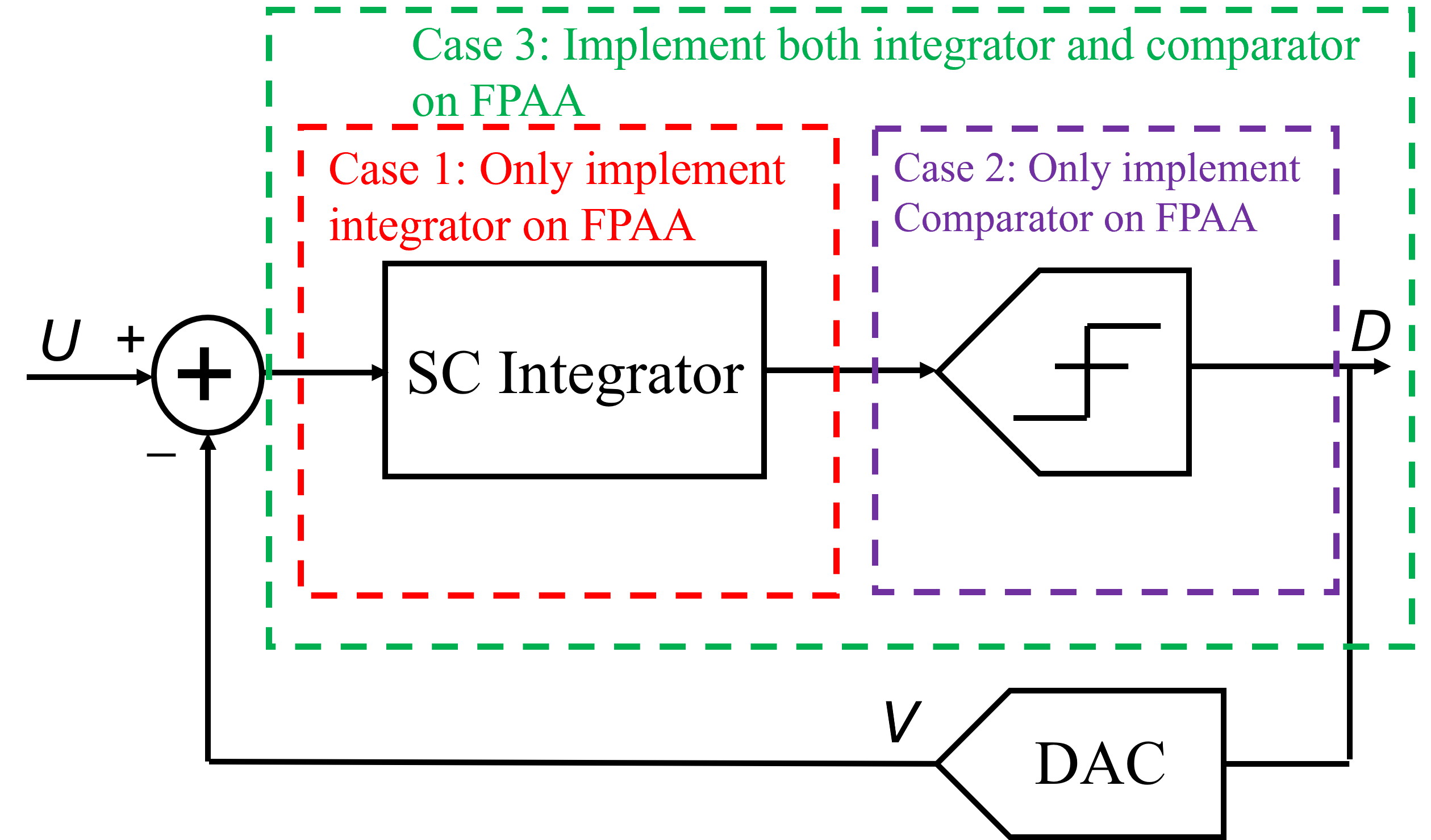}}
\caption{Block level circuit representation of a delta sigma modulator.}
\label{Delta_Modulator}
\end{figure}

The architecture of an implemented first-order delta-sigma modulator (MOD1) is shown in Fig.~\ref{Delta_Modulator}, which consists of a discrete-time integrator and a StrongARM comparator\cite{FPAA_TP2}.
The discrete-time switched-capacitor integrator is comprised of a folded-cascode op amp, a sampling capacitor, and a feedback capacitor\cite{FPAA_TP2}.
% The integrator gain is determined by the capacitance ratio of the programmable capacitors. In the delta-sigma modulator, the integrator gain is set to 1.
% The transfer function of MOD1 is given by:
% \begin{align}
%     V(Z) = Z^{-1}U(Z) + (1-Z^{-1})E(Z)
% \end{align}
% where $V(Z)$ represents the output signal, $U(Z)$ represents the input signal, $E(Z)$ represents the quantization error. 
%  The signal transfer function $STF$ and the noise transfer function $NTF$ is hence given by
% \begin{align}
%     SF &= Z^{-1}\\
%     NTF &= (1-Z^{-1})
% \end{align}
% The noise transfer function (NTF) exhibits a high-pass filter characteristic, effectively shifting the noise spectrum to higher frequencies, making it suitable for filtering by a decimation filter. Conversely, the signal transfer function (SF) indicates that the output signal is a time-delayed version of the original input signal.
Since the delta-sigma modulator consists of two sub-blocks, obfuscating the circuit requires two CABs, with each sub-block mapped onto a separate CAB. Therefore, after executing full topology obfuscation on the first-order delta-sigma modulator, the switched-capacitor (SC) discrete-time integrator and the comparator are programmed onto two separate CABs. The unprogrammed topological structure of the two CABs is identical, and, therefore, both sub-blocks are masked by the FPAA fabric. As before, no function is provided by the FPAA fabric prior to programming.
As a result, the attacker has no information on the cascading behavior of the two sub-blocks. To fully reverse engineer and deobfuscate the delta-sigma modulator, the attacker must first determine the functionality of each CAB and then configure each CAB into the correct circuit topology and with the correct sizing of devices to ensure that the required specifications are met. Specifically, for each CAB, the attacker must select the correct transistor pairs within the TP array and assign the appropriate width to each transistor. In addition, deobfuscation of the internal switch-box network of a CAB is required to determine the connections between programmable transistor pairs. 

\vspace{-5pt}

The results from SPICE characterization of a delta-sigma modulator secured with full topology obfuscation of the integrator and comparator sub-blocks are listed in Table~\ref{table_delta}. 
The clock frequency of the implemented delta sigma modulator is 15 MHz.
The three scenarios analyzed include implementing only the integrator on the FPAA,  implementing only the comparator on the FPAA, and implementing both the integrator and comparator on the FPAA, which provide an ENOB of 6.82, 8.07, and 6.7, respectively, when subjected to an input signal of 29 kHz frequency and a decimation ratio of 128.
As compared to an unobfuscated delta sigma modulator, obfuscating only the comparator decreases the ENOB by 0.25\%. Obfuscating both the integrator and comparator reduces the ENOB by 17.2\% and increases the power consumption by 43.8\%. The results indicate that the integrator is more sensitive to parasitic impedances, which introduces a greater degradation in performance than securing the comparator only.
Obfuscating both sub-blocks provides the highest level of security but also results in the highest degradation in performance, which demonstrates the trade-off between security robustness and performance.

\begin{table}[]
\setlength\tabcolsep{.006\textwidth}
\caption{Characterization of a Delta-Sigma Modulator  Implemented with Full Topology Obfuscation of Circuit Sub-blocks.}
\label{table_delta}
\begin{tabular}{|c|c|c|c|c|}
\hline
Parameter &
  Unobfuscated &
  \begin{tabular}[c]{@{}c@{}}Obfuscate\\ Integrator\end{tabular} &
  \begin{tabular}[c]{@{}c@{}}Obfuscate \\ Comparator\end{tabular} &
  \begin{tabular}[c]{@{}c@{}}Obfuscate Integrator \\ and Comparator\end{tabular} \\ \hline
ENOB  & 8.09 bits & 6.82 bits & 8.07 bits & 6.7 bits \\ \hline
Power & 1.28 mW   & 1.56 mW   & 1.64 mW   & 1.84 mW  \\ \hline
Area  & 2150 {\textmu}m\textsuperscript{2}  & 0.2782 mm\textsuperscript{2} & 0.28 mm\textsuperscript{2}  & 0.556 mm\textsuperscript{2}  \\ \hline
\end{tabular}
\end{table}

\section{Non-Ideality, Robustness, and Overhead of the FPAA 
Fabric}\label{section-non-ideal}
In this section, the non-ideality, yield, configuration latency, area and power overhead, fabrication feasibility, and scalability of the proposed hardware security technique are discussed.

\vspace{-5pt}
\subsection{Noise Analysis}
\label{section-fpaa-noise}
Noise analysis is performed on the post-layout parasitically-extracted netlist of the folded-cascode op amp obfuscated on the FPAA fabric to quantify the effect of the programmable fabric on analog noise. The input-referred noise spectral density and the integrated input-referred RMS noise over the signal band are evaluated and compared with a baseline folded-cascode op amp implemented as an ASIC. For the ASIC implementation and the FPAA-configured implementation of the folded-cascode op amp, the input-referred noise at 1 kHz is 458.855~nV/$\sqrt{\mathrm{Hz}}$ and 297.28~nV/$\sqrt{\mathrm{Hz}}$, respectively. The integrated input-referred RMS noise from 1~Hz to 1~MHz of the ASIC based and FPAA configured op amp is 56.58~{\textmu}$V_{RMS}$ and 36.92~{\textmu}$V_{RMS}$, respectively. The lower noise in the FPAA-based implementation is attributed to the larger effective device dimensions and area of the configured transistor pairs, which reduce the generated flicker-noise, and to the additional parasitic capacitances of the programmable fabric, which attenuate high-frequency noise transfer from internal high-impedance nodes. As a result, both the input-referred noise spectral density and the integrated RMS noise are reduced in the FPAA-configured implementation of the op amp.

\vspace{-10pt}
\begin{figure}[htbp]
\centering
\subfloat[]{\includegraphics[width=.23\textwidth]{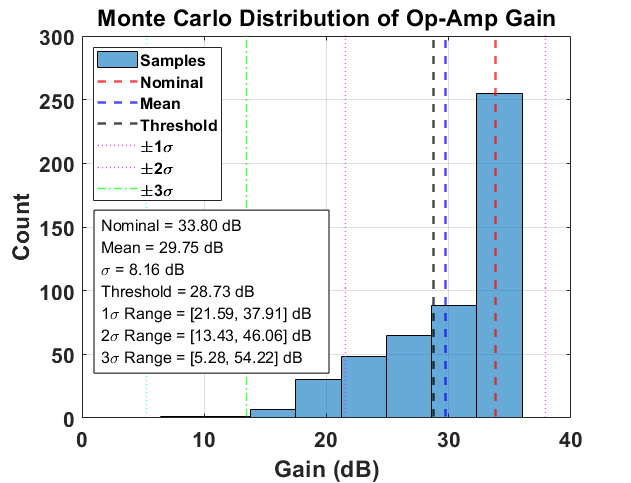}
\label{opamp_monte_gain_FPAA}}
\hfil
\subfloat[]{\includegraphics[width=.21\textwidth]{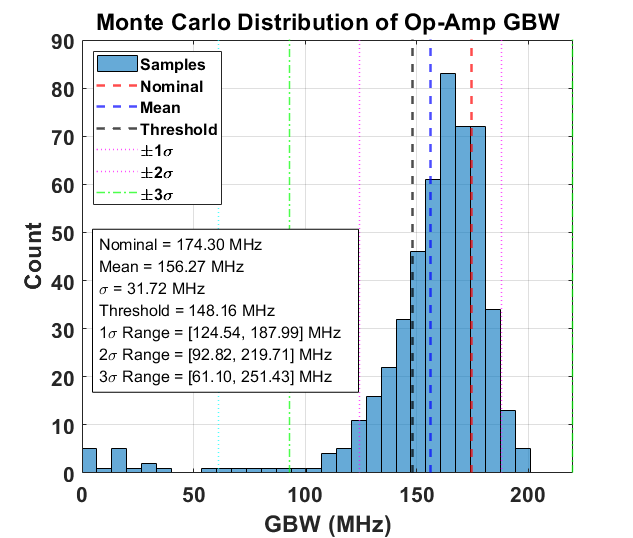}
\label{opamp_monte_GBW_FPAA}}
\hfil
\caption{Results from Monte Carlo simulation of the FPAA-configured op amp that characterize (a) gain and (b) gain-bandwidth product (GBW).}
\label{opamp_monte_gain_gbw_FPAA}
\end{figure}
\vspace{-15pt}

\subsection{Mismatch and Parametric Yield Analysis}\label{section-mismatch-yield}
Monte Carlo analysis is performed on the fully obfuscated folded-cascode op amp configured on the FPAA fabric to characterize the effect of mismatch and process variation on the performance of a programmed circuit implementation. The resulting statistical distributions of gain and gain-bandwidth product (GBW) are shown in Fig.~\ref{opamp_monte_gain_FPAA} and Fig.~\ref{opamp_monte_GBW_FPAA}, respectively. For the gain distribution, the Monte Carlo results exhibit a mean of 29.75~dB and a standard deviation of 8.16~dB, with a median of 32.47~dB. For the GBW distribution, the mean, standard deviation, and median are 156.3~MHz, 31.7~MHz, and 163.1~MHz, respectively. The results shown in Fig.~\ref{opamp_monte_gain_gbw_FPAA} indicate that both the gain and GBW remain centered near the intended operating region under statistical variation. However, the gain distribution is broader than the GBW distribution, which indicates that the gain is more sensitive to the mismatch and process variation of the FPAA fabric. To further quantify the robustness of a programmed circuit to variability, parametric yield is defined as the fraction of Monte Carlo samples that retain at least 85\% of the nominal correct-key performance. Using gain and GBW as the primary performance metrics, the corresponding yields are 71.2\% and 76.0\%, respectively. The results indicate that the FPAA-based full-topology-obfuscated implementation of the op amp remains functionally robust under realistic variations in fabrication while also providing a quantitative estimate of practical yield.

\vspace{-5pt}
\subsection{Configuration Latency}\label{section-configuration-latency}
The latency to configure the FPAA fabric is a one-time digital programming overhead incurred before circuit operation. Since the FPAA remains non-functional until the programming bitstream configures the transistor pairs, routing switches, and bias settings, the programming time to configure the FPAA is determined by the total number of bits needed to program the FPAA, which includes the clock frequency. For a single programmable CAB, the configuration keys are loaded through a 1404-bit shift-register. With a 5~kHz clock applied to the shift-register, the programming time of one CAB is 0.2808~s. The latency to configure the FPAA scales approximately linearly with the number of programmable CABs and does not affect the steady-state analog throughput after programming.

\vspace{-5pt}
\subsection{Area and Power Overhead}\label{section-area-overhead}
Partial topology obfuscation introduces relatively modest overhead. For the extracted folded-cascode op amp, obfuscating one, two, and three transistor pairs increases the area from 1925.93~{\textmu}$\text{m}^2$ to 2300.86~{\textmu}$\text{m}^2$, 2597.33~{\textmu}$\text{m}^2$, and 2932.37~{\textmu}$\text{m}^2$, and the power consumption from 1.63~mW to 1.65~mW, 1.72~mW, and 1.78~mW, respectively. When normalized to the 17 programmable bits per obfuscated transistor pair, the overhead is approximately 19.7~{\textmu}$\text{m}^2$ to 22.1~{\textmu}$\text{m}^2$ per security bit and 1.18~{\textmu}$\text{W}$ to 2.94~{\textmu}$\text{W}$ per security bit. At the system level, results from the characterization of the SAR ADC further confirm that obfuscating only the analog sub-blocks results in a limited penalty in area, with the total area increasing from 0.1302~mm$^2$ to 0.1319~mm$^2$ without inclusion of dummy transistors and to 0.134~mm$^2$ with the inclusion of dummy transistors.
In contrast, full FPAA based topology obfuscation provides stronger topological masking but incurs a significant overhead in area. A circuit mapped onto one CAB is configured by a 1404 bit shift-register, and the fully obfuscated folded-cascode op amp occupies 0.278~mm$^2$, corresponding to an overhead in area of approximately 196.5~{\textmu}$\text{m}^2$ per security bit relative to an unobfuscated ASIC baseline implementation. At the system level, the results from analyzing the delta-sigma modulator indicate that securing one or two sub-blocks with one or two CABs increases the area from 2150~{\textmu}$\text{m}^2$ to 0.278~mm$^2$ or 0.556~mm$^2$, respectively, and the power consumption from 1.28~mW to 1.56~mW or 1.84~mW, respectively, corresponding to an approximate overhead in area of 196.5~{\textmu}$\text{m}^2$ to 197.2~{\textmu}$\text{m}^2$ per security bit and an overhead in power consumption of 0.2~{\textmu}$\text{W}$ to 0.26~{\textmu}$\text{W}$ per security bit. Overall, partial topology obfuscation achieves lower area overhead per security bit, whereas full FPAA based topology obfuscation achieves stronger structural masking of a circuit, but at a higher cost in occupied area and consumed power.

\vspace{-5pt}
\subsection{Fabrication Feasibility}\label{section-fabrication-feasibility}
The feasibility of fabricating the proposed FPAA fabric is further strengthened with the use of standard CMOS-compatible circuit elements, including tunable-width transistor pairs, transmission-gate-based routing switches, programmable resistors and capacitors, and a resistor-ladder DAC that provides programmable biasing for implemented circuits. In addition, the post-layout parasitically-extracted views of the partially obfuscated circuits and fully-obfuscated circuits implemented on the FPAA fabric are evaluated. The configurable fabric is structured to allow programming of practical analog components, with a topology that permits a maximum of six stacked transistor pairs and three differential stages per CAB. Larger circuits are partitioned across multiple CABs. Together with the reported area and configuration-latency, the FPAA supports the implementation and programming of analog circuits.

\vspace{-5pt}
\subsection{Scalability}
The scalability of the proposed security framework is supported by the methodological use of partial and full topology-obfuscation. Partial topology obfuscation scales through selective replacement of only the most suitable transistor pairs in an existing ASIC implementation, where a performance-aware algorithm determines which pairs are obfuscated while satisfying a specified tolerance in degraded performance. In contrast, FPAA-based full topology obfuscation scales hierarchically by partitioning a large analog or mixed-signal system into sub-blocks, with each sub-block mapped onto one CAB and interconnections between CABs established through global routing switches. Therefore, the proposed hybrid ASIC-FPAA framework supports both low-overhead selective obfuscation and multi-CAB full topology obfuscation of larger systems.

%% file: section8.tex
\section{Evaluation of Security Robustness}\label{GG}
In this section, the proposed techniques to obfuscate the topology of circuit are evaluated using the latest analog attack algorithms, including the equation-based SMT attack, the genetic-algorithm attack, the monotonic circuit response attack, the DC nodal analysis attack, and the topology analysis attack. The results from execution of each analog attack algorithm are provided, while also accounting for the complexity of the circuit topology after obfuscation. The evaluation of the analog attack algorithms is performed on a server that includes two 12-core Intel Xeon E5 CPUs and 96 GB of DDR4 memory.

\begin{table*}[]
\caption{Results of Executing the SMT, GA, Monotonic, and DNA Attack on a Folded-cascode Op Amp, Comparator, and SAR-ADC with Different Number of Transistor Pairs Obfuscated. The “*” Symbol Indicates the Failure of Determining the Correct Key.}
\label{table_attack}
\resizebox{\textwidth}{!}{%

\begin{tabular}{|c|cccc|cccccc|cccc|cccc|}
\hline
\multirow{2}{*}{\textbf{\begin{tabular}[c]{@{}c@{}}Deobfuscation Attack\end{tabular}}} &
  \multicolumn{4}{c|}{\textbf{SMT Attack}} &
  \multicolumn{6}{c|}{\textbf{Genetic Algorithm Attack}} &
  \multicolumn{4}{c|}{\textbf{Monotonic Attack}} &
  \multicolumn{4}{c|}{\textbf{DNA Attack}} \\ \cline{2-19} 
 &
  \multicolumn{1}{c|}{Tol.} &
  \multicolumn{1}{c|}{Pairs} &
  \multicolumn{1}{c|}{Keys} &
  Time &
  \multicolumn{1}{c|}{Tol.} &
  \multicolumn{1}{c|}{Pairs} &
  \multicolumn{1}{c|}{Pop.} &
  \multicolumn{1}{c|}{Gen.} &
  \multicolumn{1}{c|}{Keys} &
  Time &
  \multicolumn{1}{c|}{Tol.} &
  \multicolumn{1}{c|}{Pairs} &
  \multicolumn{1}{c|}{Keys} &
  Time &
  \multicolumn{1}{c|}{Tol.} &
  \multicolumn{1}{c|}{Pairs} &
  \multicolumn{1}{c|}{Keys} &
  Time \\ \hline
\multirow{3}{*}{\textbf{\begin{tabular}[c]{@{}c@{}}Folded-Cascode\\ Op Amp\end{tabular}}} &
  \multicolumn{1}{c|}{10\%} &
  \multicolumn{1}{c|}{1} &
  \multicolumn{1}{c|}{55} &
  0.25 s &
  \multicolumn{1}{c|}{10\%} &
  \multicolumn{1}{c|}{1} &
  \multicolumn{1}{c|}{20} &
  \multicolumn{1}{c|}{20} &
  \multicolumn{1}{c|}{1} &
  7.5 H &
  \multicolumn{1}{c|}{10\%} &
  \multicolumn{1}{c|}{1} &
  \multicolumn{1}{c|}{1} &
  136.47 s &
  \multicolumn{1}{c|}{10\%} &
  \multicolumn{1}{c|}{1} &
  \multicolumn{1}{c|}{32} &
  255.15 s \\ \cline{2-19} 
 &
  \multicolumn{1}{c|}{10\%} &
  \multicolumn{1}{c|}{2} &
  \multicolumn{1}{c|}{4032} &
  33.13 s &
  \multicolumn{1}{c|}{10\%} &
  \multicolumn{1}{c|}{2} &
  \multicolumn{1}{c|}{20} &
  \multicolumn{1}{c|}{20} &
  \multicolumn{1}{c|}{1*} &
  7.8 H &
  \multicolumn{1}{c|}{10\%} &
  \multicolumn{1}{c|}{2} &
  \multicolumn{1}{c|}{13*} &
  411.06 s &
  \multicolumn{1}{c|}{10\%} &
  \multicolumn{1}{c|}{2} &
  \multicolumn{1}{c|}{378*} &
  734.74 s \\ \cline{2-19} 
 &
  \multicolumn{1}{c|}{10\%} &
  \multicolumn{1}{c|}{3} &
  \multicolumn{1}{c|}{200704*} &
  6949 s &
  \multicolumn{1}{c|}{10\%} &
  \multicolumn{1}{c|}{3} &
  \multicolumn{1}{c|}{20} &
  \multicolumn{1}{c|}{20} &
  \multicolumn{1}{c|}{1*} &
  7.65 H &
  \multicolumn{1}{c|}{10\%} &
  \multicolumn{1}{c|}{3} &
  \multicolumn{1}{c|}{6*} &
  509.29 s &
  \multicolumn{1}{c|}{10\%} &
  \multicolumn{1}{c|}{3} &
  \multicolumn{1}{c|}{111*} &
  1241.19 s \\ \hline
\multirow{3}{*}{\textbf{Comparator}} &
  \multicolumn{1}{c|}{10\%} &
  \multicolumn{1}{c|}{1} &
  \multicolumn{1}{c|}{12} &
  0.1 s &
  \multicolumn{1}{c|}{10\%} &
  \multicolumn{1}{c|}{1} &
  \multicolumn{1}{c|}{20} &
  \multicolumn{1}{c|}{20} &
  \multicolumn{1}{c|}{1} &
  7.81 H &
  \multicolumn{1}{c|}{10\%} &
  \multicolumn{1}{c|}{1} &
  \multicolumn{1}{c|}{1} &
  113.13 s &
  \multicolumn{1}{c|}{10\%} &
  \multicolumn{1}{c|}{1} &
  \multicolumn{1}{c|}{56} &
  23.94 s \\ \cline{2-19} 
 &
  \multicolumn{1}{c|}{10\%} &
  \multicolumn{1}{c|}{2} &
  \multicolumn{1}{c|}{253*} &
  2.5 s &
  \multicolumn{1}{c|}{10\%} &
  \multicolumn{1}{c|}{2} &
  \multicolumn{1}{c|}{20} &
  \multicolumn{1}{c|}{20} &
  \multicolumn{1}{c|}{1*} &
  7.73 H &
  \multicolumn{1}{c|}{10\%} &
  \multicolumn{1}{c|}{2} &
  \multicolumn{1}{c|}{15*} &
  448.26 s &
  \multicolumn{1}{c|}{10\%} &
  \multicolumn{1}{c|}{2} &
  \multicolumn{1}{c|}{577*} &
  227.96 s \\ \cline{2-19} 
 &
  \multicolumn{1}{c|}{10\%} &
  \multicolumn{1}{c|}{3} &
  \multicolumn{1}{c|}{1877*} &
  20 s &
  \multicolumn{1}{c|}{10\%} &
  \multicolumn{1}{c|}{3} &
  \multicolumn{1}{c|}{20} &
  \multicolumn{1}{c|}{20} &
  \multicolumn{1}{c|}{1*} &
  7.95 H &
  \multicolumn{1}{c|}{10\%} &
  \multicolumn{1}{c|}{3} &
  \multicolumn{1}{c|}{25*} &
  968.43 s &
  \multicolumn{1}{c|}{10\%} &
  \multicolumn{1}{c|}{3} &
  \multicolumn{1}{c|}{1129*} &
  443.73 s \\ \hline
\multirow{3}{*}{\textbf{SAR-ADC}} &
  \multicolumn{4}{c|}{\multirow{3}{*}{Cannot Perform Attack}} &
  \multicolumn{1}{c|}{10\%} &
  \multicolumn{1}{c|}{1} &
  \multicolumn{1}{c|}{20} &
  \multicolumn{1}{c|}{20} &
  \multicolumn{1}{c|}{1} &
  107.84 H &
  \multicolumn{1}{c|}{10\%} &
  \multicolumn{1}{c|}{1} &
  \multicolumn{1}{c|}{16} &
  4950.12 s &
  \multicolumn{4}{c|}{\multirow{3}{*}{Cannot Perform Attack}} \\ \cline{6-15}
 &
  \multicolumn{4}{c|}{} &
  \multicolumn{1}{c|}{10\%} &
  \multicolumn{1}{c|}{2} &
  \multicolumn{1}{c|}{20} &
  \multicolumn{1}{c|}{20} &
  \multicolumn{1}{c|}{1*} &
  105.15 H &
  \multicolumn{1}{c|}{10\%} &
  \multicolumn{1}{c|}{2} &
  \multicolumn{1}{c|}{48*} &
  3.72 H &
  \multicolumn{4}{c|}{} \\ \cline{6-15}
 &
  \multicolumn{4}{c|}{} &
  \multicolumn{1}{c|}{10\%} &
  \multicolumn{1}{c|}{3} &
  \multicolumn{1}{c|}{20} &
  \multicolumn{1}{c|}{20} &
  \multicolumn{1}{c|}{1*} &
  107.69 H &
  \multicolumn{1}{c|}{10\%} &
  \multicolumn{1}{c|}{3} &
  \multicolumn{1}{c|}{3319*} &
  223 H &
  \multicolumn{4}{c|}{} \\ \hline
\end{tabular}

}
\end{table*}

% \subsection{Information Required to Execute Analog Attacks}\label{threate_model}
% Knowledge of the process design kit (PDK) is required to perform an analog attack, which provides the adversary with fundamental information on the technology node utilized to fabricate an analog circuit. Parameters defined by the PDK include charge carrier mobility, gate-oxide capacitance, and threshold voltage. The PDK is essential to perform SPICE simulations on circuits reconstructed by an attacker.

% The second piece of required information is the netlist of the obfuscated circuit. The netlist of an IC is typically obtained through reverse engineering \cite{Reverse}, which involves IC depackaging, IC delayering, circuit imaging, device annotation, and netlist extraction. The extracted netlist of an obfuscated IC is needed to decrypt the key or bitstream that sets the target functionality of the circuit.

% The third component of required information is the biasing conditions of the target analog circuit, which includes the set gate voltages and/or the biasing currents provided by a current source. Determining the bias conditions is critical when reconstructing a circuit model through SMT attack. If the biasing information of an analog circuit is locked, the attacker must estimate the probable range of bias voltages based on information from the PDK, which is limited to between the threshold voltage of the devices and the supply voltage of the circuit.

\subsection{Analog Attack Techniques}\label{section_analog_attack}
Conventional SAT-based attacks are not directly applicable to the proposed FPAA-based obfuscation technique as the programmed analog circuit is not directly represented by a Boolean representation and the circuit functionality depends on topology, sizing, routing, and bias parameters. In addition, machine-learning-based attacks are hindered by the need to jointly recover active transistor pairs, switch-box routing, transistor sizes, and bias settings, which substantially increase the search space.
Various analog attack methodologies have been developed to determine the key of an obfuscated analog IC, including the equation-based satisfiability modulo theory (SMT) attack\cite{Attack_SMT,Attack_Metric}, genetic-algorithm (GA) based attack\cite{Attack_Gen, Attack_Metric}, monotonic circuit response attack\cite{Attack_Mon,Attack_Metric}, DC nodal analysis (DNA) attack\cite{Attack_DNA}, and topology attack\cite{Attack_FPAA}. 

\begin{table*}[]
\caption{Analysis of the topology complexity and the attack performance metric when executing the SMT, GA, monotonic, and DNA attacks on a folded-cascode op amp, comparator, and SAR-ADC with partial topology obfuscation. The attack performance metric is reported as a lower/upper bound after pruning structurally invalid topology assignments. The ``*'' symbol indicates failure to determine the correct key.}
\label{table_attack_metric}
\resizebox{\textwidth}{!}{%
\begin{tabular}{|c|c|c|c|ccc|ccc|ccc|ccc|}
\hline
\multirow{2}{*}{\begin{tabular}[c]{@{}c@{}}Circuit\\ Architecture\end{tabular}} &
\multirow{2}{*}{\begin{tabular}[c]{@{}c@{}}Obfuscated\\ TPs\\ $(N)$\end{tabular}} &
\multirow{2}{*}{\begin{tabular}[c]{@{}c@{}}Topology\\ Complexity Bounds\\ $(TC_{LOW}/TC_{UP})$\end{tabular}} &
\multirow{2}{*}{\begin{tabular}[c]{@{}c@{}}Ex. Time\\ of Single\\ Transient Sim.\end{tabular}} &
\multicolumn{3}{c|}{SMT Attack} &
\multicolumn{3}{c|}{Genetic Algorithm Attack} &
\multicolumn{3}{c|}{Monotonic Attack} &
\multicolumn{3}{c|}{DNA Attack} \\ \cline{5-16}
 &
 &
 &
 &
\multicolumn{1}{c|}{\begin{tabular}[c]{@{}c@{}}Attack\\ Ex. Time\end{tabular}} &
\multicolumn{1}{c|}{\begin{tabular}[c]{@{}c@{}}Candidate\\ Keys\end{tabular}} &
\begin{tabular}[c]{@{}c@{}}Attack Perf.\\ Metric\\ $(LB/UB)$\end{tabular} &
\multicolumn{1}{c|}{\begin{tabular}[c]{@{}c@{}}Attack\\ Ex. Time\end{tabular}} &
\multicolumn{1}{c|}{\begin{tabular}[c]{@{}c@{}}Candidate\\ Keys\end{tabular}} &
\begin{tabular}[c]{@{}c@{}}Attack Perf.\\ Metric\\ $(LB/UB)$\end{tabular} &
\multicolumn{1}{c|}{\begin{tabular}[c]{@{}c@{}}Attack\\ Ex. Time\end{tabular}} &
\multicolumn{1}{c|}{\begin{tabular}[c]{@{}c@{}}Candidate\\ Keys\end{tabular}} &
\begin{tabular}[c]{@{}c@{}}Attack Perf.\\ Metric\\ $(LB/UB)$\end{tabular} &
\multicolumn{1}{c|}{\begin{tabular}[c]{@{}c@{}}Attack\\ Ex. Time\end{tabular}} &
\multicolumn{1}{c|}{\begin{tabular}[c]{@{}c@{}}Candidate\\ Keys\end{tabular}} &
\begin{tabular}[c]{@{}c@{}}Attack Perf.\\ Metric\\ $(LB/UB)$\end{tabular} \\ \hline

\multirow{4}{*}{\begin{tabular}[c]{@{}c@{}}Folded-Cascode\\ Op Amp\end{tabular}} &
0 & $1/1$ & 14.8 s &
\multicolumn{1}{c|}{0.25 s} &
\multicolumn{1}{c|}{55} &
\begin{tabular}[c]{@{}c@{}}814.25 /\\ 814.25\end{tabular} &
\multicolumn{1}{c|}{7.5 H} &
\multicolumn{1}{c|}{1} &
\begin{tabular}[c]{@{}c@{}}27014.8 /\\ 27014.8\end{tabular} &
\multicolumn{1}{c|}{136.47 s} &
\multicolumn{1}{c|}{1} &
\begin{tabular}[c]{@{}c@{}}151.27 /\\ 151.27\end{tabular} &
\multicolumn{1}{c|}{255.15 s} &
\multicolumn{1}{c|}{32} &
\begin{tabular}[c]{@{}c@{}}728.75 /\\ 728.75\end{tabular} \\ \cline{2-16}

& 1 & $4/5$ & 14.8 s &
\multicolumn{1}{c|}{0.25 s} &
\multicolumn{1}{c|}{55} &
\begin{tabular}[c]{@{}c@{}}3257 /\\ 4071.25\end{tabular} &
\multicolumn{1}{c|}{7.5 H} &
\multicolumn{1}{c|}{1} &
\begin{tabular}[c]{@{}c@{}}108059.2 /\\ 135074\end{tabular} &
\multicolumn{1}{c|}{136.47 s} &
\multicolumn{1}{c|}{1} &
\begin{tabular}[c]{@{}c@{}}605.08 /\\ 756.35\end{tabular} &
\multicolumn{1}{c|}{255.15 s} &
\multicolumn{1}{c|}{32} &
\begin{tabular}[c]{@{}c@{}}2915 /\\ 3643.75\end{tabular} \\ \cline{2-16}

& 2 & $16/25$ & 23.8 s &
\multicolumn{1}{c|}{33.13 s} &
\multicolumn{1}{c|}{4032} &
\begin{tabular}[c]{@{}c@{}}1535918.72 /\\ 2399868\end{tabular} &
\multicolumn{1}{c|}{7.8 H} &
\multicolumn{1}{c|}{1*} &
\begin{tabular}[c]{@{}c@{}}449660.8 /\\ 702595*\end{tabular} &
\multicolumn{1}{c|}{411.06 s} &
\multicolumn{1}{c|}{13*} &
\begin{tabular}[c]{@{}c@{}}11527.36 /\\ 18011.5*\end{tabular} &
\multicolumn{1}{c|}{734.74 s} &
\multicolumn{1}{c|}{378*} &
\begin{tabular}[c]{@{}c@{}}155698.24 /\\ 243278.5*\end{tabular} \\ \cline{2-16}

& 3 & $64/125$ & 26.5 s &
\multicolumn{1}{c|}{6949 s} &
\multicolumn{1}{c|}{200704*} &
\begin{tabular}[c]{@{}c@{}}340838720 /\\ 665700625*\end{tabular} &
\multicolumn{1}{c|}{7.65 H} &
\multicolumn{1}{c|}{1*} &
\begin{tabular}[c]{@{}c@{}}1764255.74 /\\ 3445812*\end{tabular} &
\multicolumn{1}{c|}{509.29 s} &
\multicolumn{1}{c|}{6*} &
\begin{tabular}[c]{@{}c@{}}42770.56 /\\ 83536.25*\end{tabular} &
\multicolumn{1}{c|}{1241.19 s} &
\multicolumn{1}{c|}{111*} &
\begin{tabular}[c]{@{}c@{}}267692.16 /\\ 522836.25*\end{tabular} \\ \hline

\multirow{4}{*}{Comparator} &
0 & $1/1$ & 18.9 s &
\multicolumn{1}{c|}{0.1 s} &
\multicolumn{1}{c|}{12} &
\begin{tabular}[c]{@{}c@{}}226.9 /\\ 226.9\end{tabular} &
\multicolumn{1}{c|}{7.81 H} &
\multicolumn{1}{c|}{1} &
\begin{tabular}[c]{@{}c@{}}28134.8 /\\ 28134.8\end{tabular} &
\multicolumn{1}{c|}{113.13 s} &
\multicolumn{1}{c|}{1} &
\begin{tabular}[c]{@{}c@{}}132.03 /\\ 132.03\end{tabular} &
\multicolumn{1}{c|}{23.94 s} &
\multicolumn{1}{c|}{56} &
\begin{tabular}[c]{@{}c@{}}1082.34 /\\ 1082.34\end{tabular} \\ \cline{2-16}

& 1 & $4/5$ & 18.9 s &
\multicolumn{1}{c|}{0.1 s} &
\multicolumn{1}{c|}{12} &
\begin{tabular}[c]{@{}c@{}}907.6 /\\ 1134.5\end{tabular} &
\multicolumn{1}{c|}{7.81 H} &
\multicolumn{1}{c|}{1} &
\begin{tabular}[c]{@{}c@{}}112539.2 /\\ 140674\end{tabular} &
\multicolumn{1}{c|}{113.13 s} &
\multicolumn{1}{c|}{1} &
\begin{tabular}[c]{@{}c@{}}528.12 /\\ 660.15\end{tabular} &
\multicolumn{1}{c|}{23.94 s} &
\multicolumn{1}{c|}{56} &
\begin{tabular}[c]{@{}c@{}}4329.36 /\\ 5411.7\end{tabular} \\ \cline{2-16}

& 2 & $16/25$ & 29.9 s &
\multicolumn{1}{c|}{2.5 s} &
\multicolumn{1}{c|}{253*} &
\begin{tabular}[c]{@{}c@{}}121075.2 /\\ 189180*\end{tabular} &
\multicolumn{1}{c|}{7.73 H} &
\multicolumn{1}{c|}{1*} &
\begin{tabular}[c]{@{}c@{}}445726.08 /\\ 696447*\end{tabular} &
\multicolumn{1}{c|}{448.26 s} &
\multicolumn{1}{c|}{15*} &
\begin{tabular}[c]{@{}c@{}}14348.16 /\\ 22419*\end{tabular} &
\multicolumn{1}{c|}{227.96 s} &
\multicolumn{1}{c|}{577*} &
\begin{tabular}[c]{@{}c@{}}279684.16 /\\ 437006.5*\end{tabular} \\ \cline{2-16}

& 3 & $64/125$ & 39.5 s &
\multicolumn{1}{c|}{20 s} &
\multicolumn{1}{c|}{1877*} &
\begin{tabular}[c]{@{}c@{}}4746336 /\\ 9270187.5*\end{tabular} &
\multicolumn{1}{c|}{7.95 H} &
\multicolumn{1}{c|}{1*} &
\begin{tabular}[c]{@{}c@{}}1834207.74 /\\ 3582437*\end{tabular} &
\multicolumn{1}{c|}{968.43 s} &
\multicolumn{1}{c|}{25*} &
\begin{tabular}[c]{@{}c@{}}125179.52 /\\ 244491.25*\end{tabular} &
\multicolumn{1}{c|}{443.73 s} &
\multicolumn{1}{c|}{1129*} &
\begin{tabular}[c]{@{}c@{}}2882510.69 /\\ 5629903.7*\end{tabular} \\ \hline

\multirow{4}{*}{\begin{tabular}[c]{@{}c@{}}SAR-\\ ADC\end{tabular}} &
0 & $1/1$ & 99.7 s &
\multicolumn{3}{c|}{\multirow{4}{*}{Cannot Perform Attack}} &
\multicolumn{1}{c|}{107.84 H} &
\multicolumn{1}{c|}{1} &
\begin{tabular}[c]{@{}c@{}}388323.6 /\\ 388323.6\end{tabular} &
\multicolumn{1}{c|}{4950.12 s} &
\multicolumn{1}{c|}{16} &
\begin{tabular}[c]{@{}c@{}}6545.32 /\\ 6545.32\end{tabular} &
\multicolumn{3}{c|}{\multirow{4}{*}{Cannot Perform Attack}} \\ \cline{2-4} \cline{8-13}

& 1 & $4/5$ & 99.7 s &
\multicolumn{3}{c|}{} &
\multicolumn{1}{c|}{107.84 H} &
\multicolumn{1}{c|}{1} &
\begin{tabular}[c]{@{}c@{}}1553294.4 /\\ 1941618\end{tabular} &
\multicolumn{1}{c|}{4950.12 s} &
\multicolumn{1}{c|}{16} &
\begin{tabular}[c]{@{}c@{}}26181.28 /\\ 32726.6\end{tabular} &
\multicolumn{3}{c|}{} \\ \cline{2-4} \cline{8-13}

& 2 & $16/25$ & 101.3 s &
\multicolumn{3}{c|}{} &
\multicolumn{1}{c|}{105.15 H} &
\multicolumn{1}{c|}{1*} &
\begin{tabular}[c]{@{}c@{}}6058260.48 /\\ 9466032*\end{tabular} &
\multicolumn{1}{c|}{3.72 H} &
\multicolumn{1}{c|}{48*} &
\begin{tabular}[c]{@{}c@{}}292070.4 /\\ 456360*\end{tabular} &
\multicolumn{3}{c|}{} \\ \cline{2-4} \cline{8-13}

& 3 & $64/125$ & 106.2 s &
\multicolumn{3}{c|}{} &
\multicolumn{1}{c|}{107.69 H} &
\multicolumn{1}{c|}{1*} &
\begin{tabular}[c]{@{}c@{}}24818572.8 /\\ 48473775*\end{tabular} &
\multicolumn{1}{c|}{223 H} &
\multicolumn{1}{c|}{3319*} &
\begin{tabular}[c]{@{}c@{}}73937779.2 /\\ 144409725*\end{tabular} &
\multicolumn{3}{c|}{} \\ \hline
\end{tabular}
}
\end{table*}

\subsection{Executing Attack Algorithms on Partial Topology Obfuscation}\label{section_execute_attack}
For partial topology obfuscation, the SMT, GA, monotonic, DNA, and topology attacks are evaluated considering a simplified threat model in which either the topology of the transistor pairs or the sizing of transistors is assumed known by the adversary. The assumption permits execution of the attack algorithms and provides a common basis for benchmarking. For full circuit FPAA-based obfuscation, no known attack is available that simultaneously determines the topology, transistor sizing, routing configuration, and bias settings of a circuit masked by the FPAA fabric. In that case, the adversary must determine the active transistors, routing connections, programmed transistor-pair topologies, transistor sizes, and applied bias settings, which is equivalent to redesigning the circuit from scratch on the unprogrammed FPAA. Consequently, the reported results from execution of the attack algorithms only apply to partial topology obfuscation and do not represent the cost of a direct attack on the FPAA fabric when full topology obfuscation is implemented.

\begin{figure}[htbp]
\centering
\subfloat[]{\includegraphics[width=.24\textwidth]{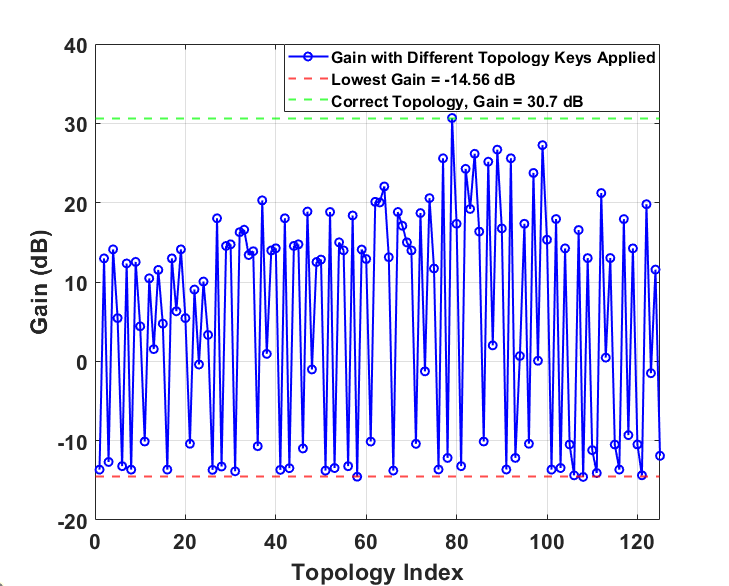}}
\label{Attack_Gain}
\hspace{-15pt}
\hfil
\subfloat[]{\includegraphics[width=.25\textwidth]{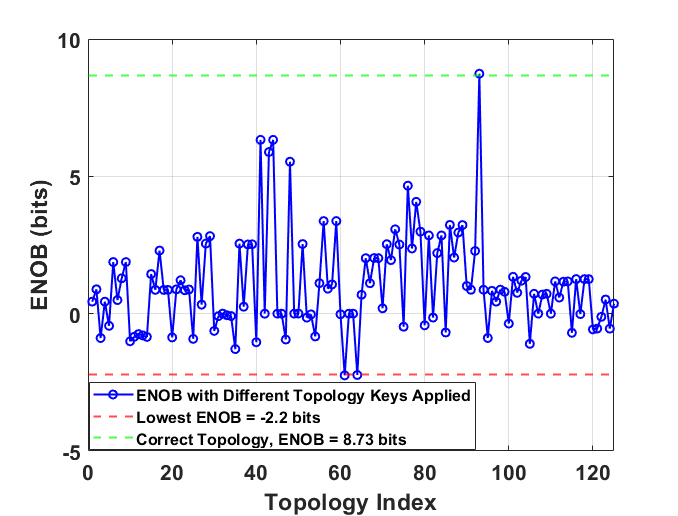}}
\label{Attack_ENOB}
\hfil

\caption{Analysis of the (a) gain of a folded-cascode op amp and  (b) ENOB of a SAR ADC when a topology attack is executed on each circuit with three transistor pairs obfuscated.}
\label{Attack_topology}
\end{figure}

The analog attacks are, therefore, executed on circuits obfuscated using the partial topology obfuscation technique.
To execute the topology attack, an adversary is assumed to possess the PDK information, the circuit netlist, and access to the programming interface of the FPAA, which permits the setting of the topology of transistor pairs. 
Unlike for the other four attacks, the sizing of the transistors is assumed known for the topology attack, which is used purely for evaluation purposes as sizing of transistors is highly dependent on topology.
The results from executing a topology attack on a folded-cascode op amp and a SAR ADC each with three transistor pairs obfuscated are shown in Fig.~\ref{Attack_topology}. For the op amp, all incorrect keys result in a minimum of 10\% deviation in the gain, with 93\% of the incorrect keys providing a minimum deviation in gain of 28\%. For the SAR ADC, all incorrect keys result in at least a 27.8\% reduction in the ENOB, with 95\% of the incorrect keys providing a reduction of at least 61.3\%.

For the equation-based SMT attack, the adversary is assumed to have access to the netlist of the obfuscated circuit, the PDK information, and the biasing voltages and currents of the circuit.
Since the formulated equations used in an SMT attack depend on the topology of the target analog circuit, the correct key that sets the topology is assumed already applied, which implies that the programmable transistor pairs are correctly configured into the proper topology. As a result, the SMT attack targets only the key bits that set the size of the programmable transistor pairs.
The results from executing the SMT attack on the obfuscated folded-cascode op amp, comparator, and SAR ADC are listed in Table~\ref{table_attack}. A tolerance of 10\% is applied to the solver when determining the correct key, where any key that results in a performance parameter that falls within 10\% of the performance of the target circuit is considered a candidate key.
For the folded-cascode op amp, when only one transistor pair is obfuscated, 55 candidate keys are returned, including the correct key that properly sizes the circuit. When two transistor pairs are obfuscated, 4032 candidate keys are returned, including the correct key. However, returning 4032 keys is effectively equivalent to a brute-force attack, as a 12-bit key provides a total of 4096 possible key combinations. In addition, when three transistor pairs are obfuscated, no correct key is returned.
For the obfuscated comparator, the correct key is only returned when one transistor pair is obfuscated.
The equation-based SMT attack does not execute on a SAR ADC as the formulated equations rely on static small-signal or DC analysis and do not describe the transient behavior of an ADC. Consequently, the ADC performance metrics including the ENOB are not determined using SMT problem formulations.

%\vspace{-10pt}

For the GA attack, the adversary is assumed to have access to the key inputs and the output ports of the oracle IC. The performance metrics evaluated on an oracle IC are measured and analyzed for different applied keys. The size of the population and the number of generations are set to 20. A tolerance of 10\% of the target performance parameter is applied when determining the correct key.
The results from executing the GA attack are listed in Table~\ref{table_attack}. For the folded-cascode op amp and comparator each with one transistor pair obfuscated, the correct key is identified after 7.5 hours and 7.81 hours, respectively. However, for the SAR ADC with one transistor pair obfuscated, the correct key is determined after 107.84 hours. The additional time required is due to the computational cost of transient simulation, which is necessary to characterize the fast fourier transform (FFT) spectrum and the ENOB of the ADC.
For all benchmark circuits, the GA attack fails to determine the key that correctly sizes the circuit when two or three transistor pairs are obfuscated.
For the monotonic attack, the adversary is assumed to have access to the input and output ports of the target circuit. The results from executing the monotonic attack are listed in Table~\ref{table_attack}. The monotonic attack successfully determines the correct key that sizes each of the three benchmark circuits when only one transistor pair is obfuscated due to the strong correlation between the key width and the output response of the circuits. However, when more than one transistor pair is obfuscated in each circuit, the monotonic attack fails to determine the correct key due to the nonlinear behavior of the circuits.
The DC nodal analysis (DNA) attack utilizes Kirchhoff’s current law (KCL) and Kirchhoff’s voltage law (KVL)  to determine the key of the obfuscated circuit. The attacker is assumed to have access to the PDK information, the netlist of the obfuscated circuit, and the biasing conditions of the circuit. The results from execution of the DNA attack are listed in Table~\ref{table_attack}. When only one transistor pair is obfuscated in each circuit, the DNA attack determines the correct key that properly sizes the folded-cascode op amp and comparator in 255.15 s and 23.94 s, respectively. The DNA attack fails to determine the correct key when two or three transistor pairs are obfuscated in each circuit.
In addition, the DNA attack cannot be executed on the SAR ADC as the DC signal behavior of the circuit is not sufficient to represent the transient behavior of the ADC. Consequently, the FFT spectrum and the ENOB of the ADC are not properly modeled using the DNA attack.

\begin{table}[]
\caption{Performance-security trade-off of analog circuits implemented with partial topology obfuscation. The increase in APM and the FOM are reported as lower bounds (LB) and upper bounds (UB) corresponding to \(TC_{LOW}=4^N\) and \(TC_{UP}=5^N\), respectively, for \emph{N} obfuscated transistor pairs.}
\label{table_FOM}
\centering
\resizebox{\columnwidth}{!}{%
\begin{tabular}{|c|c|c|c|c|cc|cc|}
\hline
\multirow{2}{*}{\begin{tabular}[c]{@{}c@{}}Circuit \\ Topology\end{tabular}} &
  \multirow{2}{*}{\begin{tabular}[c]{@{}c@{}}Num. of\\ Obfus.\\ Pairs\end{tabular}} &
  \multirow{2}{*}{\(\Delta_{\mathrm{Spec.}}\)} &
  \multirow{2}{*}{\(\Delta_{\mathrm{Power}}\)} &
  \multirow{2}{*}{\(\Delta_{\mathrm{Area}}\)} &
  \multicolumn{2}{c|}{GA Attack} &
  \multicolumn{2}{c|}{\begin{tabular}[c]{@{}c@{}}Monotonic\\ Attack\end{tabular}} \\ \cline{6-9} 
 &
   &
   &
   &
   &
  \multicolumn{1}{c|}{\begin{tabular}[c]{@{}c@{}}Increase \\ in APM\\ (LB/UB)\end{tabular}} &
  \begin{tabular}[c]{@{}c@{}}FOM\\ LB /\\ UB\end{tabular} &
  \multicolumn{1}{c|}{\begin{tabular}[c]{@{}c@{}}Increase \\ in APM\\ (LB/UB)\end{tabular}} &
  \begin{tabular}[c]{@{}c@{}}FOM\\ LB /\\ UB\end{tabular} \\ \hline

\multirow{3}{*}{\begin{tabular}[c]{@{}c@{}}Folded-\\ Cascode\\ Op Amp\end{tabular}} &
  1 &
  2.5\% &
  1.2\% &
  19.4\% &
  \multicolumn{1}{c|}{4x / 5x} &
  \begin{tabular}[c]{@{}c@{}}\(6.87\times10^4\) /\\\(8.59\times10^4\)\end{tabular} &
  \multicolumn{1}{c|}{4x / 5x} &
  \begin{tabular}[c]{@{}c@{}}\(6.87\times10^4\) /\\\(8.59\times10^4\)\end{tabular} \\ \cline{2-9} 
 &
  2 &
  4.6\% &
  5.5\% &
  34.9\% &
  \multicolumn{1}{c|}{16.6x / 26x} &
  \begin{tabular}[c]{@{}c@{}}\(1.88\times10^4\) /\\\(2.94\times10^4\)\end{tabular} &
  \multicolumn{1}{c|}{76.2x / 119.1x} &
  \begin{tabular}[c]{@{}c@{}}\(8.63\times10^4\) /\\\(1.35\times10^5\)\end{tabular} \\ \cline{2-9} 
 &
  3 &
  15\% &
  9.2\% &
  52.3\% &
  \multicolumn{1}{c|}{65.3x / 127.6x} &
  \begin{tabular}[c]{@{}c@{}}\(9.05\times10^3\) /\\\(1.77\times10^4\)\end{tabular} &
  \multicolumn{1}{c|}{282.7x / 552.2x} &
  \begin{tabular}[c]{@{}c@{}}\(3.92\times10^4\) /\\\(7.65\times10^4\)\end{tabular} \\ \hline

\multirow{3}{*}{Comparator} &
  1 &
  4\% &
  14.2\% &
  1.97x &
  \multicolumn{1}{c|}{4x / 5x} &
  \begin{tabular}[c]{@{}c@{}}\(3.58\times10^2\) /\\\(4.47\times10^2\)\end{tabular} &
  \multicolumn{1}{c|}{4x / 5x} &
  \begin{tabular}[c]{@{}c@{}}\(3.58\times10^2\) /\\\(4.47\times10^2\)\end{tabular} \\ \cline{2-9} 
 &
  2 &
  16.8\% &
  31.6\% &
  4.02x &
  \multicolumn{1}{c|}{15.9x / 24.8x} &
  \begin{tabular}[c]{@{}c@{}}\(7.44\times10^1\) /\\\(1.16\times10^2\)\end{tabular} &
  \multicolumn{1}{c|}{108.7x / 169.8x} &
  \begin{tabular}[c]{@{}c@{}}\(5.09\times10^4\) /\\\(7.96\times10^4\)\end{tabular} \\ \cline{2-9} 
 &
  3 &
  33.5\% &
  44.3\% &
  5.57x &
  \multicolumn{1}{c|}{65.2x / 127.3x} &
  \begin{tabular}[c]{@{}c@{}}\(7.88\times10^1\) /\\\(1.54\times10^2\)\end{tabular} &
  \multicolumn{1}{c|}{948.1x / 1851.8x} &
  \begin{tabular}[c]{@{}c@{}}\(1.15\times10^5\) /\\\(2.24\times10^5\)\end{tabular} \\ \hline

\multirow{3}{*}{SAR-ADC} &
  1 &
  0.35\% &
  7.8\% &
  0.3\% &
  \multicolumn{1}{c|}{4x / 5x} &
  \begin{tabular}[c]{@{}c@{}}\(4.88\times10^6\) /\\\(6.11\times10^6\)\end{tabular} &
  \multicolumn{1}{c|}{4x / 5x} &
  \begin{tabular}[c]{@{}c@{}}\(4.88\times10^6\) /\\\(6.11\times10^6\)\end{tabular} \\ \cline{2-9} 
 &
  2 &
  0.46\% &
  21.3\% &
  0.7\% &
  \multicolumn{1}{c|}{15.6x / 24.4x} &
  \begin{tabular}[c]{@{}c@{}}\(2.28\times10^6\) /\\\(3.56\times10^6\)\end{tabular} &
  \multicolumn{1}{c|}{44.6x / 69.7x} &
  \begin{tabular}[c]{@{}c@{}}\(6.50\times10^6\) /\\\(1.02\times10^7\)\end{tabular} \\ \cline{2-9} 
 &
  3 &
  1.2\% &
  73.8\% &
  1.3\% &
  \multicolumn{1}{c|}{63.9x / 124.8x} &
  \begin{tabular}[c]{@{}c@{}}\(5.55\times10^5\) /\\\(1.08\times10^6\)\end{tabular} &
  \multicolumn{1}{c|}{11296.3x / 22063x} &
  \begin{tabular}[c]{@{}c@{}}\(9.81\times10^7\) /\\\(1.92\times10^8\)\end{tabular} \\ \hline
\end{tabular}%
}
\end{table}

\subsection{Metrics to Evaluate Security}\label{section_metric}
Metrics that evaluate the security of a circuit, including the resource metric, attack performance metric, and attack evaluation metric, are provided in \cite{Attack_Metric}.
The resource metric (RM) measures the time required to gather all necessary information to perform an attack on an obfuscated analog circuit. The RM includes the number of steps needed to reverse engineer the circuit and obtain the netlist, to delayer and probe the circuit to determine the biasing information, to properly model the circuit, and to acquire PDK information.
The attack performance metric (APM) evaluates the execution time required to successfully carry out an attack, where a higher score is desired from the perspective of the defender. The APM includes the time needed to determine candidate keys and the time needed to execute a brute-force attack on the returned candidate keys to identify the correct key. The time to execute the SMT attack, genetic algorithm attack, monotonic attack, and DNA attack, is compared based on the attack performance metric score of each.
For circuits obfuscated on an FPAA fabric, the number of utilized configurable elements, including CABs, programmable transistor pairs, switch boxes, and bias voltages are used to determine the complexity of the implemented circuit topology.
% The attack evaluation metric (AM) combines the resource metric and the performance metric to assess the overall difficulty of executing an attack. AM ranks different attack techniques based on their cost, considering both the time required to obtain circuit information and the time needed to execute the attack.
The topology complexity parameter quantifies the structural search space provided by an obfuscated circuit. For circuits implemented using partial topology obfuscation, the number of obfuscated transistor pairs is used to determine the number of possible circuit topologies. Since programmable assignments in which switches \(S_1\) through \(S_5\), which are shown in Fig.~\ref{NP1_ARC} and Fig.~\ref{PP1_ARC}, are turned off produce structurally invalid circuits, such as isolated transistor pairs with floating gates, the topology complexity score is represented by lower and upper bounds rather than by a single value. For \emph{N} obfuscated transistor pairs, the upper bound is calculated as
\begin{align}
    TC_{UP}=5^N,
\end{align}
which represents the raw topology space of the TP array for the five programmable transistor pair topologies. A conservative lower bound is calculated as
\begin{align}
    TC_{LOW}=4^N,
\end{align}
which excludes structurally invalid assignments and accounts for only the four valid programmable transistor-pair topologies available. The valid topologies include the current mirror, regeneration loop, current source pair, and diode-connected pair.
Taking the topology-complexity bounds into account, the attack performance metric for a circuit implemented with partial topology obfuscation is given by
{\small
\begin{align}
    APM_{LOW} &= (T_{Attack}+NumKey\cdot T_{tran})\cdot TC_{LOW},~\text{and}\\
    APM_{UP}  &= (T_{Attack}+NumKey\cdot T_{tran})\cdot TC_{UP},
\end{align}
}where $T_{Attack}$ represents the time required to execute an attack, \emph{NumKey} represents the number of candidate keys returned by the attack algorithm, and $T_{tran}$ represents the average time required to perform transient analysis of the obfuscated analog circuit for each candidate key. 
Intuitively, the APM provides a practical means to quantify the cost of recovering the correct key as a single metric, which combines attack runtime, the cost of determining the correct key from the set of returned candidate-keys, and the search complexity due to the obfuscated topology of the circuit. The term $(T_{Attack}+NumKey \cdot T_{tran})$ captures the direct runtime and computational cost of the attack, including both the execution of the attack algorithm and the simulation-based validation of the returned candidate keys. The topology-complexity term $TC$ accounts for the number of circuit topologies that must be analyzed due to the obfuscation of the circuit. Therefore, a larger APM indicates that the attacker must spend more time, evaluate more candidate solutions, and overcome a larger topological search space to determine the correct key. In practice, increasing the tolerance of the target specifications improves the probability that the correct key is included in the set of returned candidate-keys; however, the number of candidate-keys returned increases, which results in a higher APM score.

The evaluated bounds of the attack performance metric for the folded-cascode op amp, comparator, and SAR ADC are listed in Table~\ref{table_attack_metric}.
The upper-bound of the programmable topology complexity metric is calculated as \(TC_{UP}=5^N\), with an entropy \(H_{UP}\) of \(\log_2 5^N=N\log_2 5\). After excluding structurally invalid topological assignments, a conservative lower bound for the functionally valid topology space is given by \(TC_{LOW}=4^N\), with an entropy \(H_{LOW}\) of \(\log_2 4^N=2N\). Therefore, the entropy of valid topologies lies in the range of \(2N \leq H_{\mathrm{valid}} \leq N\log_2 5\). The reported attack performance metric is consequently bounded by \(APM_{LOW}\) and \(APM_{UP}\), where the lower-bound value is smaller than the upper-bound value by a factor of \((4/5)^N\).

\begin{table}[htbp]
\setlength\tabcolsep{0.0035\textwidth}
\caption{Obfuscated Information When Implementing  Analog Security Techniques. The “*” Symbol Indicates this work.}
\label{table_obfus_info}
\begin{tabular}{|cc|cccccc|}
\hline
\multicolumn{2}{|c|}{\multirow{2}{*}{\begin{tabular}[c]{@{}c@{}}Obfuscation \\ Technique\end{tabular}}} &
  \multicolumn{6}{c|}{Obfuscated  Information} \\ \cline{3-8} 
\multicolumn{2}{|c|}{} &
  \multicolumn{1}{c|}{\begin{tabular}[c]{@{}c@{}}Num. of \\ Trans.\end{tabular}} &
  \multicolumn{1}{c|}{\begin{tabular}[c]{@{}c@{}}Size of\\ Trans.\end{tabular}} &
  \multicolumn{1}{c|}{\begin{tabular}[c]{@{}c@{}}Topology\\  of TP\end{tabular}} &
  \multicolumn{1}{c|}{\begin{tabular}[c]{@{}c@{}}Biasing \\ Voltage\end{tabular}} &
  \multicolumn{1}{c|}{\begin{tabular}[c]{@{}c@{}}Input \&\\ Output\\ Port\end{tabular}} &
  \begin{tabular}[c]{@{}c@{}}Topology \\ of\\ Circuitry\end{tabular} \\ \hline
\multicolumn{1}{|c|}{\multirow{2}{*}{\begin{tabular}[c]{@{}c@{}}TP \\ Topology\\ Obfus.*\end{tabular}}} &
  \begin{tabular}[c]{@{}c@{}}with \\ Dummy\end{tabular} &
  \multicolumn{1}{c|}{Yes} &
  \multicolumn{1}{c|}{Yes} &
  \multicolumn{1}{c|}{Yes} &
  \multicolumn{1}{c|}{No} &
  \multicolumn{1}{c|}{No} &
  Yes \\ \cline{2-8} 
\multicolumn{1}{|c|}{} &
  \begin{tabular}[c]{@{}c@{}}No \\ Dummy\end{tabular} &
  \multicolumn{1}{c|}{No} &
  \multicolumn{1}{c|}{Yes} &
  \multicolumn{1}{c|}{Yes} &
  \multicolumn{1}{c|}{No} &
  \multicolumn{1}{c|}{No} &
  Yes \\ \hline
\multicolumn{2}{|c|}{\begin{tabular}[c]{@{}c@{}}FPAA-based\\ Obfus.*\end{tabular}} &
  \multicolumn{1}{c|}{Yes} &
  \multicolumn{1}{c|}{Yes} &
  \multicolumn{1}{c|}{Yes} &
  \multicolumn{1}{c|}{Yes} &
  \multicolumn{1}{c|}{Yes} &
  Yes \\ \hline
\multicolumn{2}{|c|}{Vector-based} &
  \multicolumn{1}{c|}{No} &
  \multicolumn{1}{c|}{Yes} &
  \multicolumn{1}{c|}{No} &
  \multicolumn{1}{c|}{No} &
  \multicolumn{1}{c|}{No} &
  No \\ \hline
\multicolumn{2}{|c|}{Mesh-based} &
  \multicolumn{1}{c|}{No} &
  \multicolumn{1}{c|}{Yes} &
  \multicolumn{1}{c|}{No} &
  \multicolumn{1}{c|}{No} &
  \multicolumn{1}{c|}{No} &
  No \\ \hline
\end{tabular}
\end{table}

\vspace{-5pt}

\begin{table}[htbp]
\centering
\setlength\tabcolsep{0.0035\textwidth}
\caption{Lower-bound probability of recovering analog circuit configurations implemented on the FPAA fabric.}
\label{table_probability}
\begin{tabular}{|c|c|c|c|c|c|}
\hline
\begin{tabular}[c]{@{}c@{}}Circuit\\ Architecture\end{tabular} &
  \begin{tabular}[c]{@{}c@{}}Num. of \\ NP (\emph{N}\textsubscript{1})\end{tabular} &
  \begin{tabular}[c]{@{}c@{}}Num. of \\ PP (\emph{N}\textsubscript{2})\end{tabular} &
  \begin{tabular}[c]{@{}c@{}}Num. of\\ Biases (\emph{N}\textsubscript{DC})\end{tabular} &
  \begin{tabular}[c]{@{}c@{}}Resolution of\\  DAC (\emph{N}\textsubscript{DAC})\end{tabular} &
  Probability \\ \hline
\begin{tabular}[c]{@{}c@{}}Folded-Cascode\\ Op Amp\end{tabular} & 4   & 2   & 5 & 5-bit & 8.9 x 10\textsuperscript{-38}  \\ \hline
Comparator                                                      & 3   & 3   & 1 & 5-bit & 6 x 10\textsuperscript{-32}  \\ \hline
\begin{tabular}[c]{@{}c@{}}Delta-Sigma\\ Modulator\end{tabular} & 7 & 5 & 6 & 5-bit & 5.3 x 10\textsuperscript{-69} \\ \hline
\end{tabular}%
\end{table}

\begin{table*}[!ht]
\centering
\caption{Comparison of topology obfuscation with state-of-the-art analog and mixed-signal security techniques.}
\resizebox{\textwidth}{!}{%
\begin{tabular}{|c|c|c|c|c|c|c|c|}
\hline
Ref. &
Technique &
Obfuscated Info. &
Key Search Space &
Threats Discussed &
Wrong-key effect &
Overhead reported &
Benchmark circuits \\ \hline

\cite{Secure_vth} &
\makecell{Multi-$V_{TH}$\\ design} &
\makecell{Threshold voltage $V_{TH}$\\ (HVT, NVT, LVT)} &
\makecell{$3^{N}$\\ ($N$ protected devices)} &
\makecell{RE guess-and-validate,\\ brute-force RE effort} &
\makecell{Large performance degradation,\\ possible output clipping,\\ case dependent} &
\makecell{Device resizing overhead,\\ area-noise-security tradeoff} &
\makecell{Current mirror\\ Telescopic cascode op-amp} \\ \hline

\cite{Secure_keybased} &
\makecell{Key-based\\ parameter obfuscation} &
\makecell{Transistor sizing} &
\makecell{$2^{N}$\\ ($N$-bit sizing key)} &
\makecell{IP piracy,\\ brute-force, SMT-based\\ search} &
\makecell{Wrong key causes large\\ specification error after sizing} &
\makecell{Area and parasitic overhead,\\ BPF and op-amp spec drift\\ compensated by tuning} &
\makecell{Active-inductor BPF\\ Op-amp} \\ \hline

\cite{Secure_calibration} &
\makecell{Lock-less locking\\ by untuning} &
\makecell{Calibration settings} &
\makecell{$2^{N}$\\ ($N$-bit LUT key)} &
\makecell{RE, cloning, overbuilding,\\ remarking, recycling,\\ SAT, removal, brute-force,\\ optimization attack} &
\makecell{Complete loss of function\\ or significant degradation \\ on specification} &
\makecell{No AMS IC area or power\\ overhead, only key\\ management overhead} &
\makecell{Programmable AMS ICs\\ (calibration-driven)} \\ \hline

\cite{Table-Time} &
\makecell{Switch-mode\\ time-domain locking} &
\makecell{Time-domain\\ switching pattern} &
\makecell{$2^{N}$\\ ($N$-bit register key)} &
\makecell{IP piracy, counterfeiting,\\ RE, brute-force} &
\makecell{No meaningful operation,\\ almost 100\% variation} &
\makecell{Key-generation and control\\ support overhead} &
\makecell{Folded-cascode amplifier\\ SC bandgap reference} \\ \hline

\cite{Table-Random} &
\makecell{Randomized\\ obfuscation-circuit\\ insertion} &
\makecell{Circuit functional path} &
\makecell{$2.07e^{N}$\\ ($N$ biasing pin)} &
\makecell{Brute-force, removal, RE,\\ SMT, GA, BO,\\ Oracle-guided, Oracle-less} &
\makecell{Considerable performance\\ degradation for\\ wrong keys} &
\makecell{Inserted obfuscation circuit\\ and additional bias pins,\\ \textless 1.3\% area,\\ \textless 2.64\% power} &
\makecell{Bandgap filter\\ Op-amp\\ SAR ADC} \\ \hline

\cite{Table-FG-Security} &
\makecell{Floating-gate (FG)\\ FPAA embedded\\ security} &
\makecell{Floating-gate device mismatch, \\ threshold-voltage signature \\ PUF-like secure code} &
\makecell{Determined by FG-cell \\address and selection} &
\makecell{Physical attack,\\ transceiver port attack,\\ FPAA I/O attack,\\ host imitation,\\ always-requesting attack} &
\makecell{Wrong mismatch \\ based response, \\ wrong PUF structure} &
\makecell{FG devices, routing, \\ and shift-register \\ required by PUF} &
\makecell{Command-word classifier\\ Context-aware sensor nodes\\ Unique-function circuits} \\ \hline

\cite{Attack_FPAA} &
\makecell{FPAA fabric\\ topology obfuscation} &
\makecell{Transistor pair (TP)\\ topology} &
\makecell{$2^{N}$\\ ($N$-bit topology key)} &
\makecell{RE, topology attack,\\ brute-force} &
\makecell{Wrong configuration yields\\ wrong topology\\ or function} &
\makecell{Routing parasitic impedance,\\ 3-dB bandwidth drop of\\ about 2.3 to 3.84 GHz} &
\makecell{Op-amp\\ Biquad filter\\ Ring-oscillator\\ Frequency divider} \\ \hline

This Work &
\makecell{Hybrid ASIC-FPAA\\  partial/full\\ topology obfuscation} &
\makecell{TP topology and sizing\\ Circuit topology} &
\makecell{$2^{N+P}$\\ ($N$-bit sizing key,\\ $P$-bit topology key)} &
\makecell{SMT, GA, monotonic,\\ DNA, brute-force,\\ RE, topology attack} &
\makecell{Wrong mapping/sizing\\ breaks functionality or\\ causes strong degradation in specifications} &
\makecell{Circuit-dependent,\\ scales with number of obfuscated\\ transistor pairs and dummies} &
\makecell{Op-amp, Comparator\\ SAR ADC} \\ \hline

\end{tabular}%
}
\label{tab:sota_security}
\end{table*}

A figure of merit (FOM) given by
\begin{align}
    FOM = \frac{APM_{norm}}{\Delta_{Spec.}*\Delta_{Power}*\Delta_{Area}},
\end{align}
is utilized to quantify the trade-off between the performance and security robustness of a circuit, where \emph{APM}\textsubscript{norm} represents the attack performance metric (APM) normalized to the APM of the vector-based analog locking technique\cite{Attack_Metric}, \emph{$\Delta$\textsubscript{Spec.}} represents the deviation from the target specifications, which includes the GBW product of the folded-cascode op amp, the summation of the rise time and fall time of the comparator, and the ENOB of the SAR ADC,
\emph{$\Delta$\textsubscript{Power}} represents the overhead in power, and \emph{$\Delta$\textsubscript{Area}} represents the overhead in area. 
The calculated increase in APM and the corresponding FOM when executing a GA attack and a monotonic attack on analog circuits secured with one, two, and three obfuscated transistor pairs are listed in Table~\ref{table_FOM} as lower and upper bounds corresponding to \(TC_{LOW}=4^N\) and \(TC_{UP}=5^N\), respectively. Compared to obfuscating only the sizing of transistors, obfuscating the topology and sizing of one, two, and three transistor pairs improves the APM by at least \(4\times\), \(15.6\times\), and \(63.9\times\), respectively, while the corresponding upper-bound improves by \(5\times\), \(24.4\times\), and \(124.8\times\). For the folded-cascode op amp, obfuscating one transistor pair yields the highest calculated FOM when considering the GA attack, whereas obfuscating two transistor pairs yields the highest calculated FOM when considering the monotonic attack. For the comparator and SAR-ADC, obfuscating one transistor pair yields the highest FOM when evaluated for the GA attack, whereas obfuscating three transistor pairs yields the highest FOM for the monotonic attack.

\subsection{Security Robustness of FPAA-based Topology Obfuscation}\label{section_FPAA_robustness}
The FPAA-based obfuscation technique masks the entire target circuit topology within the fixed programmable fabric of the FPAA. The information secured when implementing the transistor pair (TP), vector, mesh, and FPAA key-based analog obfuscation techniques is listed in Table~\ref{table_obfus_info}. The FPAA secures circuit details including the number and size of transistors, the bias voltages and currents, the topology of transistor pairs, the location of input and output ports, and the overall circuit topology.
Without prior knowledge of the netlist of the target analog circuit, attacking the FPAA is equivalent to designing a circuit from scratch. The attacker must determine the number of transistors used, the routing connections between transistors, the topologies of programmed TPs,  and the sizes of the transistors.
In addition, even if the netlist of the target circuit is known, mapping the same circuit topology onto different PPs and NPs within a TP array results in deviations in the performance parameters of the circuit, as indicated by results shown in Fig.~\ref{FPAAOpampAC}, which further improves the security robustness of the implemented circuits.

To evaluate the security robustness of an analog circuit implemented on the FPAA fabric, a lower-bound probability of recovering the correct circuit configuration is calculated under a simplified assumption that only the keys used to program the transistor pairs and DC bias voltages are considered, while the keys associated with the switch boxes and the routing between the transistor pairs are ignored (assumed known). Under such assumption, the lower-bound probability is given by
\begin{align}
P_{FPAA} &= P_{PMOS}\cdot P_{NMOS}\cdot P_{Bias}\cdot P_{TP}\cdot P_{Width} \\
&= \frac{1}{C^{N_1}_{N_{FP1}} C^{N_2}_{N_{FP2}}}
\cdot \left(\frac{1}{2}\right)^{N_{DAC}\cdot N_{DC}} \notag\\
&\quad \cdot \left(\frac{1}{5}\right)^{N_1+N_2}
\cdot \left(\frac{1}{2^{12}}\right)^{N_1+N_2},
\end{align}
where \emph{P}\textsubscript{PMOS} and \emph{P}\textsubscript{NMOS} represent the probabilities of activating the correct PMOS and NMOS transistor pairs within the programmable transistor-pair array, respectively, \emph{N}\textsubscript{1} and \emph{N}\textsubscript{2} denote the number of PMOS and NMOS transistor pairs in the target analog circuit, respectively, and \emph{N}\textsubscript{FP1} and \emph{N}\textsubscript{FP2} denote the number of PMOS and NMOS transistor pairs available in the programmable transistor-pair array, respectively. Therefore, the probabilities of activating the correct PMOS and NMOS transistor pairs are given by the inverse binomial coefficients $C^{N_1}_{N_{FP1}}$ and $C^{N_2}_{N_{FP2}}$, respectively. \emph{P}\textsubscript{Bias} represents the probability of assigning the correct bias voltages generated by the digital-to-analog converters programmed on the FPAA, where \emph{N}\textsubscript{DAC} denotes the DAC resolution and \emph{N}\textsubscript{DC} denotes the number of DC bias voltages required by the programmed circuit. \emph{P}\textsubscript{TP} represents the probability of configuring the correct topology on each activated transistor pair, and \emph{P}\textsubscript{Width} represents the probability of assigning the correct unique 12-bit key that sets the width of each activated transistor pair.

The probability of deobfuscating the topology of a folded-cascode op amp, a comparator, and a delta-sigma modulator implemented on the FPAA is $8.9 \times 10^{-38}$, $6 \times 10^{-32}$, and $5.3 \times 10^{-69}$, respectively,  as listed in Table~\ref{table_probability}. Note that for the delta-sigma modulator that is obfuscated across two CABs, the probability is calculated as the product of the probability of determining each sub-circuit implemented within separate individual CABs.

%% file: section9.tex
\section{Comparison with Prior Work on Analog Security }\label{section-comparison-table}
A comparison with prior analog security techniques is provided in Table~\ref{tab:sota_security}. As compared to multi-threshold-voltage obfuscation \cite{Secure_vth} and key-based parameter obfuscation \cite{Secure_keybased}, which preserve the original topology while masking device parameters, the proposed method secures the structural implementation of the circuit at the transistor-pair level. As compared to calibration-based locking \cite{Secure_calibration}, the proposed technique is applicable to AMS circuits and does not require a calibration algorithm. As compared to switch-mode time-domain locking \cite{Table-Time}, the proposed technique provides topology-level obfuscation of the analog circuit itself, rather than relying on hidden time-domain switching patterns during switch-mode operation of the circuit.
In \cite{Table-Random}, the obfuscation circuit is inserted into the analog current path, whereas the proposed technique avoids modifying the original signal path. As a result, the proposed technique preserves the structure of the original signal-path and avoids the need to optimize an added obfuscation block to maintain the original functionality. The FG-FPAA embedded security technique described in \cite{Table-FG-Security} highlights the opportunities and vulnerabilities of ultra-low-power SoC FPAAs that utilize floating-gate-mismatch to generate unique functions that secure embedded circuit operation. In this work, a hybrid ASIC-FPAA topology-obfuscation framework that masks the topology of the transistor pairs, the sizing of devices, the circuit biasing, and transistor-level circuit mapping is evaluated quantitatively.
As compared to the FPAA-based obfuscation described in \cite{Attack_FPAA}, the proposed technique extends topology obfuscation from CAB-level mapping to a fine-grained transistor-level security methodology, which is further strengthened through the use of dummy transistor pairs and partial and full topology obfuscation. Another strength of the proposed methodology is that the security provided by the ASIC-FPAA technique is evaluated across a broader set of analog deobfuscation attacks, which includes the topology, SMT, GA, monotonic, and DNA attacks. In addition, the resulting performance-security tradeoff across multiple benchmark circuits is also characterized. The primary limitation is that the proposed circuit obfuscation technique requires a programmable fabric and introduces circuit-dependent overhead, with an increasing degradation in performance as a greater percentage of the circuit is obfuscated, which is particularly true for full-topology obfuscation.

\section{Conclusion}\label{HH}
Obfuscation techniques are proposed to secure the topology of an analog circuit and protect against reverse engineering, IC piracy, and counterfeiting. The developed partial topology obfuscation technique secures the structure of the selected transistor pairs and masks the number of utilized transistor pairs with dummy transistor pairs.
The proposed full topology obfuscation technique masks the entire structure of an analog circuit within the fixed fabric of the FPAA, which provides the strongest security guarantees possible. The implementation of performance-driven obfuscation is characterized on a folded-cascode op amp, a StrongARM comparator, and a 9-bit SAR ADC, while the implementation of full topology obfuscation is characterized on a folded-cascode op amp and a delta-sigma modulator. The security robustness is evaluated through the latest analog attack algorithms, which includes the topology attack,  SMT attack,  GA attack,  monotonic attack, and the DNA attack.
In addition, the complexity of the programmed circuit is evaluated through a topology attack, which indicates a probability of deobfuscation ranging from $5.3 \times 10^{-69}$  to $6 \times 10^{-32}$.
The attack performance metric is developed to quantify the security robustness. When one, two, and three transistor pairs are obfuscated, a minimum increase of 4x, 15.6x, and 63.9x in the APM score is observed, respectively, as compared to the APM score of vector-based locking.